\documentclass[11pt]{article}
\usepackage{jcappub}
\usepackage{array}
\usepackage{graphicx} 
\usepackage{amsmath}
\usepackage{amssymb}
\usepackage{subcaption}
\usepackage{xcolor,multirow}
\usepackage{float} 
\newcommand{\CLASS}{\texttt{CLASS}}

\title{\boldmath Dark Matter-Radiation Scattering Enhances CMB Phase Shift through Dark Matter-loading
}

\author[a,b]{Subhajit Ghosh,}
\author[b]{Daven Wei Ren Ho,}
\author[b]{and Yuhsin Tsai}

\affiliation[a]{Texas Center for Cosmology and Astroparticle Physics, Weinberg Institute,
Department of Physics, The Unversity of Texas at Austin, Austin, TX 78712, USA}
\affiliation[b]{Department of Physics and Astronomy, University of Notre Dame, IN 46556, USA}
\emailAdd{sghosh@utexas.edu}
\emailAdd{dho2@nd.edu}
\emailAdd{ytsai3@nd.edu}
\preprint{UT-WI-17-2024}

\abstract{A phase shift in the acoustic oscillations of cosmic microwave background (CMB) spectra is a characteristic signature for the presence of non-photon radiation propagating differently from photons, even when the radiation couples to the Standard Model particles solely gravitationally. It is well-established that compared to the presence of free-streaming radiation, CMB spectra shift to higher $\ell$-modes in the presence of self-interacting non-photon radiation such as neutrinos and dark radiation. In this study, we further demonstrate that the scattering of non-photon radiation with dark matter can further amplify this phase shift. We show that when the energy density of the interacting radiation surpasses that of interacting dark matter around matter-radiation equality, the phase shift enhancement is proportional to the interacting dark matter abundance and remains insensitive to the radiation energy density. Given the presence of dark matter-radiation interaction, this additional phase shift emerges as a generic signature of models featuring an interacting dark sector or neutrino-dark matter scattering. Using neutrino-dark matter scattering as an example, we numerically calculate the amplified phase shift and offer an analytical interpretation of the result by modeling photon and neutrino perturbations with coupled harmonic oscillators. This framework also explains the phase shift contrast between self-interacting and free-streaming neutrinos. Fitting models with neutrino-dark matter or dark radiation-dark matter interactions to CMB and large-scale structure data, we validate the presence of the enhanced phase shift, affirmed by the linear dependence observed between the preferred regions of the sound horizon angle $\theta_s$ and interacting dark matter abundance. An increased $\theta_s$ and a suppressed matter power spectrum is therefore a generic feature of models containing dark matter scattering with abundant dark radiation.
}

\begin{document}

\maketitle

\section{Introduction}

Cosmology provides a distinct opportunity to investigate new physics with minimal interactions with Standard Model (SM) particles. Whether the interaction is solely gravitational, the existence of new particles with substantial energy density in the early universe can modify the anisotropy of photons and baryons. This impact on the cosmic microwave background (CMB) and large-scale structure (LSS) provides valuable insights into the propagation and clustering behavior of these particles, shedding light on their interactions. 

The ability to probe interactions between particles that couple weakly to the SM is crucial in studying neutrino and dark sector physics. It is shown that an efficient neutrino self-scattering could persist during the CMB time to modify temperature and polarization perturbations while satisfying existing bounds from collider searches~\cite{Kreisch:2019yzn,Lancaster:2017ksf,Oldengott_2017,Blinov_2019,Brinckmann:2020bcn}.
Dark sector models containing new massless non-abelian gauge bosons \cite{Buen-Abad:2015ova,Lesgourgues:2015wza,Buen-Abad:2017gxg} or dark photons interacting with other massless dark sector particles~\cite{Chacko:2016kgg} provide candidates for interacting DR that can also change the CMB perturbations. The idea of self-interacting neutrinos and dark radiation has drawn much attention in addressing the $H_0$ problem~\cite{Lancaster:2017ksf,Brust:2017nmv,Blinov_2019,Kreisch:2019yzn,RoyChoudhury:2020dmd,Oldengott_2017,Ghosh:2021axu,Brinckmann:2022ajr}

A phase shift in the acoustic oscillations of CMB spectra, compared to free-streaming radiation scenarios, is a generic signature of these interacting radiation models~\cite{Bashinsky:2003tk,Baumann:2015rya,Pan:2016zla,Baumann:2017lmt,Choi:2018gho,Ghosh:2019tab,Ghosh:2021axu,Baumann:2017gkg,Baumann:2019keh,Green:2020fjb,Green:2019glg,Gerbino:2022nvz,Berryman:2022hds,Dvorkin:2022bsc}
Fluctuations in the radiation background, whether arising from cosmic neutrinos or dark radiation, are known to induce phase shifts in the acoustic peaks of the CMB. Compared to  $\Lambda$CDM model without neutrinos, the presence of free-streaming neutrinos (or additional free-streaming dark-radiation) shifts the phase towards lower $\ell$ modes, and the effect has been observed in the Planck data~\cite{Follin:2015hya,Baumann:2015rya}. On the contrary, the acoustic peaks experience a shift towards higher $\ell$ modes when the neutrino or additional dark radiation propagates as a perfect fluid. As discussed in~Ref.~\cite{Baumann:2015rya}, this phase shift in the CMB spectrum can only come from two origins: either due to the different propagation sound speeds of radiations or due to radiations carrying isocurvature perturbations. In models with adiabatic perturbations, the limited origins of the phase shift make it a distinct signature for identifying radiations with exotic propagation properties. 

Beyond self-interactions, both dark radiation and neutrino exhibit delayed propagation in presence of scattering with dark matter particles. Such interactions exist in scenarios like interacting dark matter~\cite{Buen-Abad:2015ova,Lesgourgues:2015wza,Chacko:2016kgg,Raveri:2017jto,Buen-Abad:2017gxg,Rubira:2022xhb,Buen-Abad:2023uva} and atomic dark matter~\cite{Kaplan:2009de,Kaplan:2011yj,Cline:2012is,Cyr-Racine:2012tfp,Cyr-Racine:2013fsa,Fan:2013tia,Chacko:2018vss,Curtin:2020tkm,Bansal:2021dfh, Cyr-Racine:2021oal, Blinov:2021mdk,Bansal:2022qbi,Zu:2023rmc,Greene:2023cro,Hughes:2023tcn,Greene:2024qis} models, often proposed to address issues such as the $H_0$~\cite{Riess:2021jrx} and $S_8$~\cite{Hildebrandt:2018yau,MacCrann:2014wfa,Joudaki:2016mvz,Joudaki:2019pmv,KiDS:2020suj,KiDS:2021opn,Heymans:2020gsg,DES:2021wwk,Philcox:2021kcw,Abdalla:2022yfr} tensions, the Higgs hierarchy problem, and similarity between the baryon and dark matter abundance~\cite{Foot:2002iy,Foot:2003jt}. Similar setups are considered for neutrino models featuring dark matter scatterings to also tackle challenges like the Hubble tension and small-scale structure anomalies~\cite{Wilkinson_2014,Escudero:2015yka,DiValentino:2017oaw,Ghosh:2019tab,Akita:2023yga}. While existing cosmological discussions on these dark sector models predominantly focus on signals such as additional light degrees of freedom $\Delta N_{\rm eff}$ or matter power spectrum suppression due to the dark acoustic oscillation process~\cite{Cyr-Racine:2013fsa,Buckley:2014hja}, we emphasize that the propagation of interacting radiation within the dark matter medium also presents another distinctive signal in the form of CMB phase shift. 
    
In this study, we demonstrate that a loading effect from dark matter (DM) on radiation propagation, akin to the baryon-loading effect that slows down the photon sound speed, results in a positive shift in $\ell$-modes  and surpasses that of self-interacting radiation scenario. As we show, the phase shift corresponds to a larger sound horizon angle $\theta_s$ from fitting the CMB data. An observation of non-zero $\Delta N_{\rm eff}$, a suppressed matter power spectrum ($S_8$), and a larger $\theta_s$ compared to the fit from the self-interacting radiation model will be strong evidence of the presence of DM-dark radiation scattering. 
While previous studies such as Ref.~\cite{Ghosh:2019tab,Bansal:2021dfh} have discussed the presence of a phase shift in models with radiation-DM scatterings, the contribution from DM-loading, leading to an additional phase shift in these models, has been overlooked. Here, we use the interacting neutrino scenario to observe the extra phase shift due to the DM-loading effect, and the discussions on the parametric dependence of the signal can be extended to scenarios with dark radiation-DM interactions.

By comparing results from full numerical calculations and semi-analytical approximations, we demonstrate that the phase shift between interacting neutrinos with and without DM-loading, referred to as DM-loading neutrinos (DL-$\nu$) and self-interacting neutrinos (SI-$\nu$)~\footnote{In this work, by SI-$\nu$ we mean all the neutrinos are coupled till today. This is equivalent to neutrinos behaving as perfect-fluid. Although, if the neutrinos are coupled only till recombination we get a very similar effect on CMB compared with SI-$\nu$. In the literature, SI-$\nu$ sometimes refers to `Strongly-Interacting Neutrino mode' where, due to finite interaction strength, neutrinos are only coupled approximately till matter-radiation equality~(e.g. Ref.~\cite{Lancaster:2017ksf})}, can be approximated by treating photon and neutrino perturbations as harmonic oscillators coupled via gravity perturbations. This approximation reveals that the observed phase shift arises from the slowing down of the neutrino sound speed due to DM-loading, with the shift's magnitude proportional to the scattering dark matter abundance. Detecting this additional phase shift provides a means to measure the DM-loading effect and uncover the medium effect on neutrino and dark radiation propagations. The analytical discussion of the DL-$\nu$ phase shift can also be applied to the phase difference between SI-$\nu$ and free-streaming neutrino (FS-$\nu$), offering a simple picture of coupled harmonic oscillators to explain the origin of the phase shift.

The paper is organized as follows. In Sec.~2, we introduce the interacting neutrino scenario with DM-loading and define its parameters. In Sec.~3, using calculations from the Cosmic Linear Anisotropy Solving System (\CLASS)~\cite{Diego_Blas_2011,lesgourgues2011cosmic}, we illustrate examples of the phase shift between DL-$\nu$ and SI-$\nu$ scenarios in the CMB TT and EE spectra. We demonstrate the linear dependence of the phase shift on the interacting dark matter abundance and its insensitivity to the interacting neutrino abundance. Sec.~4 explains how the evolutions of photon and neutrino perturbations can be described by coupled harmonic oscillators, providing insights into the origin and parametric dependence of the phase shift. We discuss the generalization of the results to models with $N$ secluded dark sectors with interacting DR. In Sec.~5, we conduct a Markov Chain Monte Carlo (MCMC) analysis, demonstrating that the fitting result with current cosmological data does reveal a phase shift in the form of the sound horizon angle $\theta_s$ that linearly increases with the scattering dark matter abundance. We conclude in Sec.~6.

\section{Interacting radiation with DM-loading}\label{sec.param}
In general, a relativistic fluid can be formed in the dark sector from the efficient scattering of dark radiation (DR) with dark matter. We aim to study how the presence of dark matter interaction in such a fluid before recombination can produce an observable phase shift in the CMB power spectrum, and derive a simple mechanism that captures the physics behind this DM-loading effect.

As a concrete example of DM-loading effects, we study the DM-$\nu$ interaction. Massless neutrinos act identically as free-streaming DR as far as cosmological observables are concerned. Therefore, for concreteness, we begin the discussion by focusing on a particular example where neutrinos play the role of DR and introduce interaction with a fraction of dark matter.
Cosmology provides the best constraints on the DM-$\nu$ interaction parameter for asymmetric dark matter models through its effects on the CMB and matter power spectrum. In the following, we will summarize the essential features of the interaction model. 

The DM-loading effects are also present in DM-DR scattering. As we will see later, the effects of phase shift due to DM-loading are independent of the DR energy density. Thus the effects are visible in the CMB even when a small amount of DR interacts with the dark matter. Due to their similarities, the expressions given below for DM-$\nu $ interactions are also applicable to DM-DR interactions as well.  

\subsection{Models with DM-radiation interactions}
\begin{figure}[t!]
    \centering
    \includegraphics[width=\linewidth]{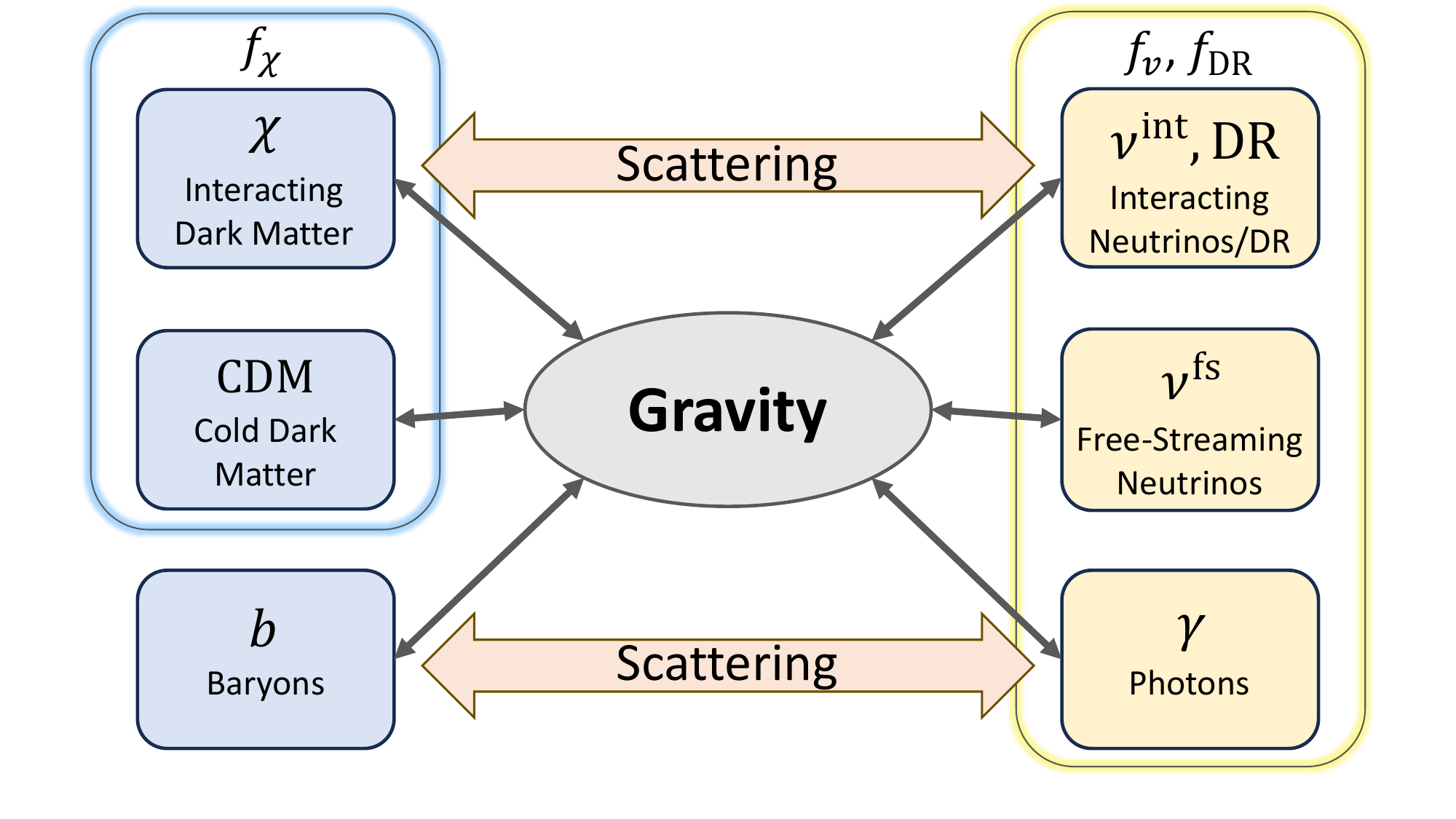}
    \caption{Schematic of multi-component dark matter and radiation system with interactions shown. The interacting dark matter fraction $f_\chi$ of the total DM, and interacting neutrino (dark radiation) fraction $f_\nu$ ($f_{\rm DR}$) of the total radiation are denoted.}
    \label{Multi}
\end{figure}

Consider multi-component dark matter and neutrino sectors, where a fraction of the neutrinos $\nu$ scatters with a fraction of the dark matter $\chi$, with all remaining dark matter as cold dark matter (CDM) and all remaining neutrinos free-streaming. The multi-component neutrino sector serves as a proxy for either scattering DR in the presence of free-streaming neutrinos or flavor-specific DM-Neutrino interaction. The fraction of interacting dark matter is quantified by the parameter $f_\chi = \rho_\chi/\rho_\mathrm{DM}$ where $\rho_\mathrm{DM} = \rho_\chi + \rho_\mathrm{CDM}$, while the fraction of interacting neutrinos that participate in the scattering (out of the total radiation energy) is denoted by $f_\nu = \rho_{\nu}^{\mathrm{int}}/(\rho_\gamma + \rho_\nu)$  where $\rho_\nu = \rho_{\nu}^{\mathrm{int}} + \rho_{\nu}^{\mathrm{fs}}$ is the total neutrino density consisting of both scattering and free-streaming neutrinos. The strength of the coupling between $\chi$ and the interacting neutrinos $\nu$ is quantified by the interaction parameter $y$, which will be defined below. A schematic of this multi-component system is provided in Fig.~\ref{Multi}.  

In the case of multicomponent DR-DM interaction, $f_\chi$ still denotes the fraction of total dark matter interacting with DR. The parameter $f_\nu$, in that case, will denote $f_{\rm DR}$ which is the ratio of the scattering DR energy density to the total radiation energy density. In summary,
\begin{equation}
    \label{eq:f_def}
    f_\chi = \cfrac{\rho_\chi}{\rho_{\rm DM}},\quad f_\nu = \cfrac{\rho_{\nu}^{\mathrm{int}}}{\rho_{\rm rad,tot}} = 
    \begin{cases}
        \cfrac{\rho_{\nu}^{\mathrm{int}}}{\rho_\gamma + \rho_{\nu}^{\mathrm{int}} + \rho_{\nu}^{\mathrm{fs}}}~\hspace{4mm}: ~\nu-{\rm DM}~\\
        \cfrac{\rho_{\rm DR}^{\mathrm{int}}}{\rho_\gamma + \rho_{\nu}^{\mathrm{fs}} + \rho_{\rm DR}^{\mathrm{int}}}~\hspace{4mm}: ~{\rm DR-DM}
    \end{cases}
\end{equation}





Many beyond the Standard Model (BSM) scenarios involve interactions between dark matter and SM neutrinos as has been discussed in the introduction. If such $\nu$-DM  scattering significantly alters neutrino propagation for redshift $z \lesssim 10^5$, it can imprint a DM-loading effect on the CMB spectrum. Our cosmological analysis will maintain a general approach to $\nu$-DM   scattering, avoiding focus on a particular BSM model. In Appendix.~\ref{app.model}, we present a specific example of $\nu$-DM scattering induced by a dimension-five operator $({\bar L \tilde{H} \psi \chi})/{\Lambda}$~\cite{Ghosh:2017jdy,Ghosh:2019tab}, discussing the necessary mass and coupling for efficient scattering, along with associated collider constraints. 

The $\nu$-DM scattering rate can depend differently on the neutrino temperature $T_\nu$. For concreteness, let us consider the case where the velocity averaged scattering cross-section $\sigma=\langle \sigma_{\chi\nu} v \rangle$ is independent of $T_\nu$~\cite{Wilkinson_2014,Ghosh:2017jdy}. As is discussed in Appendix~\ref{app.model}, the scenario can exist if the mass difference between dark matter and the mediator that generates the scattering is smaller than the neutrino temperature. In this case, the comoving $\nu$-DM scattering rate (equivalently for DR-DM scattering) can be parameterized as, 
\begin{equation}
    \label{eq:Gamma}
   \dot\kappa_{\rm DR-DM}\equiv a\,\sigma f_\chi\left(\frac{\rho_{\rm DM}}{m_\chi}\right)=2.4\cdot10^{-2}\,y\left(\frac{f_\chi}{0.01}\right) \left(\frac{\omega_{\rm DM}}{0.12}\right)\left(\frac{z}{4\cdot10^4}\right)^2\,{\rm Mpc}^{-1}\,,
\end{equation}
where the interaction parameter $y$ is defined as,
\begin{equation}
    \label{eq:u_def}
    y = \left(\frac{\sigma}{{\rm GeV}^{-2}}\right)\left(\frac{{\rm GeV}}{m_\chi}\right)\,.
\end{equation}
This is related to the $u_\chi$ parameter often used in the dark matter neutrino interaction literature by $y= 17.08u_\chi$~\cite{Wilkinson_2014,Escudero:2015yka,DiValentino:2017oaw,Ghosh:2017jdy,Ghosh:2019tab,Mosbech:2020ahp}. CMB power spectrum at $\ell\sim 10^3$ is mainly supported by perturbation modes with comoving wavenumbers $k_{\ell\sim10^3}\sim 0.1$~Mpc$^{-1}$, which enter horizon at comoving time $k_{\ell\sim10^3}^{-1}$ around $z=4\cdot10^4$. The scattering remains efficient, meaning $\dot{\kappa}_{\rm DR-DM} k_{\ell\sim10^3}^{-1}\gtrsim1$, if the cross-section and dark matter mass are below the GeV scale. 

When discussing cosmological perturbations, we work in the conformal Newtonian gauge where $\psi$ and $\phi$ characterize the two scalar perturbations on the background metric~\cite{Ma:1995ey}
\begin{equation}
	ds^2 = a^2(\tau) [ - ( 1 + 2 \psi ) d \tau^2 + ( 1 - 2 \phi ) \delta_{ij} dx^i dx^j ]  \; ,
\end{equation}
where $\tau$ denotes conformal time. We use $\mathcal{H}$ to denote the Hubble parameter in $\tau$. The Boltzmann equations describing the motion of the interacting $\nu$ (will be denoted by the subscript DR since those are equivalent systems) and $\chi$ components resemble that of the photons and baryons in fiducial cosmology. 
\begin{eqnarray}
\dot\delta_{{\rm DR}}+\frac{4}{3}\theta_{{\rm DR}}-4\dot\phi&=&0\,,
\\
\dot{\theta}_{\rm DR}+k^2(\sigma_{\rm DR}-\frac{1}{4}\delta_{\rm DR})-k^2\psi&=&\dot\kappa_{{\rm DR}-{\rm DM}}(\theta_{{\rm DR}}-\theta_\chi)\,,
\\
\dot{F}_{{\rm DR},\ell}+\frac{k}{2\ell+1}((\ell+1)F_{{\rm DR},\ell+1}-\ell F_{{\rm DR},\ell-1})&=&\alpha_\ell\dot{\kappa}_{{\rm DR}-{\rm DM}}F_{{\rm DR},\ell}\,,\quad(\ell\geq2)
\end{eqnarray}
The overhead dot $( ^{\cdot})$ represents the derivative with respect to conformal time ${d} / {d\tau}$. $\delta_{\rm DR} \equiv \delta \rho_{\rm DR}/\bar{\rho}_{\rm DR}\,$, $\theta_{\rm DR}\equiv  \partial_i v^i_{\rm DR} $, and $\sigma_{\rm DR}$ are the density perturbation, velocity divergence, and shear stress of DR, respectively. $F_{{\rm DR},\ell}$ is the $\ell$th moment of DR perturbation. For the example of the temperature-independent cross-section $\sigma$ that we are considering, the scattering rate $\dot{\kappa}_{\rm DR-DM}$ is given in Eq.~(\ref{eq:Gamma}), and $\alpha_{\ell\geq 2} = 1$ if the scattering comes from a tree level scalar mediator. On the other hand, the Boltzmann equations governing the interacting dark matter perturbations are
\begin{align}
	\dot{\delta}_{\chi} + \theta_{\chi} - 3 \dot{\phi} &= 0, \\
	\dot{\theta}_{\chi} + \frac{\dot{a}}{a}\theta_{\chi} - c^2_{\chi} k^2 \delta_{\chi}  - k^2 \psi &= 
	- \frac{4 \rho_{\rm DR}}{3 \rho_{\chi}} \dot{\kappa}_{\rm DR-DM} (\theta_{\chi} - \theta_{\rm DR} )\,.
\end{align}
where $c_{\chi} \ll 1$ represents the adiabatic sound speed of the non-relativistic $\chi$. \CLASS~computes $c_{\chi}$ when generating power spectra, but we do not consider this sound speed in the semi-analytical discussions in Sec.~\ref{sec.analy}. 

The cosmologically relevant quantity from the DR-DM interaction model is the scattering rate which appears in the Boltzmann equations above. Similar to the photon-baryon interaction in $\Lambda$CDM, this governs the rate of momentum transfer between particles and is independent of the attractive or repulsive nature of the interaction\footnote{For the particular DL-$\nu$ interaction model considered in Appendix.~\ref{app.model}, the interaction rate would be proportional to the fourth power of the coupling and would thus be insensitive to the sign of coupling.}. The DR-DM scattering is efficient to modify the DR perturbation when $\tau \dot{\kappa}_{\rm DR-DM} \gtrsim 1$. In the \emph{tightly coupled} regime $\tau \dot{\kappa}_{\rm DR-DM} \gg 1$, scattering events occur frequently enough compared to the Hubble rate to maintain momentum equilibrium between the radiation and dark matter, so that the radiation and dark matter perturbations have the same velocity divergence and propagate together as a fluid. More specifically, the slip term $\dot\kappa_{{\rm DR}-{\rm DM}}(\theta_{{\rm DR}}-\theta_\chi)$ dominates the $\theta_{\rm DR}$ equation, such that $\theta_{\rm DR}\approx\theta_\chi$ and the higher $\ell\geq 2$ moments of the DR perturbation get turned off. As in the standard treatment of baryon-photon plasma, we can approximate the DR-DM system in the tightly coupled regime by re-writing the perturbation equations into a second-order DR equation~\cite{2020moco.book.....D}
\begin{align}
\ddot{\delta}_{\rm DR} + \mathcal{H}\frac{R_{\rm DR}}{1+R_{\rm DR}}\dot{\delta}_{\rm DR} + \frac{k^2}{3(1+R_{\rm DR})}\delta_{\rm DR} = 4\ddot{\phi} + 4\mathcal{H}\frac{R_{\rm DR}}{1+R_{\rm DR}}\dot{\phi} - \frac{4k^2}{3}\psi\,.
\end{align}
Here $R_{\rm DR} = 3\rho_\chi/4\rho_{\rm DR}$, and the frequency divided by $k$ can be identified as the sound speed of tightly coupled fluid
\begin{equation}
c_{\rm DR}^2 = \frac{1}{3(1+R_{\rm DR})}\,.
\label{eqn:cs}
\end{equation}
Just as the presence of baryons in the photon-baryon plasma led to a baryon loading effect that suppressed the photon sound speed over time (with an increasing $R_\gamma$), there is an analogous suppression effect in the DL-$\nu$ or DL-DR fluid due to the presence of $\rho_\chi$. We therefore refer to the $R_{\rm DR}$ suppression of the DR sound speed as \textit{dark-matter loading}. As a reference case, we will also consider the self-interacting SI-$\nu$ and SI-DR cases which, in the efficient scattering regime, can be interpreted as the $f_\chi \rightarrow 0$ limit of the DL-DR fluid with negligible DM-loading.



In Table~\ref{tab:models}, we summarize the models that are considered in this paper. Models 2-5 describe scenarios with non-photon radiation scattering. In the SI-x cases (models 2 and 4) the scattering is coming only from the self-interaction of the radiation, while in the DL-x cases (models 3 and 5) it is coming only from scattering with DM. These are compared against model 1 when finding the phase shift with respect to the base $\Lambda$CDM model with all $\nu$ free-streaming. When isolating the phase shift enhancement due specifically to the DM-loading effect in the scattering radiation  fluid, we compare model 3 against model 2 for the DL-$\nu$ case and model 5 against model 4 for the DL-DR case.

For the DL-$\nu$ discussion, we will first focus on very strongly interacting DL-$\nu$ $\dot\kappa_{\rm DR-DM} \gg aH$ for the entirety of the neutrino evolution. Later, in the MCMC section, we will relax this assumption and allow $y$ (and hence $\dot\kappa_{\rm DR-DM}$) to vary. However, as will be seen later, the effects of the DM-loading are still visible with DL-$\nu$ decoupling at a later time. We also show results for the DL-DR scenario from the MCMC study. An efficient DR-DM scattering is easier to model build since all the dynamics occur in the dark sector, and there is no coupling with SM besides gravity.

\begin{table}
    \centering
    \begin{tabular}{|c|c|c|c|c|c|}\hline
          & Model&   Dark Matter&Neutrinos&  DR ($\Delta N_{\rm eff}$)&  Radiation Scattering\\\hline
          1&$\Lambda$CDM &   CDM&Free-Streaming&  --&  --\\\hline
          2&SI-$\nu$&   CDM&Int($f_\nu$) + FS&  --&  $\nu$ self-interaction\\
          3&DL-$\nu$&   Int($f_\chi$) + CDM&Int($f_\nu$) + FS&  --&  $\nu$ interacting with DM\\\hline
          4&SI-DR&   CDM&Free-Streaming&  Int($f_{\rm DR}$)&  DR self-interaction\\
          5&DL-DR&   Int($f_\chi$) + CDM&Free-Streaming&  Int($f_{\rm DR}$)&  DR interacting with DM\\\hline
    \end{tabular}
    \caption{Summary of models, parameterized by the $f_\chi$ and $f_\nu$ (or $f_{\rm DR}$) fractions. Here, we denote Int = ``Interacting" and FS = ``Free-Streaming". Variations in the interaction parameter $y$ will be studied in Sec.~\ref{sec:MCMC-NUDM} for DL-$\nu$ with later time decoupling.  This work mainly focuses on the enhanced phase shift between models 3 and 2, as well as between models 5 and 4. In general matter-loading effects will enhance phase-shift where radiation is scattering with non-relativistic species compared to self-scattering radiation with equivalent interaction strength.}
    \label{tab:models}
\end{table}

\subsection{Implementation in \CLASS}


For numerical calculations, we use the built-in interacting DM-DR (\textbf{`idm\_idr'}) module in \CLASS~to implement both the DL-$\nu$ and DL-DR scenarios. This implementation is based on an effective theory of structure formation, known as ETHOS~\cite{Cyr-Racine:2015ihg}. 

We make use of the following \textbf{`idm\_idr'} parameters: Firstly, the \textbf{f\_idm} parameter is the energy density ratio of the interacting dark matter component out of the total DM, which corresponds directly to the dark matter fraction $f_\chi$ defined in Eq.~(\ref{eq:f_def}). In terms of \CLASS~variables and using the ETHOS parametrization, \textbf{N\_ur} is the number of free-streaming radiation species, which is equal to the number of neutrinos in the base $\Lambda$CDM cosmology, while \textbf{N\_idr} is the amount of interacting DR. We use the \textbf{N\_idr} parameter to denote the interacting neutrino species for the DL-$\nu$ model and the interacting DR in the DM-DR model. Since the radiation species in both cases do not have self-interaction, we set the self-interaction parameter \textbf{b\_idr} = 0. We treat the neutrinos to be massless and, therefore, they can be described by either \textbf{N\_ur} or \textbf{N\_idr} when they have interaction. Since in the DL-$\nu$ case neutrino free-streaming properties get modified when neutrinos are relativistic, neutrino mass has a negligible effect on the mechanism that we will be discussing.  To implement the DL-$\nu$ system, we fix \textbf{N\_idr} + \textbf{N\_ur} = 3.046, the total number of neutrinos\footnote{Recent studies have found the contribution of the SM neutrinos to the relativistic degrees of freedom to be 3.044~\cite{Gariazzo:2019gyi,Bennett:2020zkv,Froustey:2020mcq,Akita:2020szl}. This difference will have no impact on our results.}. The ratio of interacting neutrinos out of the total radiation energy is then parameterized by 
\begin{equation}
\label{eq:fnu}
  f_\nu = \left(\frac{N\_\mathrm{idr}}{N\_\mathrm{idr}+N\_\mathrm{ur}}\right)\frac{\rho_{\nu,{\rm total}}}{\rho_\gamma+\rho_{\nu,{\rm total}}} \,,
  \end{equation}
where $\rho_{\nu,{\rm total}}/(\rho_\gamma+\rho_{\nu,{\rm total}})= 0.41$ for \textbf{N\_idr} + \textbf{N\_ur} = 3.046. For the subsequent implementation of the DL-DR system, the DR fraction is defined as
\begin{equation}
\label{eq:fdr}
  f_{\rm DR} = \left(\frac{N\_\mathrm{idr}}{N\_\mathrm{idr}+3.046}\right)\frac{\rho_{\rm DR}+\rho_{\nu,{\rm total}}}{\rho_\gamma+\rho_{\rm DR}+\rho_{\nu,{\rm total}}} \, ,
  \end{equation} 
where the SM free streaming neutrinos are fixed to \textbf{N\_ur} = 3.046. The \textbf{a\_idm\_dr} parameter in ETHOS corresponds to the scattering cross-section between the interacting radiation and dark matter components in both DL-$\nu$ and DM-DR. In the discussions of Sec.~\ref{sec.result} and~\ref{sec.analytical}, we simply set the \textbf{a\_idm\_dr} parameter to a very large value $10^4 - 10^6$ to ensure efficient scattering for all time with no decoupling. For the subsequent MCMC analysis in Sec.~\ref{sec.MCMC}, we also consider varying \textbf{a\_idm\_dr} where the DL-$\nu$ system decouples before recombination. 
We focus on the example with temperature-independent scattering cross section and map the ETHOS parameters to Eqs.~\eqref{eq:Gamma} and \eqref{eq:u_def} as
\begin{equation}
\label{eq:ethosmap}
{\rm {\textbf{a\_idm\_dr}}} = 1.13\times 10^6 \,y ,\quad {\rm{\textbf{ nindex\_idm\_dr}}} = 2,\quad {\rm {\textbf{alpha\_idm\_dr}}} = 1.
\end{equation}

   %


The \textbf{idr\_nature} flag determines whether the radiation component is free-streaming or propagates as a perfect fluid. We set this to \emph{free-streaming} for both the DL-$\nu$ and DM-DR scattering cases so that the effects come only from the scattering with DM. We set the choice to \emph{fluid} only when implementing the purely self-interacting radiation fluid case, whether this is composed of $\nu$ in the SI-$\nu$ scenario or of DR in the SI-DR scenario. For this pure radiation reference case, we additionally set \textbf{a\_idm\_dr} = 0. Finally, we use the fixed $\Lambda$CDM parameters
\begin{equation}
(100\theta_s,\,\omega_{b},\,\omega_{\rm cdm},\,10^9A_s,\,n_s,\,\tau_{\rm reio})=(1.0453,\,0.02238,\,0.1201,\,2.1006,\,0.9661,\,0.05431)
\end{equation}
when studying the origin of the phase shift in Sec.~\ref{sec.result} and~\ref{sec.analytical} and will allow the parameters to flow when performing the MCMC study in Sec.~\ref{sec.MCMC}.

\section{Enhanced phase shift: numerical calculations}\label{sec.result}
In this section, we show the results of CMB phase shift between the DL-$\nu$ and SI-$\nu$ models obtained from \CLASS~and discuss the parametric dependence of the phase shift result. We will discuss the origin of the phase shift and explain the parametric dependence of the result in Sec.~\ref{sec.analytical}. 

\subsection{Locating acoustic peak positions}
To quantify the phase shift in the calculated CMB power spectrum, we determine how the positions of the acoustic peaks change relative to different models ($\Lambda$CDM, DL-$\nu$, or SI-$\nu$). However, one difficulty in extracting the location of acoustic peaks in either $D_\ell^{TT,EE}$ spectra or the transfer function $T_\gamma(k)$ is that the diffusion damping, together with the discrete $\ell$ and $k$ points output from \CLASS, makes it tricky to determine the peak location at high $\ell$ (or high $k$) with precision $\delta\ell\sim 1$ ($\delta k\sim10^{-5}$~Mpc$^{-1}$) for seeing the signal. We hence need to implement the following scheme to fit the peak locations: 
\newline

\noindent 1. For $D_\ell^{TT,EE}$, we include the lensing correction in the spectra when finding the peak locations. Since $\ell$ is an integer, the digitized peak positions are found by searching for the values of $\ell$ that correspond to local maxima in the $D_\ell$'s, with the condition that these points be larger than all other points within an interval of $[\ell -10, \ell +10]$. As the peak structure gets less well-defined at larger $\ell$'s due to diffusion damping, the positions of the larger $\ell$ peaks are excluded as unreliable. The resulting shift in peak positions $\Delta\ell$, which are only resolved up to integer level, may experience some numerical fluctuation due to rounding uncertainty. Instead of looking at the individual $\Delta\ell$ points, we analyze the mean and standard deviation of several points to quantify the average phase shift in the $D_\ell$'s for each model for the high-$\ell$ peaks. Further investigation of the phase shift signal would involve fitting the CMB spectra with a template designed to capture the shifts in $\ell$-peaks~\cite{benspaper}, which we leave for future work.
\newline

\noindent 2. The peak structure of the photon transfer function $T_\gamma(k)$ allows us to monitor the generation of photon phase shift at different redshifts and is important for understanding the origin of the phase shift. We adjust the \CLASS~settings to ensure a high density of points for the $T_\gamma(k)$ output. However, due to the limitation of the computation time, we can only obtain the spectrum down to step size $\Delta k\sim2\times10^{-4}k_p$, where $k_p$ represents the peak location, 
and find the precise peak positions by fitting with a Gaussian function $f(k) = A\exp(-(k-k_p)^2/2\sigma^2) + b$. The fitting is done over a range of $100$ to $400$ points on each side of the maxima to extract the mean value $k_p$ as the true peak position -- as many points as possible are included in the fit to obtain a symmetric profile for the peak region while aiming for the variance in the fitting of the mean to satisfy a threshold of  $\sqrt{{\rm Var}[k_p]}/k_p \lesssim 10^{-6}-10^{-5}$.\footnote{Here we define ${\rm Var}[k_p]$ as the diagonal element of the covariant matrix for the mean value of the Gaussian, obtained as an output from fitting with \textbf{scipy curve\_fit} in Python~\cite{2020SciPy-NMeth}.} As damping at larger $k$  smears out the peak structure, the Gaussian fit for the larger $k$-peaks with $k\tau\gtrsim 45$ is excluded as unreliable. 

\subsection{Phase shift enhancement in the CMB power spectra}

\begin{figure}
  \centering
\includegraphics[width=12cm]{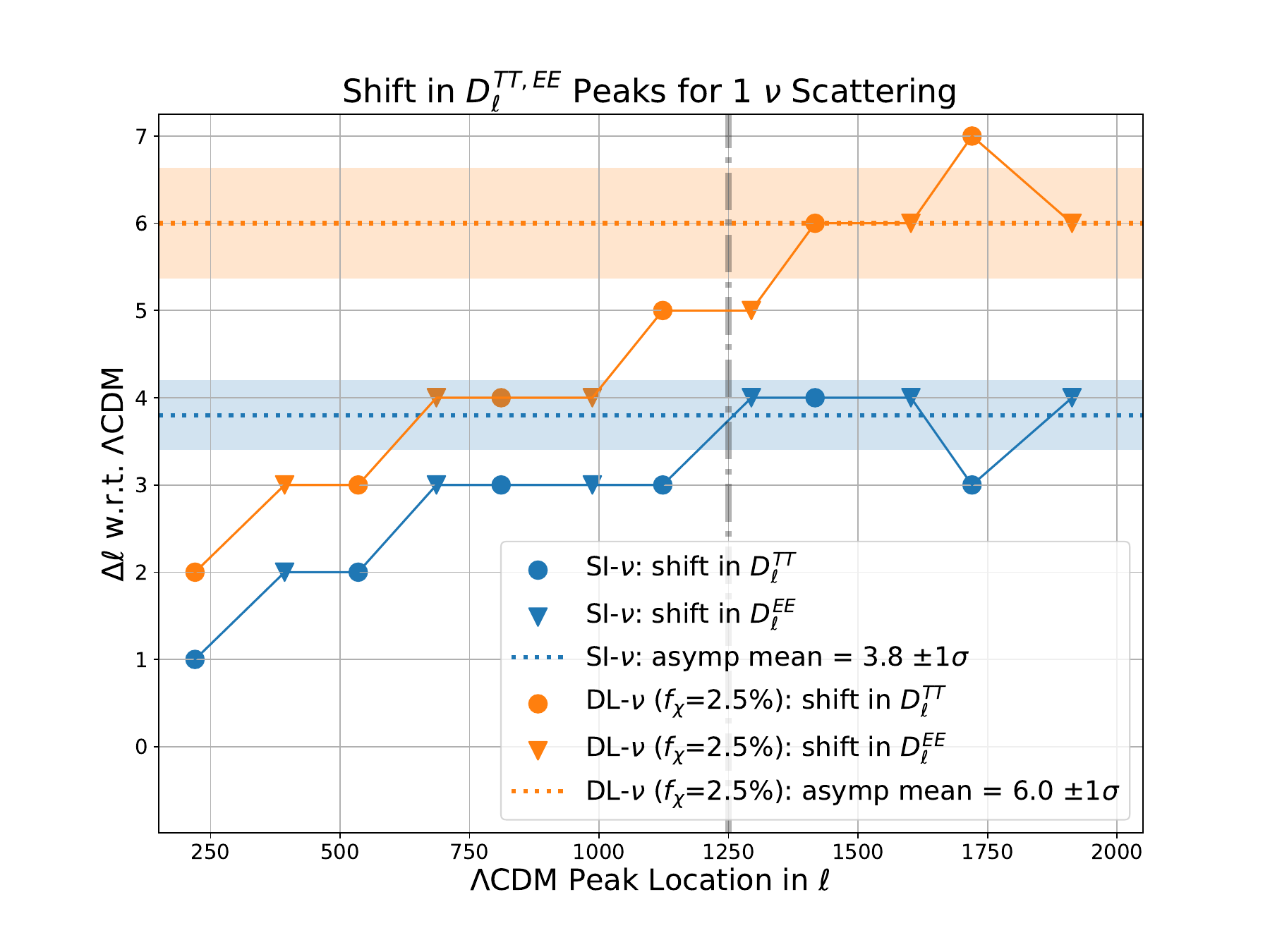}
\\
\includegraphics[width=12cm]{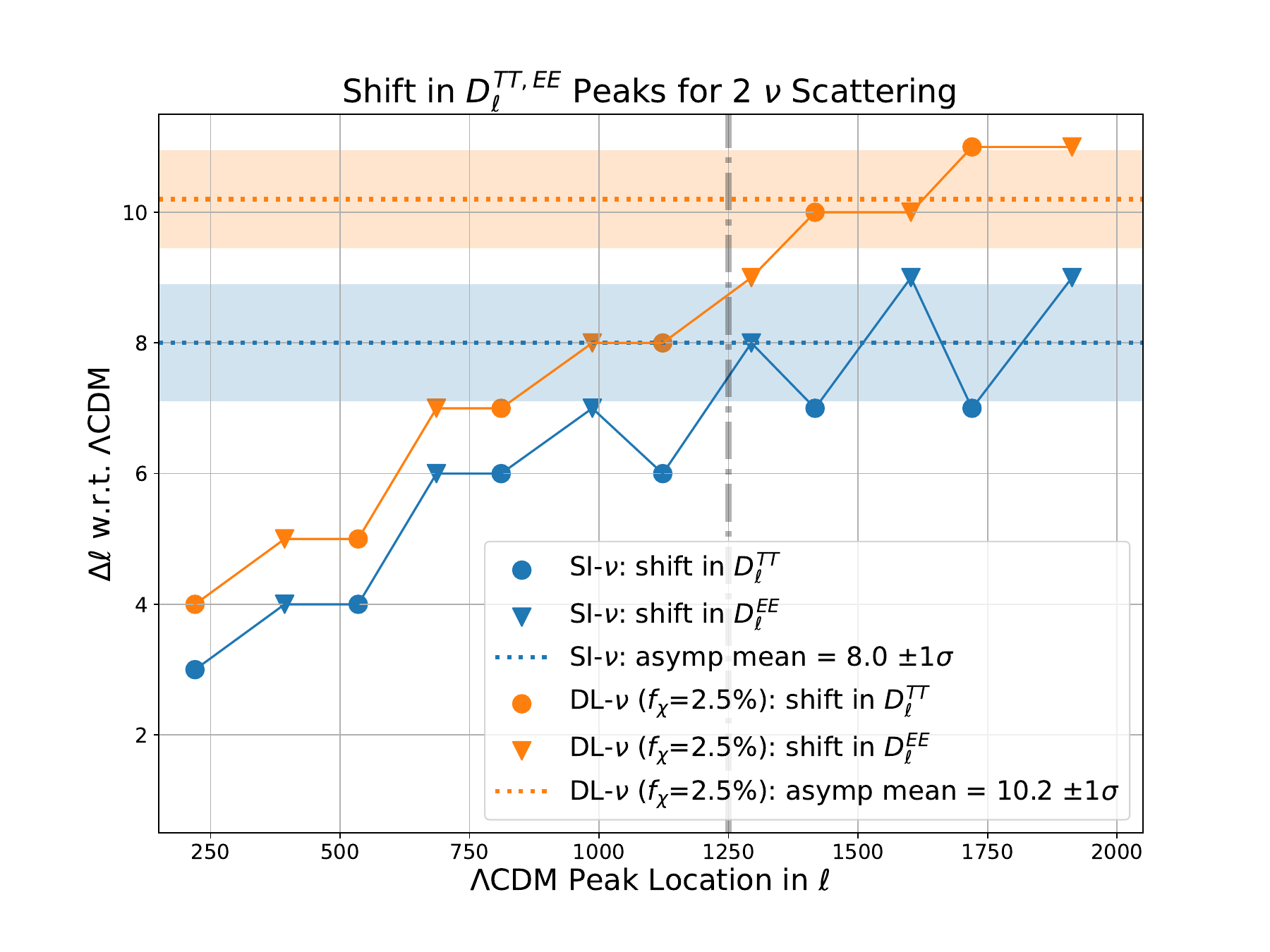}
\caption{CMB phase shift enhancement from \CLASS~for one (upper) and two (lower) species of scattering neutrinos. The $\Delta\ell$ shift in peaks in the $D_\ell^{TT}$ (circle) and $D_\ell^{EE}$ (triangle) lensed spectra for SI-$\nu$ (blue) and $f_\chi = 2.5\%$ DL-$\nu$ (orange) fluids were taken with respect to $\Lambda$CDM, where $\ell$ is an integer. Lines connecting the points have been provided to help guide the eye, but should not be directly interpreted as having physical meaning due to the integer-level rounding fluctuations of the points. For the five peaks larger than $\ell=1250$ (black vertical line), the asymptotic mean (``asymp mean") $\overline{\Delta\ell}$ and standard deviation $\sigma$ of the $\Delta\ell$'s in the deep radiation era were calculated. The $\overline{\Delta\ell}\pm 1\sigma$ band was plotted for each case.}
\label{CLTTEE}
\end{figure}
\begin{figure}
  \centering
\includegraphics[width=12cm]{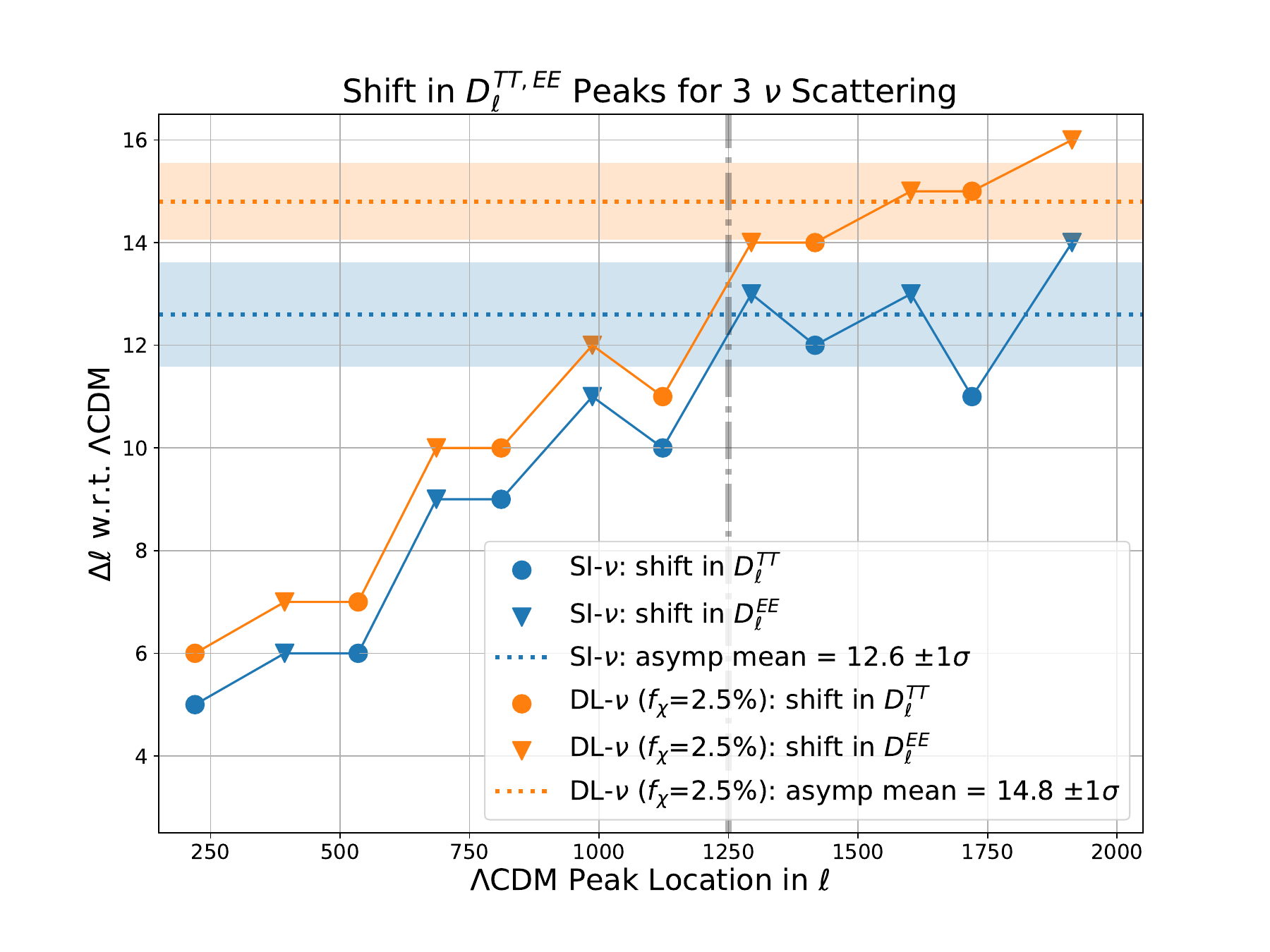}
\caption{Same as Fig.~\ref{CLTTEE} but for $3$ interacting neutrino species.}
\label{CLTTEE2}
\end{figure}
\begin{figure}
    \centering
\includegraphics[width=11.8cm]{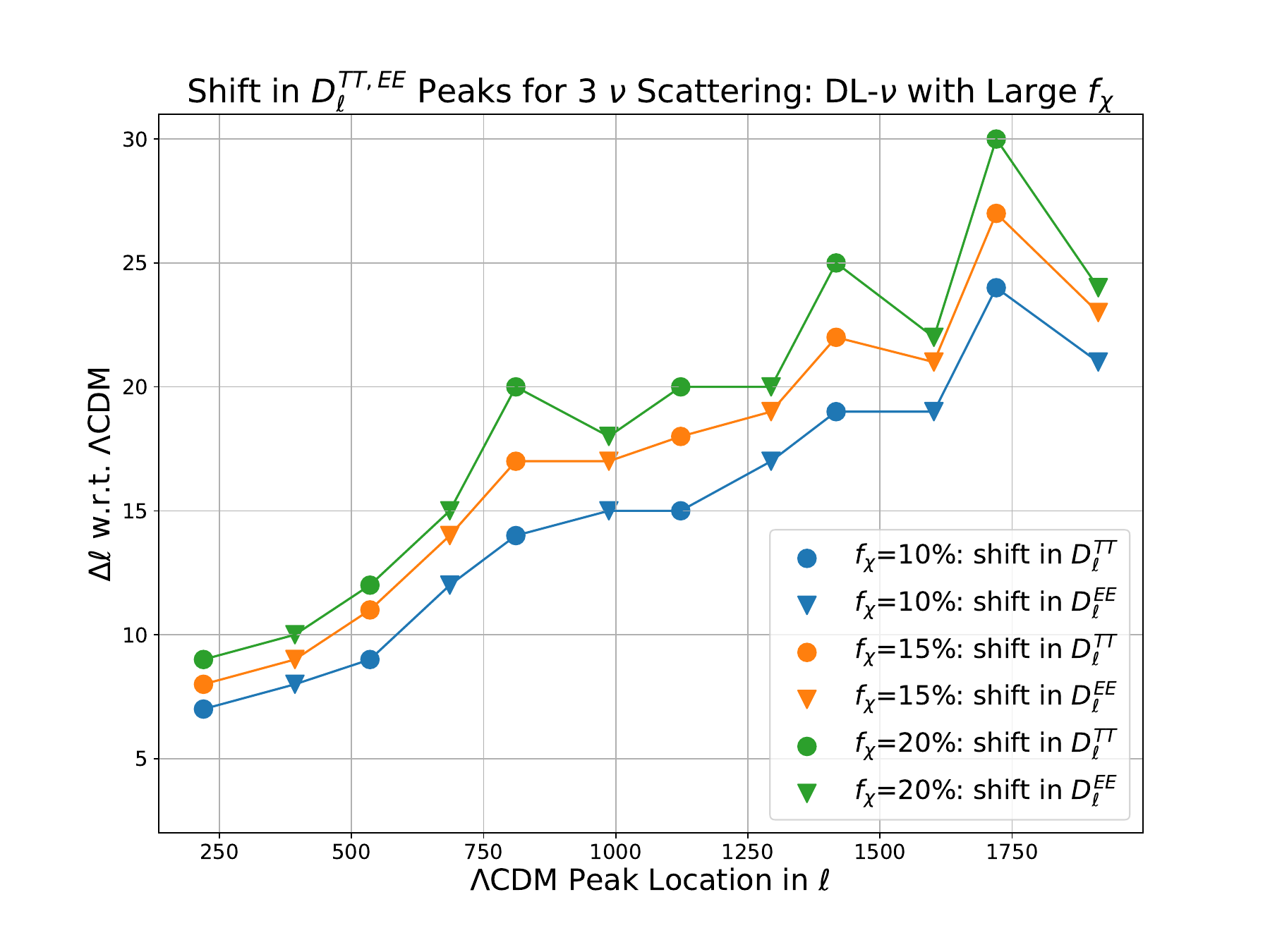}
    \caption{$\Delta\ell$ shift in peaks in the $D_\ell^{TT}$ (circle) and $D_\ell^{EE}$ (triangle) lensed spectra for large $f_\chi = 10\%-20\%$ DL-$\nu$ fluid with all three neutrinos scattering. The shifts were taken with respect to $\Lambda$CDM. The integer-valued $\Delta\ell$ points are joined by lines for visual clarity.}
    \label{largefx}
\end{figure}
\begin{figure}
    \centering
    \includegraphics[width=12cm]{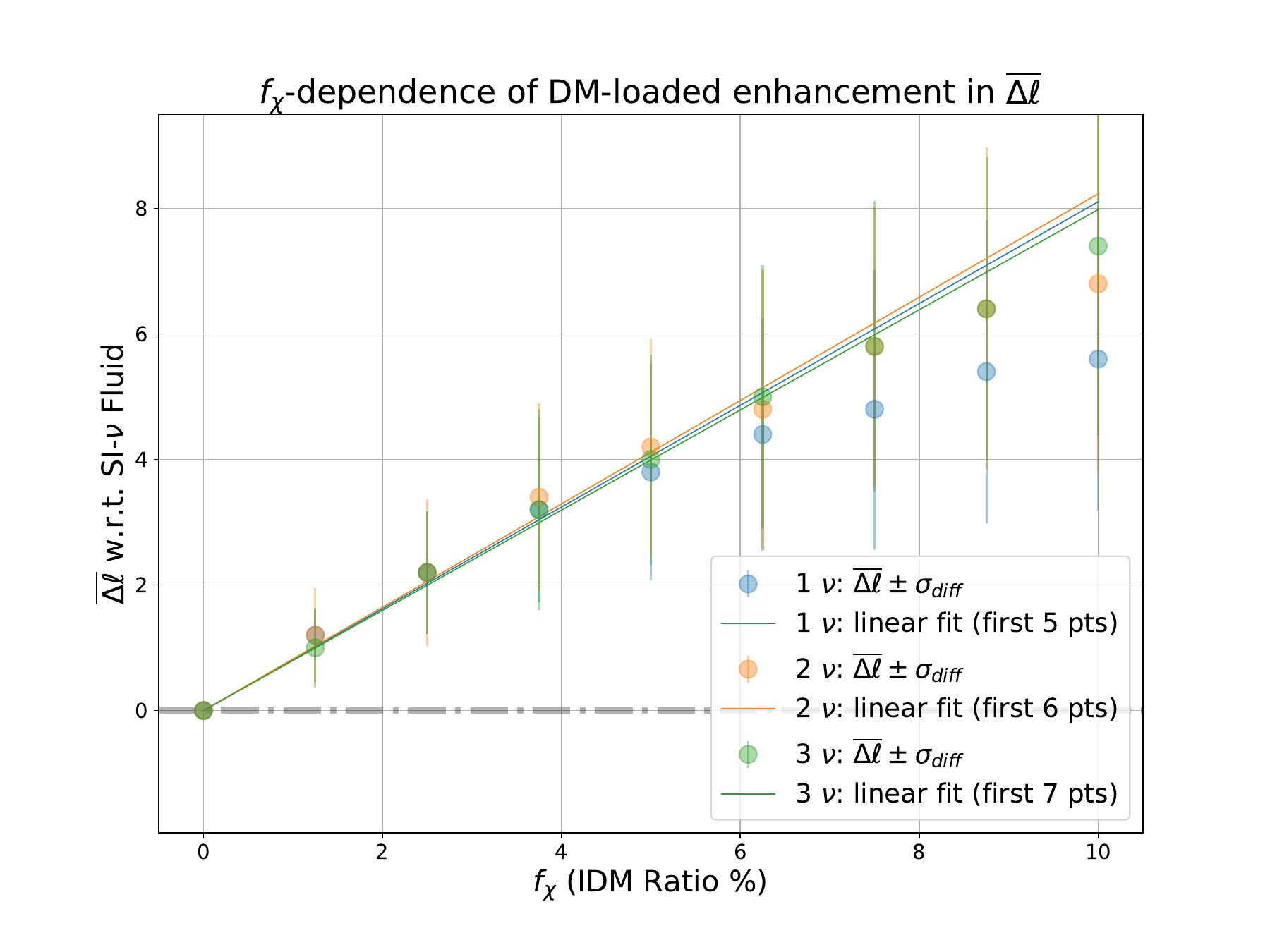}
    \caption{ $f_\chi$-dependence of the $\overline{\Delta\ell}$ enhancement on top of SI-$\nu$  ($f_\chi = 0$) for respective numbers of scattering $\nu$'s, where $\overline{\Delta\ell}$ is the mean value of the $\Delta\ell$'s for the five peaks larger than $\ell=1250$. Error bars correspond to the standard deviation $\sigma_{\mathrm{diff}}$ in the differences between the computed $\Delta\ell$'s, quantifying the relative integer rounding fluctuation of points with respect to SI-$\nu$. A linear fit to the first 5-7 points in $f_\chi$ is provided for each $f_\nu$. 
    }
    \label{fxCl}
\end{figure}

In Fig.~\ref{CLTTEE}, we show the $\Delta\ell$ shift for each acoustic peak in the lensed $D_\ell^{TT}$ (circle) and $D_\ell^{EE}$ (triangle) spectra of the SI-$\nu$ (blue) and DL-$\nu$ scenarios compared to the $\Lambda$CDM model. Among the total neutrino number $3.046$, we assume either one-third (labeled as $1\nu$), two-thirds (labeled as $2\nu$), or all of the neutrinos (labeled as $3\nu$) are interacting. We see certain general features in the CMB phase shift by plotting the $\Delta\ell$'s as a function of the $\Lambda$CDM peak positions in $\ell$. First, in both interacting neutrino models, there is a positive $\Delta\ell$ that grows with $\ell$ across both the $D_\ell^{TT}$ and $D_\ell^{EE}$ spectra, indicating that there is a phase shift to higher $\ell$-modes from neutrino scattering. The phase shift gets more pronounced for modes entering the horizon deeper in the radiation era.

Conversely, $\Delta\ell$ goes to zero as $\ell$ goes to zero, which is to be expected since the low-$\ell$ modes only enter the horizon in the deep matter-dominated era. At this time, the metric perturbations would be dominated by non-oscillating contributions coming from the clumping of cold dark matter, making the contributions of the SI-$\nu$ and DL-$\nu$ acoustic oscillations to the metric negligible in comparison. We do not have a clear analytic understanding of the interim period between the deep radiation domination and deep matter domination eras, although we can reasonably expect the $\Delta\ell$ values to interpolate between the two extremes. We see this in the full  $\CLASS$ result (which remain valid in the interim period) where  $\Delta\ell$ steadily declines as we go from larger to smaller $\ell$ modes. 
To study the phase shift in $D_\ell$, we focus on the deep radiation era and consider only $\Delta\ell$ values for larger $\ell$ modes, as this is where analytical understanding is feasible and numerical calculations show the largest phase shift. 


As shown in Figs.~\ref{CLTTEE} and~\ref{CLTTEE2}, when comparing models with the same number of scattering neutrinos, there is a noticeable positive $\Delta\ell$ enhancement from the DL-$\nu$ model (orange) on top of the SI-$\nu$ (blue). We show this additional $\ell$-peak shift in Fig.~\ref{largefx}, which increases with the fraction of interacting dark matter $f_{\chi}$ and continues to grow for $f_\chi\geq15\%$ when the sound speed of the DL-$\nu$ is slower than the photon sound speed. We will discuss the relationship between neutrino sound speed and phase shift in Sec.~\ref{sec.Cs}. To better quantify this additional $\Delta\ell$ shift produced in the deep radiation-dominated era, we focus on the modes $\ell>1250$ (black vertical line) mainly contributed by perturbations entering the horizon before matter density becomes significant. To account for numerical fluctuations due to the integer-level rounding of the peak positions, we average the shift $\overline{\Delta\ell}$ of the larger $\ell$ peaks $(\ell>1250)$ and obtain the standard deviation $\sigma$ of the $\Delta\ell$ points about this average. The significance of this effect is indicated by the separation of the $\pm1\sigma$ bands in Fig.~\ref{CLTTEE}.

In Fig.~\ref{fxCl}, we show the average and standard deviation of DL-$\nu$'s $\Delta\ell$ shift relative to the SI-$\nu$ case against $f_\chi$, where $f_\chi = 0$ corresponds to the SI-$\nu$ case. A clear linear dependence of the average $\overline{\Delta\ell}$ in $f_\chi$ is observed until $f_\chi$ gets larger $\approx 6\%$. 
However, as we show in Sec.~\ref{sec.MCMC}, current CMB data favors $f_\chi\lesssim 1\%$, restricting us to the regime where the interacting neutrino density dominates over the interacting dark matter density around matter-radiation equality, and the linear dependence in $f_\chi$ holds well. Despite an increasing fraction of interacting neutrinos $f_\nu$ leading to a larger phase shift compared to the $\Lambda$CDM model (Fig.~\ref{CLTTEE} and \ref{CLTTEE2}), the additional phase shift compared to the SI-$\nu$ model remains insensitive to the fraction of interacting neutrinos (Fig.~\ref{fxCl}) in this regime.

\section{Enhanced phase shift: analytic understanding}\label{sec.analytical}
By solving the Boltzmann equations, we find that neutrino-DM scattering further shifts the acoustic oscillation peaks to higher $\ell$-modes. In this section, we show that the mechanism underlying the DM-induced phase shift can be explained by coupled harmonic oscillators between photon and neutrino perturbations. This insight helps us to understand the linear dependence on $f_{\chi}$ and the insensitivity to $f_{\nu}$ in the additional phase shift.

As we explain below, the phase shift is determined by perturbation evolution right after horizon re-entry. For perturbations with $\ell\gtrsim 10^3$, we can assume a radiation-dominated Universe in the following discussion. 

\subsection{Sound speeds of the cosmic fluids}
\label{sec.Cs}


%
\begin{figure}
    \centering
    \includegraphics[width=12cm]{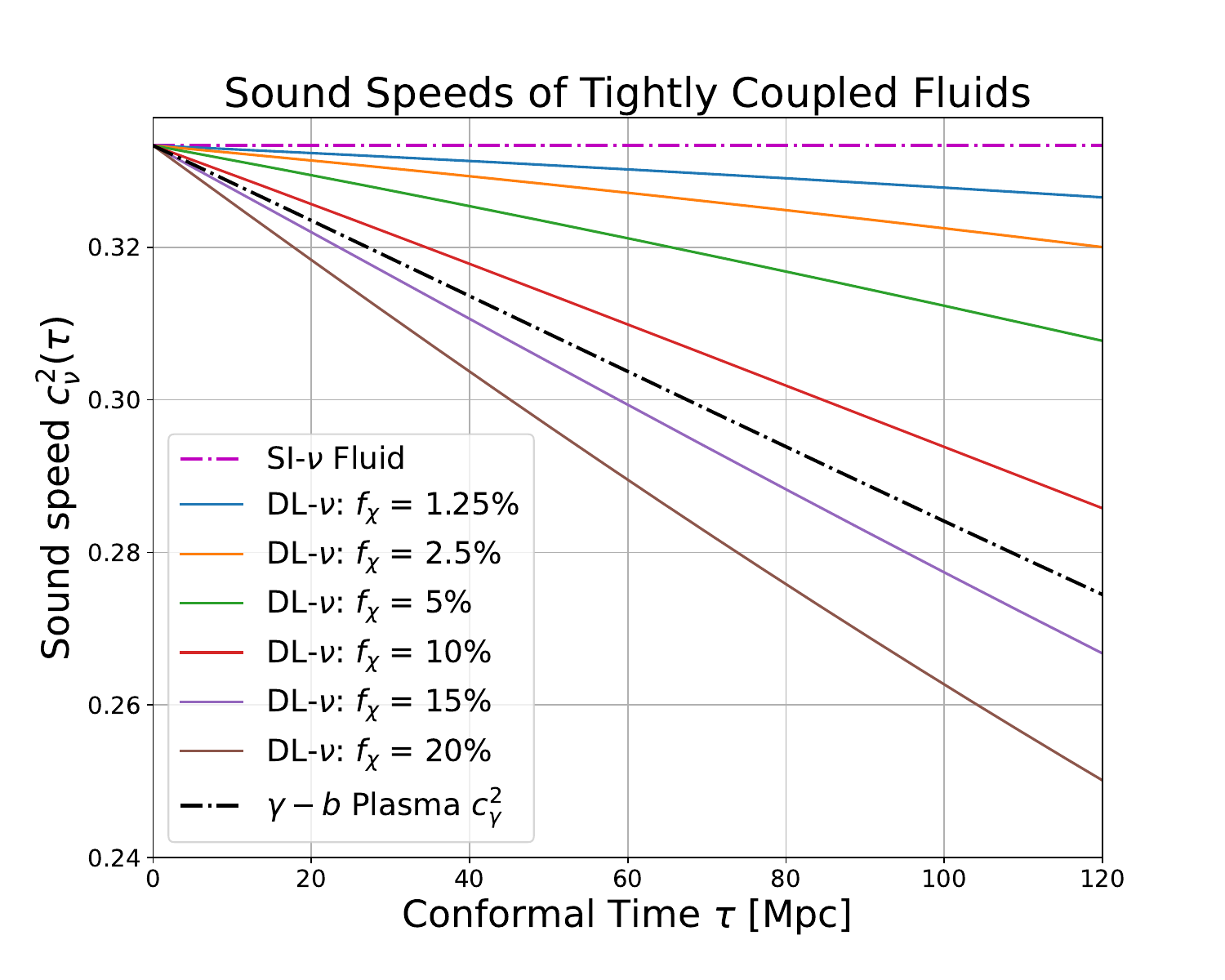}
    \caption{Illustration of the time-dependence of the tightly-coupled radiation sound speeds in the presence of matter loading. The curves were obtained using the sound speed formula~(\ref{eqn:cs}) and the background densities from \CLASS~for all neutrinos scattering and a range of $f_\chi$. The corresponding sound speed of the photon-baryon plasma $c_\gamma^2$ is included for reference.}
    \label{CsPlot}
\end{figure}
Earlier studies on the CMB phase shift~\cite{Bashinsky:2003tk,Baumann:2015rya} have shown that a phase shift is generated in the photon fluid if we have non-photon radiation with sound speed larger than photons, or if the radiation carries isocurvature fluctuations. These earlier studies were done in the context of a radiation-dominated universe with negligible matter content. In this paper, we focus on adiabatic fluctuations and consider the phase shift due to the propagation behavior of radiation inside the horizon when matter-loading effects are taken into account. The key quantity of interest is the adiabatic sound speed $c_g^2 = \delta P_g/\delta\rho_g$ describing the speed at which the fluctuations propagate, where $g$ is a general index that may denote any particular component or combination of components. This can be related to the equation of state $\omega_g = P_g/\rho_g$ by the general relation~\cite{Hu_1998}:
\begin{align}
c_g^2 = \omega_g + \rho_g\frac{d\omega_g}{d\rho_g} =\omega_g - \frac{\dot{\omega_g}}{3\mathcal{H}(1+\omega_g)}\,.
\label{eqn:ad}
\end{align}

In a radiation-dominated universe, the overall equation of state is a constant $\omega = 1/3$. The total sound speed reduces to $c_s^2 = \omega$ and Ref.~\cite{Baumann:2015rya} shows that there is no phase shift if $c_s^2 < c_\gamma^2$. In a universe containing both radiation and matter, $\omega$ would instead exhibit a time-dependence:
\begin{align}
    \omega &= \frac{1}{3}\frac{\rho_{\rm rad}}{\rho_{\rm total}} = \frac{1}{3}\frac{1}{1+\frac{a(\tau)}{a_\mathrm{eq}}}\,,
\end{align}
where $a_{\rm eq}$ is the scale factor at matter-radiation equilibrium. This results in a non-zero $\dot{\omega}$ and the total sound speed becomes:
\begin{align}
    c_s^2 = \frac{1}{3}\frac{1}{1+\frac{3}{4}\frac{a(\tau)}{a_\mathrm{eq}}}\,. 
\end{align}
While $c_s^2 = \omega$ holds true only at $a(\tau)\ll a_{\rm eq}$, time evolution due to the presence of matter in the background leads us into the $c_s^2\neq\omega$ regime discussed in Ref.~\cite{Baumann:2015rya} where a phase shift can be generated, albeit without isocurvature.

To understand the mechanism behind the phase shift effect we obtained from \CLASS, it is important to consider the relative sizes of the individual radiation sound speeds. Applying Eq.~(\ref{eqn:ad}) to a cosmic fluid consisting of tightly coupled radiation and matter components yields the sound speed formula~(\ref{eqn:cs}) from before. The $\tau$-dependence of the sound speed for various tightly coupled fluids can be obtained by plugging in the background densities from \CLASS~into the energy ratio in Eq.~(\ref{eqn:cs}), with $R_\gamma = 3\rho_b/4\rho_\gamma$ for the photon-baryon plasma and $R_\nu = 3\rho_\chi/4\rho_\nu$ for the efficiently scattering DL-$\nu$  fluid. These are plotted in Fig.~\ref{CsPlot} for a range of $f_\chi$ for $\tau \leq 120$ Mpc when the tight-coupling approximation for the photon-baryon plasma holds well. 

As expected, the DL-$\nu$ fluids exhibit a DM-loaded suppression in the sound speed over time compared to the pure SI-$\nu$ fluid (purple dash-dotted line). For the fraction of scattering dark matter $f_\chi \lesssim 10\%$, the sound speeds follow the order:
\begin{align}
c_\gamma < c_\nu^{{\rm DL}-\nu} < c_\nu^{{\rm SI}-\nu} < c_\nu^{{\rm FS}-\nu}\,,
\end{align}
where $c_\nu^{{\rm SI}-\nu}={1}/{\sqrt{3}}$ for the  SI-$\nu$ fluid with zero DM-loading \footnote{Beginning at horizon re-entry, the evolution of the FS-$\nu$ perturbation diverges from that of the SI-$\nu$ perturbation as the shear perturbation begins to grow. In Section~\ref{sec.analy}, we will show that the sound speed $c_\nu^{{\rm FS}-\nu}$ in the FR-$\nu$ perturbation equation starts around $\approx {1}/{\sqrt{3}}$ at horizon re-entry and accelerates to $\approx 1$ through diffusion damping.}. Notably, the ordering in the neutrino sound speeds $c_\nu$ coincides with the relative sizes of the phase shift calculated in Sec.~\ref{sec.result}, with slower sound speeds corresponding to larger positive shifts in $\ell$. Comparing Figs.~\ref{largefx} and~\ref{CsPlot}, the phase shift induced by DM-loading continues to rise for $f_\chi\gtrsim15\%$, even when the corresponding DL-$\nu$ sound speed now follows the ordering $c_\nu^{DL-\nu} < c_\gamma$. While this implies that the total sound speed $c_s < c_\gamma$ in the absence of any other radiation component, we reiterate that this is not in contradiction with the discussion in Ref.~\cite{Baumann:2015rya} since $c_s^2 \neq \omega$.

\subsection{A toy model analysis}
To obtain a qualitative understanding of the enhanced phase shift, we construct a toy model mimicking the evolution of photon and neutrino perturbations from the \CLASS~calculation. We first examine the scenario where all neutrinos undergo scattering with a fraction of DM, then discuss the case where some neutrinos remain free-streaming.

The temperature perturbation of photons $\delta T_\gamma/\bar{T}_\gamma$ results from a combination of energy density perturbation $\delta_\gamma\equiv\delta\rho_\gamma/\bar\rho_\gamma$ and metric perturbation $\phi$. However, in the radiation-dominated era, $\phi$ decays upon entering the horizon and contributes significantly less to temperature perturbations, especially for the high-$\ell$ modes we are considering. As a result, we can simplify the discussion by focusing on $\delta_\gamma$ instead of the gauge invariant $\delta T_\gamma/\bar{T}_\gamma$. As we discussed in more details in Appendix.~\ref{app.toymodel}, under the tight coupling approximation, the perturbations of fluids in the $\gamma-b$ and DL-$\nu$ (or SI-$\nu$) systems can be described by a pair of gravitationally coupled oscillators
\begin{align}
\ddot{\delta_\gamma}(\tau) + k^2c_\gamma^2(\tau)\delta_\gamma(\tau) &= \frac{4\mathcal{H}^2(\tau)}{1 + \frac{a(\tau)}{a_\mathrm{eq}}}[f_\gamma\delta_\gamma(\tau) + f_\nu\delta_\nu(\tau)]\,,
\label{toyeqn1}
\\
\ddot{\delta_\nu}(\tau) + k^2c_\nu^2(\tau)\delta_\nu(\tau) &= \frac{4\mathcal{H}^2(\tau)}{1 + \frac{a(\tau)}{a_\mathrm{eq}}}[f_\gamma\delta_\gamma(\tau) + f_\nu\delta_\nu(\tau)]\,.
\label{toyeqn2}
\end{align}
$c_{\gamma,\nu}$ are the photon and neutrino sound speeds $c_{\gamma,\nu}^2 =[3(1+R_{\gamma,\nu}(\tau))]^{-1}$ with 
\begin{align}
\label{Ratio}
R_\gamma(\tau) = \frac{3}{4}\frac{f_b}{f_\gamma }\frac{a(\tau)}{a_{eq}}\,,\quad R_\nu(\tau) = \frac{3}{4}f_\chi\frac{f_{\rm DM}}{f_\nu }\frac{a(\tau)}{a_{eq}}\,.
\end{align}
$f_i = {\rho}_i/{\rho}_\mathrm{rad}$ for $i = \gamma,\nu$, and $f_i = {\rho}_i/{\rho}_\mathrm{\rm m}$ for $i = b,\rm DM$, are the background density ratios of the components in the radiation and matter respectively. The background density ratios $f_i$ were taken from the \CLASS~background output. To the two decimal places, they are $f_{\rm DM}=0.84$, $f_b = 0.16$, $f_\gamma = 0.59$ and $f_\nu = 0.41$ for all neutrinos scattering. Eqs.~(\ref{toyeqn1}) and (\ref{toyeqn2}) are derived based on four assumptions described below that offer reasonable approximations of $\delta_{\gamma,\nu}$ evolution right after horizon re-entry. This is when $\delta_{\gamma}$ obtains a phase shift, as we will show. The coupled oscillators can effectively reproduce several qualitative features observed in the \CLASS~results. Nevertheless, these approximations have limitations, and deviations between this toy model and the \CLASS~results are to be expected.

\begin{itemize}
\item \textbf{(Assumption 1)} The system does not contain free-streaming radiation, hence there is no anisotropic stress and metric perturbations $\phi=\psi$. The toy model cannot capture diffusion-damping effects in the CMB, especially near recombination, where photon and baryon perturbations decouple from each other starting from the higher $k$-modes. 
Deviations from \CLASS~are hence anticipated near recombination. When considering some neutrinos as free-streaming, as discussed at the end of Sec.~\ref{sec.analy}, since the shear perturbation $\sigma$ has not undergone significant growth in the early evolution, the toy model still captures the main features of the full result that we focus on.

\item \textbf{(Assumption 2)} The matter loading gives a minor correction to photon and neutrino propagation, with $R_{\gamma,\nu} \ll 1$ when a perturbation mode enters the horizon. This condition holds true for the perturbation modes we consider when $f_\chi < f_\nu$ at matter-radiation equality. This assumption allows for the neglect of Hubble damping and simplification of the Einstein equations, as discussed in Appendix~\ref{app.toymodel}. 

\item \textbf{(Assumption 3)} The toy model focuses on the evolution of perturbations inside the horizon ($k\tau>1$), where $\phi\propto\tau^{-2}$ holds in the radiation-dominated era. It does not accurately represent the behavior close to horizon re-entry ($k\tau \lesssim 1$), leading to expected deviations from \CLASS~for the location of low-$k$ peaks. 

\item \textbf{(Assumption 4)} The energy density perturbation is dominated by photons and neutrinos, a valid approximation for modes entering the horizon during the radiation-dominated era. This assumption simplifies the setup as a closed system with $\delta_{\gamma,\nu}$. However, the toy model cannot capture contributions to the phase shift arising from dark matter acoustic oscillations (DAO), which become more pronounced at later times. 
Consequently, the toy model can underestimate the phase shift compared to the \CLASS~result, with the underestimation amplifying over time and with increasing $f_\chi/f_\nu$. In this work, we concentrate on scenarios where $f_\chi/f_\nu < 1$ until recombination, and the DAO produce a minor correction to the phase shift.  
\end{itemize}

To validate the toy model, we compare the phase shift in $\delta_\gamma(k)$ with the results obtained from \CLASS~using Hubble expansion rates and energy densities $\rho_i$ ($i=\gamma,\nu,b,{\rm DM}$) from \CLASS~output. We set the initial conditions $\delta_{\gamma,\nu}(\tau_{in}) = 1$ and $\dot{\delta}_{\gamma,\nu}(\tau_{in}) = 0$ for $k\tau_{in} = 1$ to approximate the \CLASS~perturbations starting from horizon re-entry. When studying the phase shift signal, the choice of initial conditions only determines the initial amplitude and phase of the oscillation, which is identical among DL-$\nu$ and SI-$\nu$ scenarios. The difference in the phase shift is therefore insensitive to the choice of initial conditions. To obtain transfer functions in $k$ modes, we solve the toy model across a range of $k$-values, evaluating until redshifts $z_*=10800$, $3400$, and $1070$, corresponding to the deep radiation-dominated era ($\tau\approx40$~Mpc), the matter-radiation equality ($\tau\approx 113$ Mpc), and recombination ($\tau\approx280$~Mpc), respectively. The transfer function is obtained from $\delta_\gamma(\tau(z_*))$ values over $k$.

\begin{figure}[t!]
  \centering
\includegraphics[width=12cm]{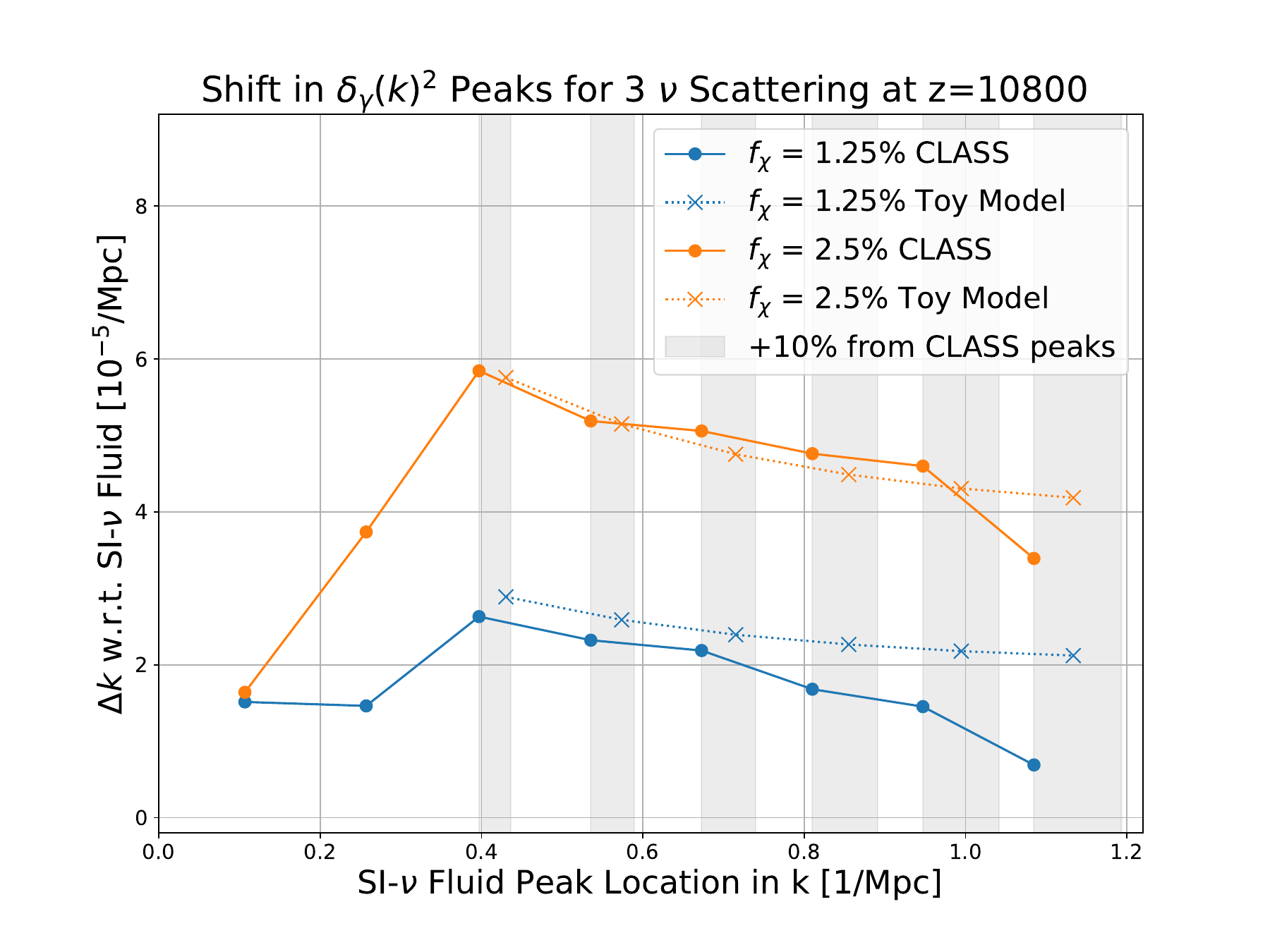}
\caption{ 
\CLASS~(solid) vs toy model (dotted) comparison of $\Delta k$ for $f_\chi = 1.25\%$ (blue) and $2.5\%$ (orange) DL-$\nu$ with respect to SI-$\nu$ for 3 scattering neutrinos in the radiation-dominant era ($\tau\approx 40$~Mpc). The peak locations of the toy model (cross points) lie within $+10\%$ of the peaks from \CLASS~(solid dots), as demarcated by the grey vertical bands. 
}
\label{dknu3RAD}
\end{figure}

\begin{figure}[t!]
\centering
\includegraphics[width=12cm]{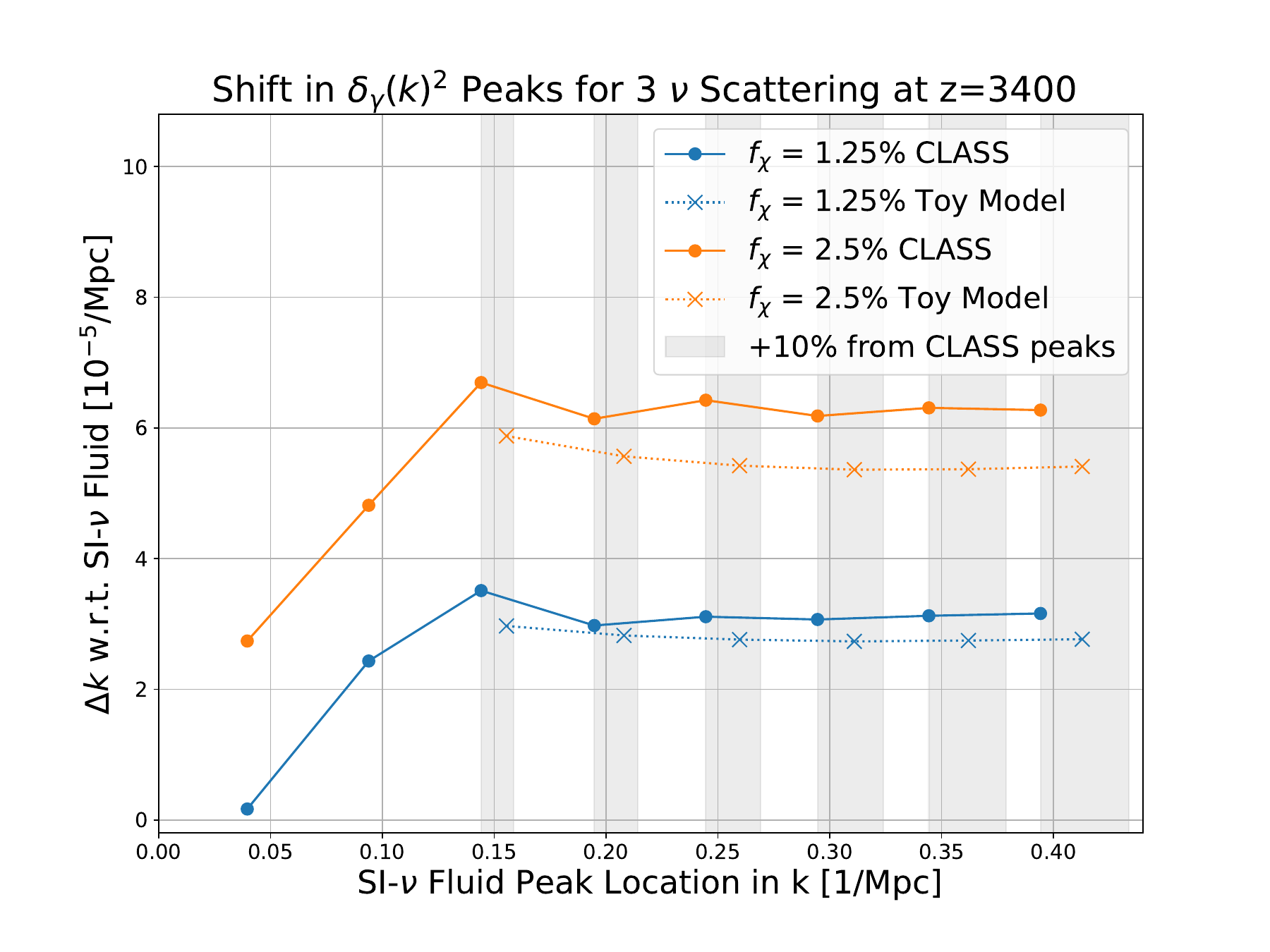}
\\
\includegraphics[width=12cm]{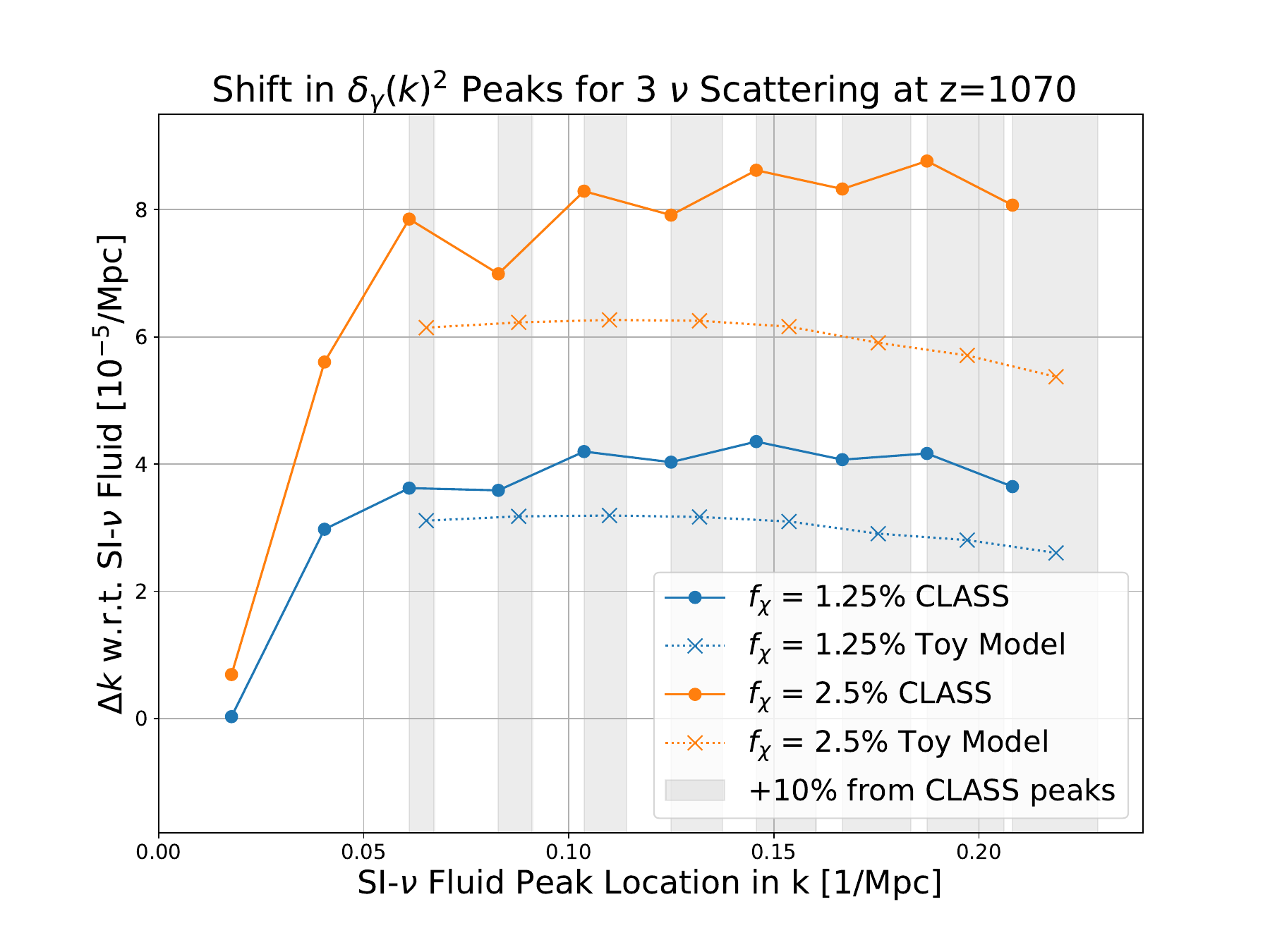}
\caption{ 
Same as Fig.~\ref{dknu3RAD} but with $\tau\approx 110$~Mpc at matter-radiation equality (upper) and $\tau\approx 280$~Mpc at recombination (lower).
}
\label{dknu3EQB}
\end{figure}

\begin{figure}[t!]
\centering
\includegraphics[width=12cm]{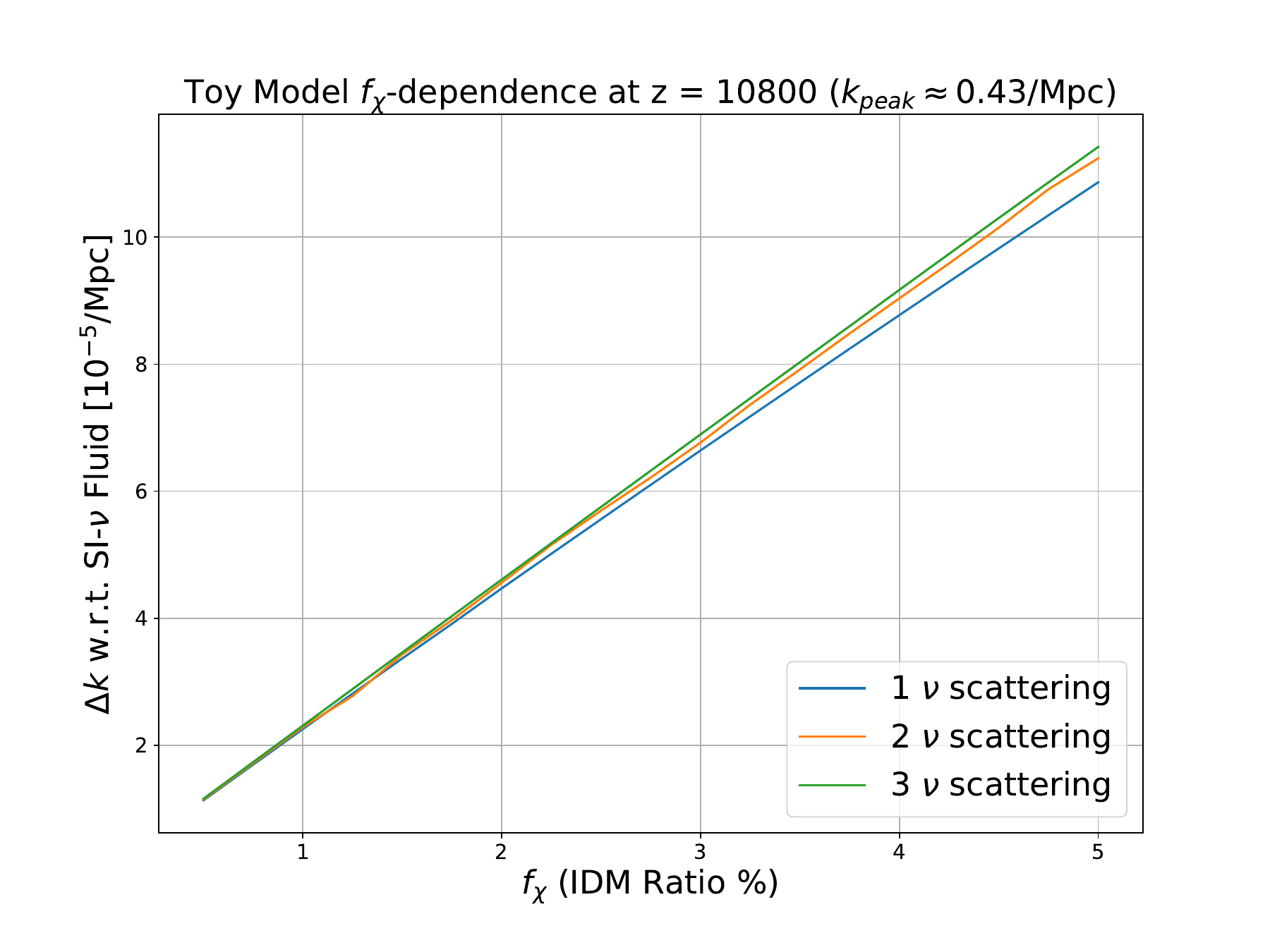}
\caption{ 
Toy model $f_\chi$-dependence of the shift in the peak at $k\approx 0.43/\rm Mpc$, obtained in the radiation-dominant era ($\tau\approx 40$~Mpc) for different numbers of scattering neutrinos. The lines were obtained by varying $f_\chi$ continuously from $0-5\%$ (no linear fit).
}
\label{dkfx}
\end{figure}

In Figs.~\ref{dknu3RAD} and \ref{dknu3EQB}, we present the shift of  $\delta_\gamma(k)$ peaks relative to the scattering neutrino scenario ($f_\chi=0$ and $c^2_\nu=\frac{1}{3}$), calculated by the toy model (dotted) with $f_\chi = 1.25\%$ (blue) and $2.5\%$ (orange). For comparison, we include results from the \CLASS~calculation (solid), with connected points for visual clarity. The toy model predicts peak positions within $+10\%$ deviation from the respective \CLASS~peaks in all cases (shown as the grey vertical bands). This discrepancy mainly arises from the initial evolution of the metric perturbation $\phi$ and the neglect of matter perturbations \textbf{(Assumptions 3 \& 4)}, which affect the time-dependence of $\phi$ and hence the $\delta_\gamma$ oscillations in the complete Boltzmann equations\footnote{The deviation in the peak location is consistent with the discussion in Ref.~\cite{2020moco.book.....D}, where an analytic approximation of the acoustic oscillations obtained by ignoring contributions from the time-dependence of $\phi$ predicted peak locations at higher $k$-modes that lie within $10\%$ from the full numerical solution.}. However, since we are only concerned with the relative phase shift between DL-$\nu$ and SI-$\nu$ scenarios, deviations in the peak locations relative to \CLASS~that appear in both scenarios, in the same way, would cancel and hence do not matter to the discussion.


The toy model solution reasonably approximates the enhanced phase shift obtained in the full result, demonstrating a linear dependence on $f_{\chi}$ and insensitivity to $f_\nu$ as in Fig.~\ref{dkfx}. As expected from \textbf{(Assumption 3)}, the toy model does not replicate the horizon entry behavior precisely enough to reproduce low-$k$ modes' phase shift to the $\Delta k/k_{\rm peak}\sim 10^{-5}$ level precision. We therefore only consider $\delta k$ results starting from the second oscillation peak in $\tau$ so that the perturbation modes fully enter the horizon. \textbf{(Assumption 4)} also suggests increasing deviations between toy model and \CLASS~at later times, exacerbated for larger $f_\chi$. At recombination, a deviation grows with $k$, as anticipated from \textbf{(Assumption 1)}. We leave additional plots for the DL-$\nu$ scenarios with one and two interacting neutrinos in Appendix.~\ref{ap.1nu}.

Given the toy model's reasonable approximation and simultaneous reproduction of $f_{\chi}$ and $f_\nu$ dependence, it serves for analytical understanding of the full result.

\subsection{Parametric dependence from the toy model analysis}\label{sec.analy}
Eqs.~(\ref{toyeqn1}) and (\ref{toyeqn2}) describe two harmonic oscillators $\delta_\gamma$ and $\delta_\nu$ with natural frequencies $k c_\gamma$ and $k c_{\nu}$ and couple to each other via gravitational interaction 
\begin{equation}\label{eq.Fk}
F_{\rm driv}(\tau) \equiv \frac{4\mathcal{H}^2(\tau)}{1 + \frac{a(\tau)}{a_\mathrm{eq}}}\approx\frac{4}{\tau^2(1+\frac{\tau}{\tau_{eq}})}\,
\end{equation}
with $\mathcal{H}=\tau^{-1}$ for being in the deep radiation-dominated era. To focus on the parametric dependence on $f_\chi$ for the DM-loading effect, we ignore baryon-loading ($f_b/f_\gamma\ll1$) in $c_\gamma$ to simplify the discussion, such that $c_\gamma^2=\frac{1}{3}$. While baryon-loading changes $\delta_\gamma$ oscillation and corrects its peak positions, the resulting phase shift shows up equally in the DL-$\nu$ and SI-$\nu$ scenarios. When comparing phase shifts between the two scenarios, the baryon-loading corrections cancel out, leaving the enhanced phase shift that we focus on.

Considering small DM-loading with $f_\chi\ll f_\nu$ (\textbf{Assumption 2}), or equivalently $\rho_\chi\ll\rho_\nu$ at matter-radiation equality, we have $c_{\nu}= c_\gamma-\delta c$, where 
\begin{align}\label{eq.Dcgamma}
\delta c(\tau) = R_\nu(\tau)\frac{c_\nu^2(\tau)}{c_\gamma+c_\nu(\tau)}\ll c_\gamma\,.
\end{align}
DM-loading slows down $c_\nu$ as the universe expands, which drives the oscillation frequencies between $\delta_\gamma$ and $\delta_\nu$ apart over time.
\begin{figure}[t!]
\centering
  \includegraphics[width=13cm]{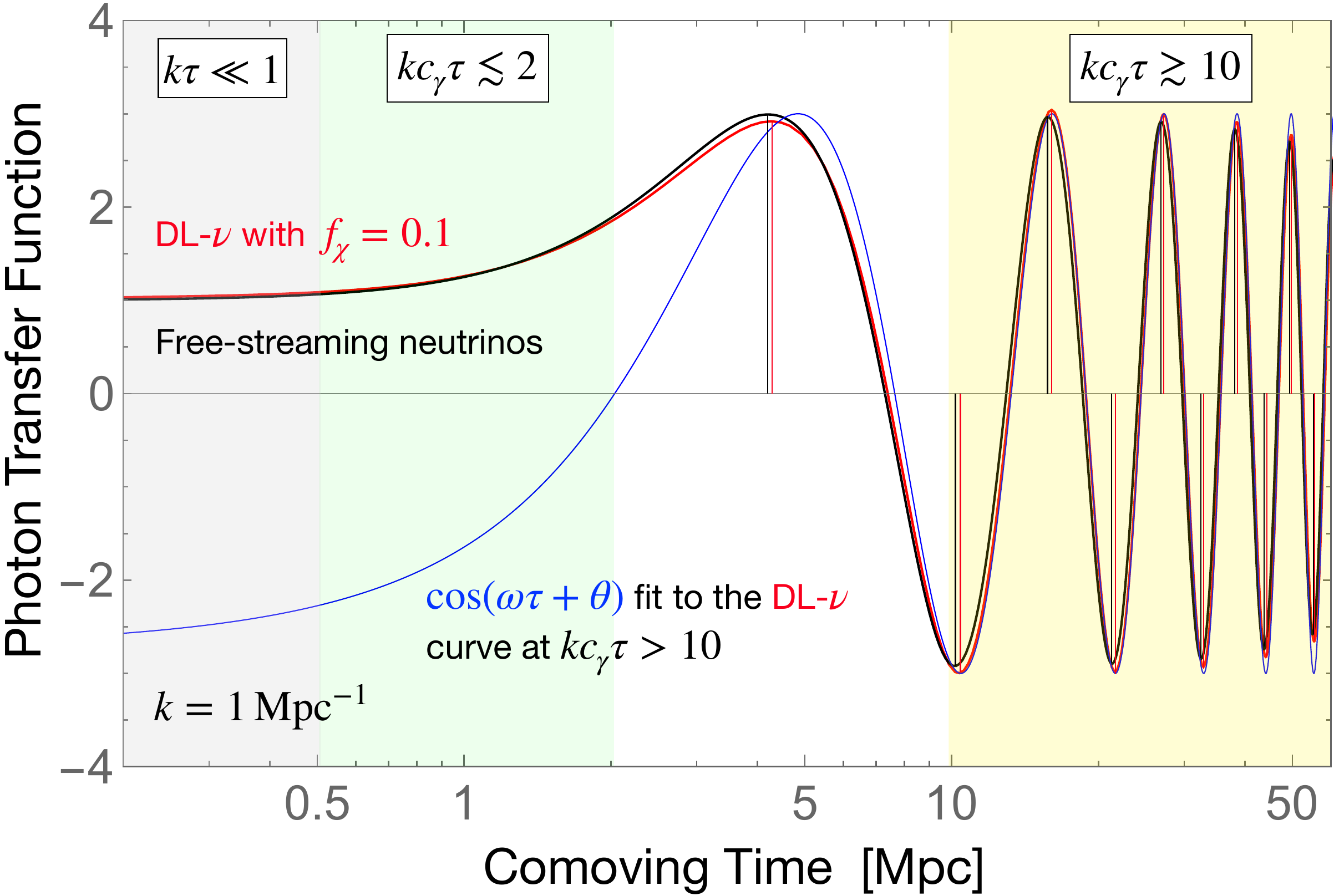}
  \caption{Evolution of the photon transfer function from \CLASS~calculations, comparing the free-streaming neutrino scenario (black) to the DL-$\nu$ scenario with $f_\chi=0.1$ (red). Perturbations grow mildly outside the horizon (grey region). The two curves evolve similarly when $kc_\gamma\lesssim\frac{2}{\tau}$ (Eq.~(\ref{eq.Fk}), green region) until perturbations begin oscillating around the peak in the white region. As $k c_\gamma\tau\gg 2$ (yellow), oscillations resemble freely propagating waves. The blue curve represents a plane wave fit to the DL-$\nu$ curve in the yellow region, characterized by a constant $\omega = \frac{0.97}{\sqrt{3}}k$. The phase shift between the black and red curves is determined within the white time window, where the driving force $F_{\rm driv}(\tau)$ approaches $k^2 c_{\gamma,\nu}^2$. Vertical lines mark peak locations in the DL-$\nu$ (red) and free-streaming (black) scenarios, showing a positive phase shift in the DL-$\nu$ curve. 
  }
  \label{cartoon}
\end{figure}
As depicted in Fig.~\ref{cartoon} obtained using \CLASS, the evolution of the photon's transfer function unfolds in three steps. Initially, the perturbations only grow mildly before fully entering the horizon (grey region). Between $k\tau\approx 1$ and $kc_\gamma\tau\approx 2$ (green region), the perturbations start to grow, and the evolutions of $\delta_{\gamma,\nu}$ are governed by the same gravitational interaction $F_{\rm driv}(\tau)$ in Eqs.~(\ref{toyeqn1}) and (\ref{toyeqn2}). This gives identical evolution for all fluids, whether neutrinos are free-streaming or interacting with or without DM-loading. Conversely, gravitational interaction becomes negligible as $k c_\gamma \tau\gg2$ (yellow region), and $\delta_{\gamma,\nu}$ oscillates with their natural frequencies $kc_{\gamma,\nu}$. There is no further change in the phase shift upon this point. As a result, we can fit the oscillation frequency and phase in the interacting neutrino scenario with constants (blue curve). As depicted in the white region, $\delta_\gamma$ with DM-loading (red) develops a positive phase shift compared to the free-streaming neutrino case (black) when $k c_\gamma \tau\approx 2$. 

In the radiation-dominated era, $F_{\rm driv}\approx k^2c_\gamma^2$ happens when $kc_\gamma\tau\approx 2[1-(kc_\gamma\tau_{eq})^{-1}]$. Since the decoupling of the driving force $F_{\rm driv}(\tau)$ happens right after this time scale, to simplify the discussion, we define a fixed $\mathcal{O}(1)$ number $\alpha$ such that the phase shift is fixed around time $\tau_\alpha$
\begin{equation}
kc_\gamma\tau_\alpha=2\alpha\left(1-\frac{1}{kc_\gamma\tau_{eq}}\right)\,. 
\end{equation}
As discussed in Appendix~\ref{app.toymodel}, we find $\alpha\approx1.5$ by matching the oscillation phase of the numerical solutions of Eqs.~(\ref{toyeqn1}) and (\ref{toyeqn2}) to harmonic oscillator solutions. Given $\tau_{eq}\approx 110$~Mpc, when considering modes with $k=\mathcal{O}(0.1)$~Mpc$^{-1}$, the $k$-dependent term in the parenthesis only introduces an $\mathcal{O}(10)\%$ percent level correction to the result. The interplay between the weakening of $F_{\rm driv}$ and the growth of $\delta c$ dictates the evolution of the phase shift. Despite the toy model offering a more concise approximation of the \CLASS~result, the time-dependence in $F_{\rm driv}(\tau)$ and $\delta c(\tau)$ still complicates obtaining analytical insights on the parametric dependence of the phase shift. To further simplify the system, we exploit the fact that at the timescale $\tau_\alpha$ we focus on, both the changing rate of $|\dot{F}_{\rm driv}/F_{\rm driv}|\approx\tau_\alpha^{-1}=(2\alpha)^{-1}k c_\gamma$ and the difference in the oscillation frequency $k\delta c $ are slower than the natural oscillation frequency $\approx kc_\gamma$. We then assume $F_{\rm driv}$ and $k\delta c$ to be constants in time when deriving corrections to the $\delta_\gamma$ oscillation. While this assumption is somewhat crude, given that $\tau_\alpha^{-1}$ is not significantly smaller than $kc_\gamma$, it serves as a useful approximation for identifying the parametric dependence of the phase shift from otherwise intricate equations.

In the SI-$\nu$ scenario ($\delta c=0$), Eqs.~(\ref{toyeqn1}) and (\ref{toyeqn2}) around $\tau_\alpha$ can be approximated as\footnote{As discussed below Eq.~(\ref{sec:SInuAP}), the approximation should come with an initial phase $\phi_{\rm in}=2\alpha+\tan^{-1}\left(\frac{1+\sqrt{33}}{4\alpha}\right)$ in the oscillation. The phase is determined by matching the initial power-law solution of Eq.~(\ref{toyeqn1}), when $F_{\rm driv}\gg k^2c_\gamma^2$, to the cosine solution starting at $\tau_\alpha$. Given our interest solely in the phase difference among different models, the overall phase $\phi_{\rm in}$ becomes irrelevant for comparison. We hence omit $\phi_{\rm in}$ in the subsequent discussion and concentrate on the additional phase generated from DM-loading.}
\begin{equation}\label{sec:SInu}
\delta_\gamma(k,\tau) =\delta_\nu(k,\tau) \approx  \cos(\omega\tau)\,,\quad \omega=kc_\gamma\,.
\end{equation}
See Appendix~\ref{app.toymodel} for more discussions. Once the neutrino sound speed decreases due to DM-loading, $\delta c$ slows down the neutrino's oscillation frequency to $\omega-k\delta c$, hence altering the driving force of the photon oscillation:
\begin{eqnarray}\label{eq.theta}
F_{\rm driv}\left[ f_\gamma \cos(\omega\tau) + f_\nu \cos((\omega - k\delta c) \tau) \right]\approx F_{\rm driv}\left[\cos(\omega\tau) + f_\nu k \,\delta c\,\tau \sin(\omega\tau)\right]\,.
\end{eqnarray}
We use $f_\gamma+f_\nu=1$ and the small angle approximation, incorporating $\sin(k\delta c\tau)\approx k\delta c\tau$ in the second equation. As the oscillation from $k\delta c$ is much slower than $\omega$, we treat $k\tau=k\tau_\alpha$ as static. The presence of the $\sin(\omega\tau)$ oscillation generates a phase shift $\cos(\omega\tau+\Delta\phi_{\rm load})$ to the initially cosinusoidal driving force with 
\begin{equation}
\Delta\phi_{\rm load} \approx -f_\nu\left(\frac{\delta c}{c_\gamma}\right)2\alpha\left(1-\frac{1}{kc_\gamma\tau_{eq}}\right)
= -\frac{3\alpha^2 f_\chi f_{\rm DM}}{kc_\gamma\tau_{eq}}\left(1-\frac{1}{kc_\gamma\tau_{eq}}\right)^2
\frac{c_\nu^2}{c_\gamma(c_\gamma+c_\nu)}\,,
\end{equation}
where we use the correction to the sound speed in Eq.~(\ref{eq.Dcgamma}). 
 When absorbing the angle as a shift to the oscillation, the solution $\cos(\omega\tau+\Delta\phi_{\rm load})$ still satisfies Eq.~(\ref{toyeqn1}) under the approximation of static $F_{\rm driv}$ and $c_\gamma$. The phase shift becomes
\begin{equation}
\label{eq:theta_approx}
\Delta\phi_{\rm load}\approx-\frac{3\alpha^2f_{\rm DM}}{2c_\gamma\tau_{eq}}\frac{f_\chi}{k+a}\,,\qquad a=\frac{1}{c_\gamma\tau_{eq}}\left(2+\frac{3}{4}\frac{\alpha f_{\rm DM}}{f_\nu}f_\chi\right)\,.
\end{equation}
When considering modes with $k=\mathcal{O}(0.1)$~Mpc$^{-1}$, the estimate gives a phase shift between DL-$\nu$ vs. SI-$\nu$ with size $\Delta\phi_{\rm load}\sim -0.1f_\chi$, and the shift of the $k$ pole $a\sim 10^{-2}$~Mpc$^{-1}$. 

At a later time $\tau$, the acoustic peaks in the perfect fluid scenario exist when $\omega\tau\approx k_{\rm peak}c_\gamma\tau=n\pi$. In the radiation-dominated era, the peak location in $k$ obtains a positive shift with size
\begin{equation}\label{eq.dkresult}
\delta k\approx\frac{-\Delta\phi_{\rm load}}{c_\gamma\tau}\approx \frac{3\alpha^2f_{\rm DM}}{2c_\gamma^2\tau_{eq}}\frac{f_\chi}{(k+a)\tau}\approx 0.07 f_\chi(k\tau)^{-1}\,{\rm Mpc}^{-1}
\end{equation}
if taking $\alpha=1.5$ in the last expression. Considering comoving time and wave number with $k\tau\approx 30$ in Fig.~\ref{dknu3RAD}, the estimated $\delta k\approx 2\times10^{-3}f_\chi$~Mpc$^{-1}$ aligns with the order of magnitude of the \CLASS~results (solid curves). Although in the analytical approximation we treat the photon sound speed $c_\gamma$ in Eq.~(\ref{eq.dkresult}) as a constant at later times, it decreases by about $10\%$ from the deep radiation-dominated era to $\tau_{eq}$ in the \CLASS~calculation. We hence expect $\delta k$ to increase over time by $\approx20\%$ when comparing results with the same $k\tau$ between Figs.~\ref{dknu3RAD} and~\ref{dknu3EQB}.
Nevertheless, the estimate from Eq.~(\ref{eq.dkresult}) effectively captures the order of magnitude for $\delta k$ and explains the linear dependence on $f_\chi$.
\begin{figure}[t!]
    \centering
   \includegraphics[width=12cm]{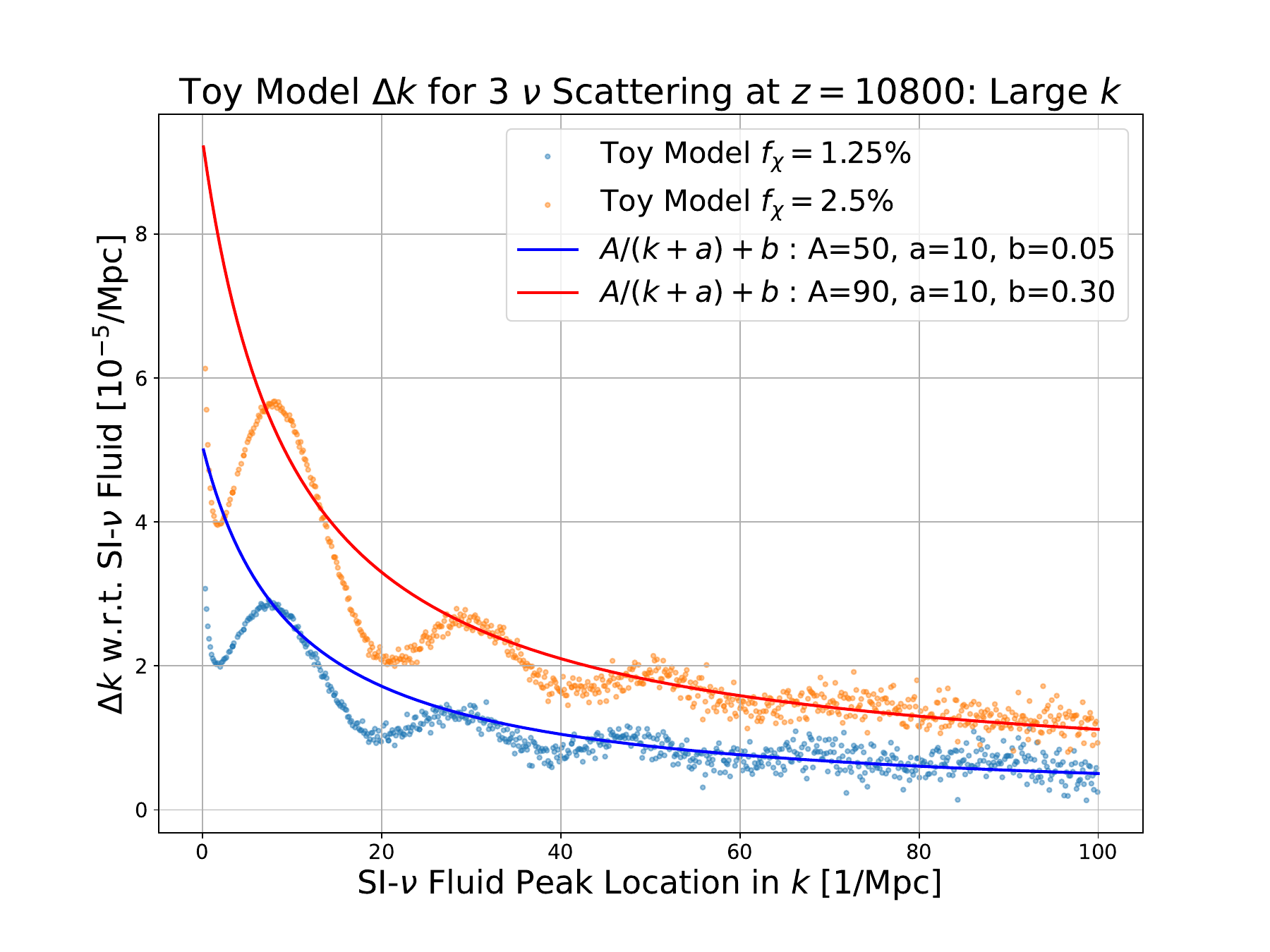}
    \caption{Toy model: large $k$ dependence of $\Delta k$ for $f_\chi = 1.25\%$ (blue) and $2.5\%$ (orange) DL-$\nu$ with respect to SI-$\nu$ for 3 scattering neutrinos. Points are obtained by solving the toy model numerically up to $k=100/\rm Mpc$ at redshift $z=10800$. Functions $A/(k+a)-b$, where $A, a, b$ are constant parameters, are plotted to show the inverse $k$ dependence at large $k$.}
    \label{ToyLargeK}
\end{figure}

The $k^{-1}$ dependence in Eq.~(\ref{eq.dkresult}) suggests a decoupling limit for $\delta k$ to vanish at large $k$. This limit is anticipated because $\delta c$ is smaller at an earlier time when a perturbation mode with higher $k$ enters the horizon. While it is challenging to obtain $\delta k$ from \CLASS~for large enough $k$-modes due to the severe diffusion damping of $\delta_\gamma$, the toy model provides a glimpse of the $k$-suppression. In Fig.~\ref{ToyLargeK}, we show the phase shift obtained by numerically solving the toy model at $\tau=40$~Mpc for large $k$-modes, and the $\delta k\propto (k+a)^{-1}$ dependence in Eq.~(\ref{eq.dkresult}) does capture the suppression pattern of the phase shift. Our analytical approximation does not reproduce the oscillation in the phase shift. We suspect it comes from the time dependence in $c_\gamma(\tau)$ since the oscillations disappear when setting $c_\gamma$ as a constant. We leave a more detailed understanding of the large-$k$ dependence for future work. Considering CMB measurements, we also do not anticipate observing the phase shift at such large $k$-modes.

In the flat-sky approximation, the shift in the acoustic peaks of $C_\ell^{\rm TT,EE}$ spectra is related to the $k$-modes as $\delta\ell/\Delta\ell_{\rm peak}\approx\delta k/\Delta k_{\rm peak}\approx \frac{0.07c_\gamma}{\pi k}f_{\chi}$, where $\Delta\ell_{\rm peak}$ and $\Delta k_{\rm peak}$ are the distance between oscillation peaks at recombination. For modes entering the horizon in the radiation-dominated era, $\Delta\ell_{\rm peak}\approx 330$, our analytical estimate gives $\delta\ell\approx 120 f_\chi$ with $k\approx 0.1$~Mpc$^{-1}$, which is not far from fitting the \CLASS~result $\delta\ell\approx 80(1\pm0.5)f_\chi$ in Fig.~\ref{fxCl}.  
\begin{figure}[t!]
\centering
\includegraphics[width=10.3cm]{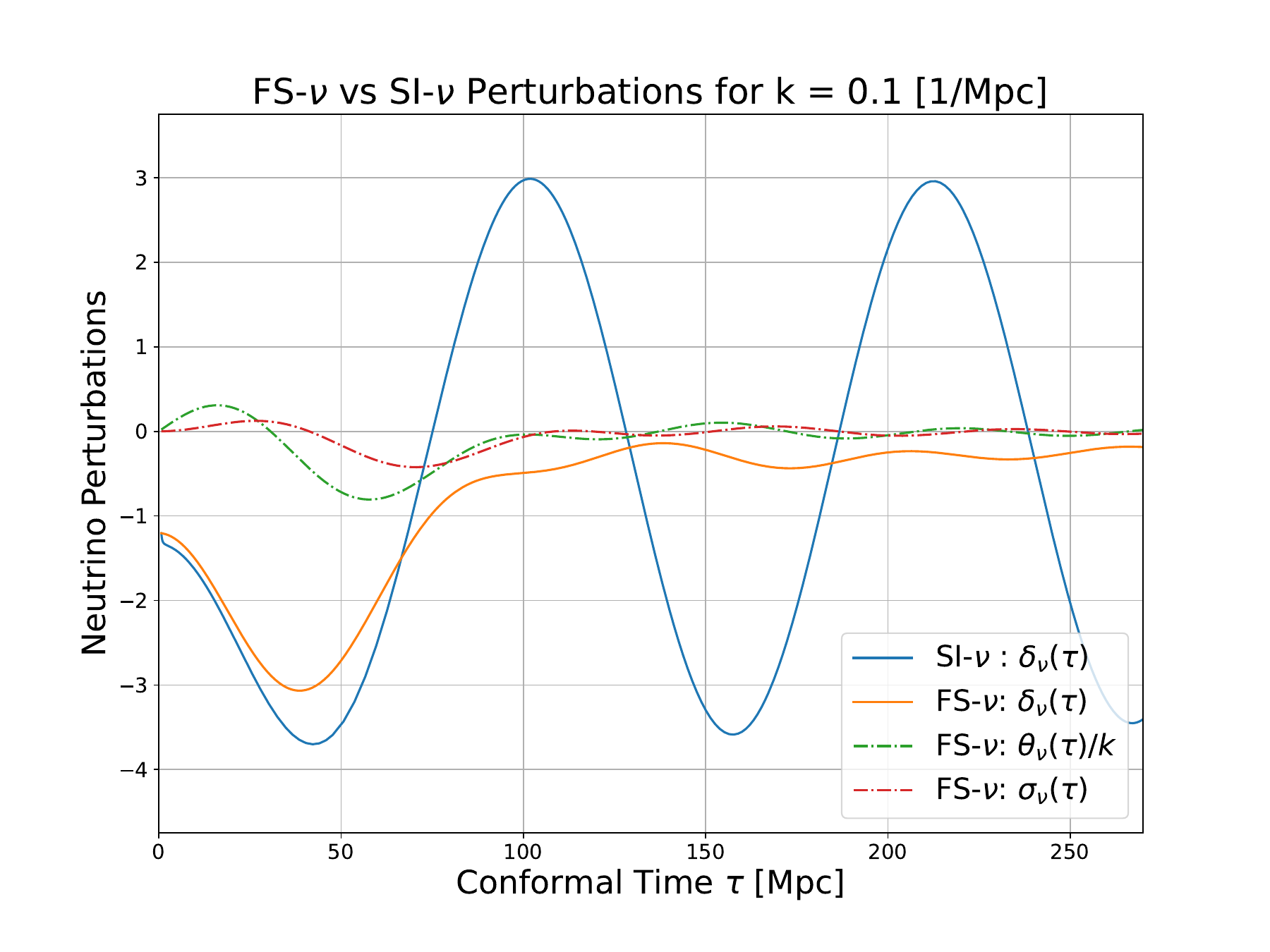} 
\vfil
\centering
\includegraphics[width=10.3cm]{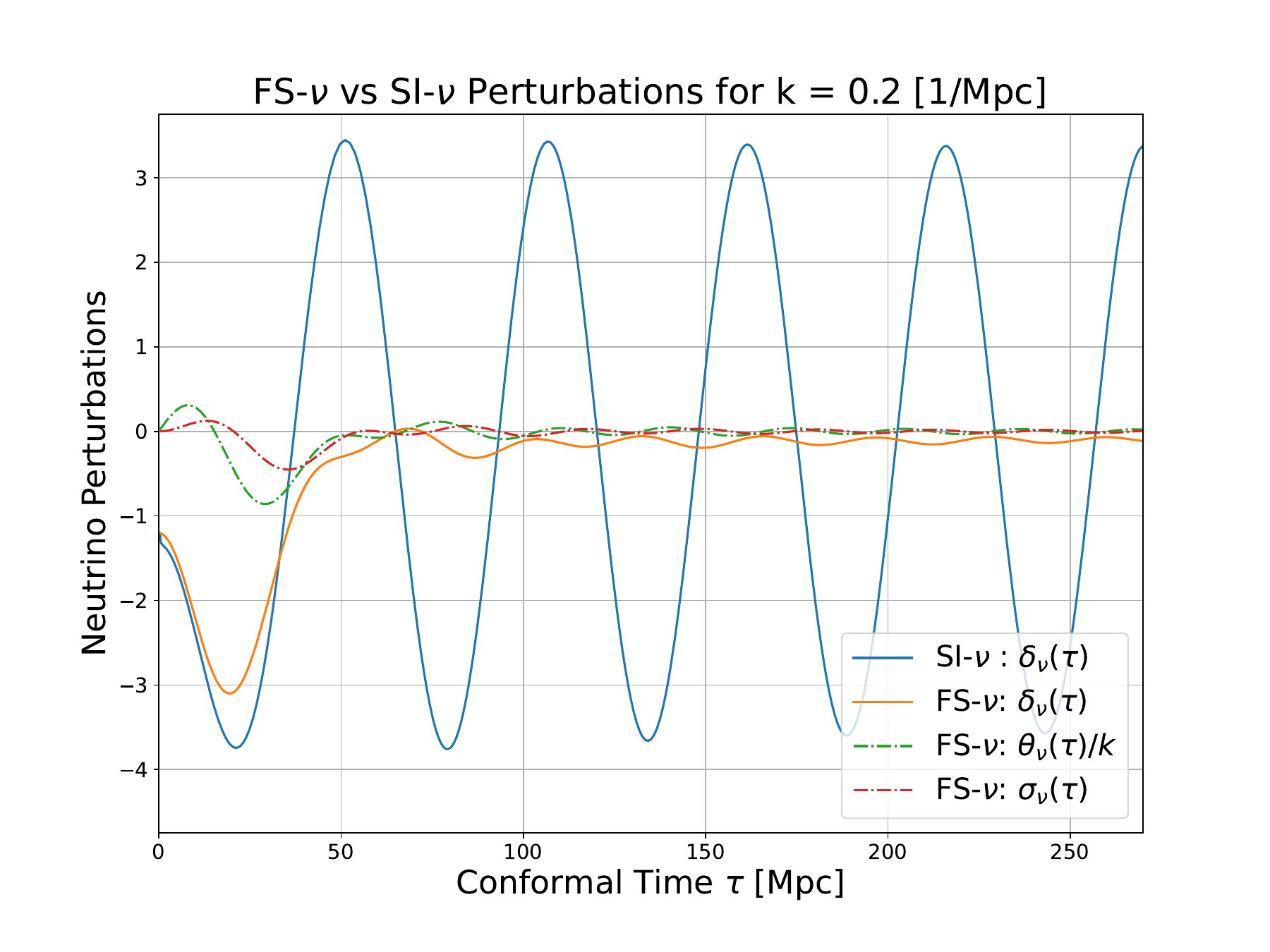}
\caption{\CLASS: Evolution of $\delta_\nu$ perturbations in conformal time for FS-$\nu$ and SI-$\nu$ at two different $k$-modes, with $k = 0.1/\rm Mpc$ on the top and $k = 0.2/\rm Mpc$ below. For comparison, the dipole $\theta_\nu/k$ and shear $\sigma_\nu$ for FS-$\nu$ are included.}
\label{FsFld}
\end{figure}

When considering scenarios where only a fraction $f_\nu^{\rm int}$ of neutrinos are interacting, and $f_\nu^{\rm fs}$ fraction are free-streaming (with $f_\nu^{\rm int}+f_\nu^{\rm fs}=f_\nu$), the modified driving force for photon oscillation is given by:
\begin{eqnarray}\label{eq.FSnu}
F_{\rm driv}\left[ f_\gamma A\cos(\omega\tau) + f_\nu^{\rm int} A\cos(\tilde\omega_{\rm int}\tau) + f_\nu^{\rm fs} A\cos(\tilde\omega_{\rm fs}\tau) \right],
\end{eqnarray}
where $\tilde\omega_{\rm int}\approx\omega-k\delta c$ as before. The free-streaming neutrino has $\tilde\omega_{\rm fs}\to 1$ at later evolution, significantly different from the $\omega$ of interacting radiation. However, as illustrated in Fig.~\ref{FsFld}, before $\delta_\nu$ (orange) reaches its first oscillation peak at $kc_\nu\tau\approx\pi$, the shear $\sigma_\nu$ (red dotted) has not increased significantly. The diffusion damping has not started yet, and $\tilde\omega_{\rm fs}\approx\omega$. Since the phase shift is also fixed right before $kc_\nu\tau_\alpha=2\alpha\approx\pi$, we can write $\tilde\omega_{\rm fs}=\omega+\delta\omega_{\rm fs}$ with $\omega\gg\delta\omega_{\rm fs}>0$. The driving force around the $\tau_\alpha$ time we consider approximates
\begin{eqnarray}
F_{\rm driv}A\left[\cos(\omega\tau) + (f_\nu^{\rm int} k \,\delta c- f_\nu^{\rm fs} \delta\omega_{\rm fs})\tau_\alpha \sin(\omega\tau)\right]\,
\end{eqnarray}
with $f_\gamma+f_\nu=1$. When considering the phase shift between DL-$\nu$ and SI-$\nu$ with the same fraction $f_\nu^{\rm int}$ of interacting neutrinos, the phase shift from $f_\nu^{\rm fs} \delta\omega_{\rm fs}$ cancels out, and the enhanced phase shift is still approximated as Eq.~(\ref{eq.theta}) if $f_\chi\ll f_\nu^{\rm int}$. The phase shift between the two scenarios therefore remains insensitive to $f_\nu^{\rm int}$ as shown in Figs.~\ref{fxCl} and~\ref{dkfx}. Applying the same approximation, the phase shift of FS-$\nu$ compared to SI-$\nu$, commonly discussed in the literature, can be approximated as $\cos(\omega\tau+\Delta\phi_{\rm int})$ for an initially cosinusodial wave with
\begin{equation}\label{eq.fsnu}
\Delta\phi_{\rm int}\approx\alpha\left(\frac{\delta\omega_{\rm fs}}{kc_\nu}\right)f_\nu^{\rm fs}=\mathcal{O}(0.1)\times f_\nu^{\rm fs}\,.
\end{equation}
This FS-$\nu$ shift is in the opposite direction compared to the DL-$\nu$ shift in Eq.~(\ref{eq:theta_approx}).
In the second equality we use the fact that $\alpha\approx 1.5$, $\omega\approx kc_\nu$, and $(\delta\omega_{\rm fs}/\omega)=\mathcal{O}(0.1)$ from the numerical solution depicted in Fig.~\ref{FsFld}. This linear dependence is similar to the approximation $\Delta\phi_{\rm int}\approx 0.191\pi f_\nu^{\rm fs}$ derived in~\cite{Baumann:2015rya}, up to first order in $f_\nu^{\rm fs}$. However, to determine the pre-factor, a more precise estimation of $\delta\omega_{\rm fs}$ from the diffusion damping process is necessary, as discussed in Ref.~\cite{Bashinsky:2003tk,Baumann:2015rya}.


\subsection{Phase shift from $N$-copies of DR-DM system}\label{sec.DMDR}

DR-DM scattering has been extensively explored in studies of dark sector cosmology. Our analysis of the phase shift in the DL-$\nu$ scenario is readily applicable to these DR-DM models. In this context, one can substitute $f_\nu^{\rm int}$ in Eq.~(\ref{eq.FSnu}) with $f^{\rm int}_{\rm DR}=\rho^{\rm int}_{\rm DR}/(\rho_\gamma+\rho^{\rm fs}_\nu+\rho^{\rm int}_{\rm DR})$, representing the fraction of interacting DR. The enhanced phase shift due to DM-loading emerges as a generic feature across dark sector models featuring DR-DM scattering.

\begin{figure}
    \centering
    \includegraphics[width=12cm]{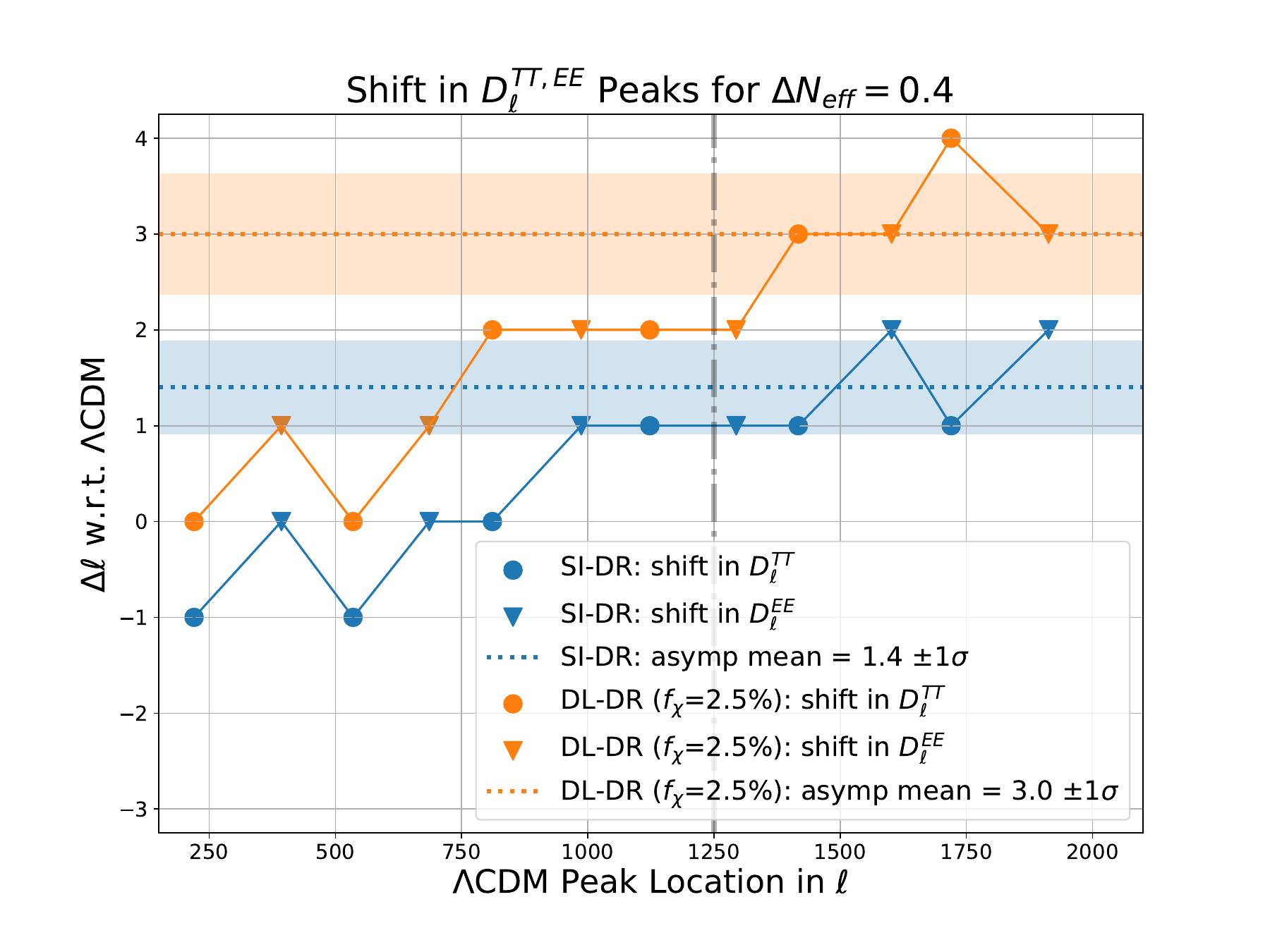}
    \vfil
    \includegraphics[width=12cm]{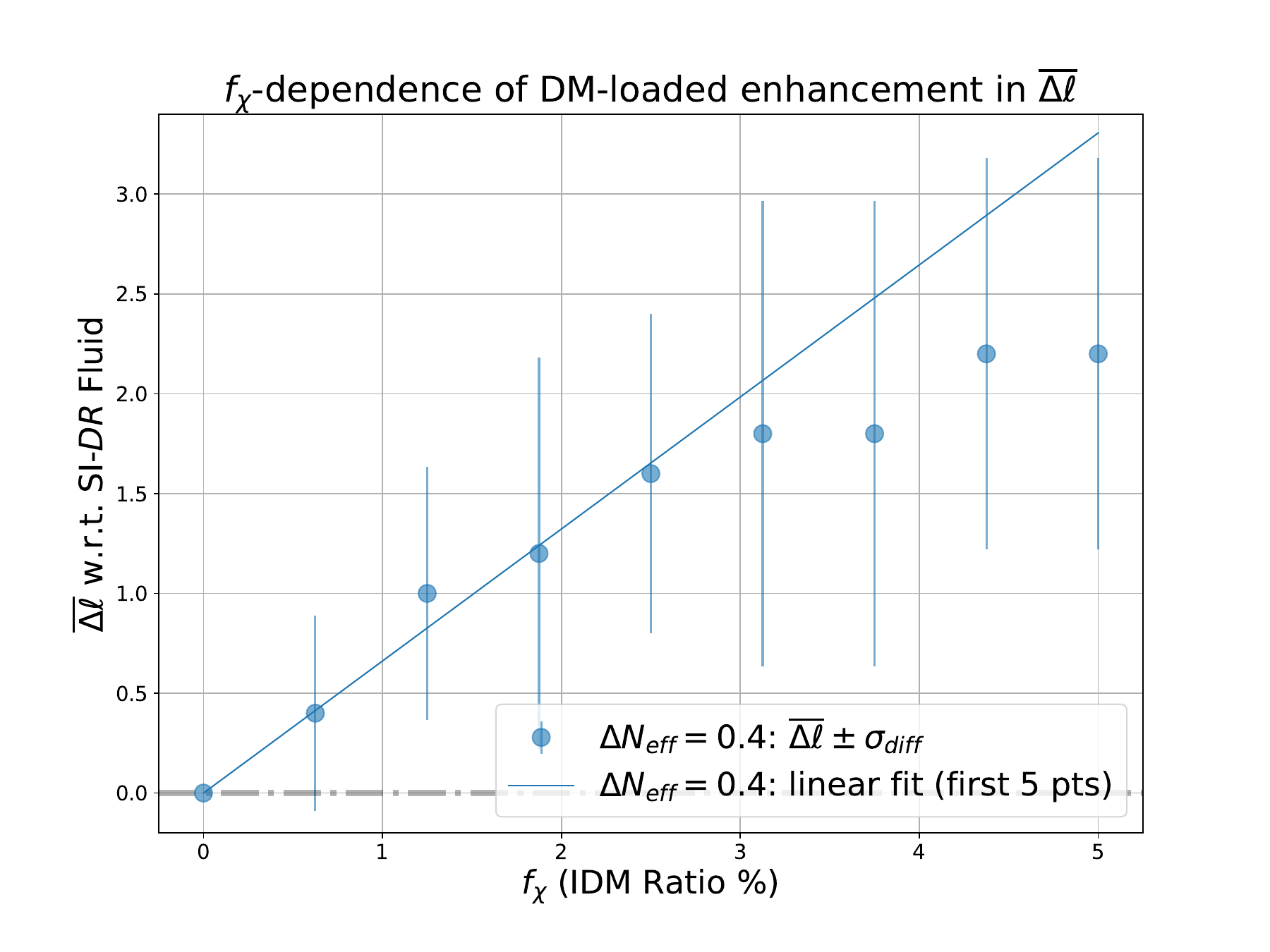}
    \caption{CMB phase shift enhancement from \CLASS~for interacting DR-DM, with $N_{\rm DR}=0.4$ and all neutrinos free-streaming. \emph{Top}: $\Delta\ell$ shift in peaks in $D_\ell^{TT}$ and $D_\ell^{EE}$ lensed spectra for SI-DR (blue) and $f_\chi = 2.5\%$ DL-DR (orange), with respect to $\Lambda$CDM. Lines connecting the points have been provided to help guide the eye. As in Fig.~\ref{CLTTEE}, the asymptotic mean (``asymp mean") $\overline{\Delta\ell}$ and standard deviation $\sigma$ of the $\Delta\ell$'s in the deep radiation era are shown. \emph{Bottom}: $f_\chi$-dependence of the $\overline{\Delta\ell}$ enhancement on top of SI-DR  ($f_\chi = 0$). Error bars correspond to the standard deviation $\sigma_{\mathrm{diff}}$ in the differences between the computed $\Delta\ell$'s, quantifying the relative integer rounding fluctuation of points with respect to SI-DR. A linear fit to the first 5 points is provided.}
    \label{CLTTEE_dNeff}
\end{figure}

To examine the phase shift in the DR-DM system, we incorporate interacting DR within the ETHOS framework in \CLASS~\cite{Cyr-Racine:2015ihg}. We assume all SM neutrinos as free-streaming and introduce an extra DR component that undergoes scattering with $\chi$ for the DM-loading DR (DL-DR) scenario. This is compared to the self-interacting DR (SI-DR) for an additional phase shift. In Fig.~\ref{CLTTEE_dNeff} \emph{left}, we show the phase shift of DL-DR (orange) and SI-DR (blue) models relative to the free-streaming DR (FS-DR) scenario, assuming a large $\Delta N_{\rm eff}=0.4$ that has been considered to address the $H_0$ problem~\cite{Planck:2018vyg,Blinov:2020hmc} and $f_\chi=2.5\%$. The enhanced phase shift in the DL-DR model is evident. Similar to the interacting neutrino case, the extra phase shift from DR-DM scattering is proportional to $f_\chi$ (Fig.~\ref{CLTTEE_dNeff} \emph{right}), as long as $f_\chi\ll f^{\rm int}_{\rm DR}\approx 5\%$ in this case. 

The study of the DL-$\nu$ and DL-DR gives us insights into how a `loaded' fluid can generate an additional phase in the photon acoustic peaks, and we can generalize the idea for $N$ isolated dark sectors, each with their own interacting DR-DM system. For concreteness, let us consider multiple efficiently scattering relativistic fluids DR$_i - \chi_i$, indexed by $i = 1,...,N$, where the DR component DR$_i$ can include scattering $\nu$. The toy model then describes $N+1$ coupled oscillators for the photon-baryon fluid and the $N$ dark sector fluids. Consider again the solution with initial conditions $\delta_{\gamma,{\rm DR}}(\tau_{\rm in})=1$, $\dot\delta_{\gamma,{\rm DR}}(\tau_{\rm in})=0$, and drop the same initial phase $\phi_i$ and amplitude $A$
\begin{equation}
\delta_\gamma(\tau) = \cos( kc_\gamma\tau)\,,\quad\delta_{{\rm DR}_i}(\tau) = \cos(kc_{{\rm DR},i}\tau)   
\end{equation}
with DR sound speed $c_{{\rm DR},i} = c_\gamma - \delta c_i$. Assuming $\delta c_i\ll c_\gamma$ for small deviations from photon oscillations for all $i$, the driving force contribution in Eq.~(\ref{toyeqn1}) is given by
\begin{align}
F_{\rm driv}[f_\gamma\delta_\gamma(\tau) + \sum_i f_{{\rm DR},i}\delta_{{\rm DR}_i}(\tau)]\approx F_{\rm driv}[\cos(kc_\gamma\tau) + k\tau_\alpha\left(\sum_if_{{\rm DR},i}\delta c_i\right) \sin(kc_\gamma\tau)]\,,
\end{align}
where we apply the small angle approximation for $k\tau_\alpha\delta c_i=2\alpha\left(\frac{\delta c_i}{c_\gamma}\right)\ll1$ with $\alpha$ being an order one number. Given that $2\alpha\left(\frac{\delta c_i}{c_\gamma}\right)\approx\frac{3\alpha f_{\chi,i}f_{\rm DM}}{2f_{{\rm DR},i}}\frac{\tau_\alpha}{\tau_{eq}}$, the total phase shift from the dark sectors becomes
\begin{equation}
\Delta\phi_{\rm total}\approx\frac{3\alpha^2}{kc_\gamma\tau_{eq}}\sum_if_{\chi,i}\,.
\end{equation}
Therefore, in the case where all dark sectors are dominated by the DR energy, i.e., $f_{{\rm DM},i}\ll f_{{\rm DR},i}$, the additional phase shift compared to the SI-DR should be linearly proportional to the sum of interacting dark matter energy density, not the individual dark sectors. To numerically confirm the statement with \CLASS, it is necessary to generalize the interacting dark matter module to include multiple copies of DR-DM system, and we defer the task to future studies.

\section{MCMC Analysis}\label{sec.MCMC}

In this section, we investigate the impact of the phase shift induced by DM-loading on cosmological data. As illustrated in Figs.~\ref{fxCl} and \ref{CLTTEE_dNeff}, the variation in the sound speed due to DM-loading leads to a shift in CMB multipole peaks ($\Delta \ell$), exhibiting a linear relationship with $f_\chi$ when $f_\chi \lesssim f_{\nu,{\rm DR}}$ at matter-radiation equality. Additionally, for a fixed $f_\chi$, we observe an approximately linear relation between $\Delta \ell$ and $\ell$ up to $\ell \approx 1200$, beyond which $\Delta \ell$ begins to plateau or grow slower than $\ell$ at higher $\ell$ modes as can be seen from Fig.~\ref{CLTTEE} and \ref{CLTTEE_dNeff}. Although we cannot extract the $\Delta \ell$ shifts for high-$\ell$ modes due to diffusion damping, the intuition from the toy model analysis (Fig.~\ref{ToyLargeK}) suggests that the shift should vanish for very high $k$ which corresponds to high-$\ell$ modes.


One signature of modification of the acoustic phase shift in CMB is a shift in the angular sound horizon $\theta_s$~\cite{Ghosh:2019tab}. The locations of the acoustic peaks of the CMB correspond to the extremas of the photon transfer function which can be approximated as $\cos (kr_s + \phi)$ where $k$ is the corresponding wavenumber, $r_s$ is the sound horizon and $\phi$ is the phase shift. In the multipole space, the location of the $m$-th acoustic peak corresponds to,
\begin{equation}
    \label{eq:l_peak}
    \ell_m = k_m D_A = \cfrac{m\pi - \phi}{r_s}D_A = \cfrac{m\pi - \phi}{\theta_s}\;,
\end{equation}
where $D_A$ is the angular diameter distance and $\theta_s = r_s / D_A$. Since the observed peak positions are very well measured in CMB experiments, Eq.~\eqref{eq:l_peak} implies that a change in the phase shift $\Delta\phi$ will result in a shift, $\Delta\theta_s/\theta_s\approx-\Delta\phi/m\pi$, for a good fit of the CMB data. Models that modify neutrino induced phase of the CMB, such as DL-$\nu$ interaction and neutrino-self interaction, all predict a significant shift in the $\theta_s$ from its $\Lambda$CDM value~\cite{Ghosh:2019tab,Choi:2018gho,Ghosh:2021axu}.


Compared to the typically assumed FS-$\nu$ scenario in the $\Lambda$CDM model, the DL-$\nu$ or DL-DR scenarios experience phase shifts from two sources.
\begin{enumerate}
    \item \underline{Phase shift from self-interacting neutrinos}: The dominant modification of the acoustic phase shift originates because the $\nu$ scattering stops neutrinos from free-steaming. 
    The modification of the phase shift w.r.t. the $\Lambda$CDM with FS-$\nu$ is,
    \begin{equation}
        \Delta\phi_{\rm int}-\Delta\phi_{\Lambda{\rm CDM}}
        \propto \frac{\rho_{\nu}^{\rm fs}}{\rho_\gamma+\rho_\nu^{\rm int}+\rho_\nu^{\rm fs}}-\frac{\rho_\nu}{\rho_\gamma+\rho_\nu} \sim (f_\nu^{\rm fs}-0.41)\;,
    \end{equation}
    where $f_\nu^{\rm fs}$ is the amount of scattering radiation as defined in Eq.~(\ref{eq.fsnu}). Thus, this effect is proportional to the amount of scattering radiation fraction i.e, $f_\nu$.
    Note that, this is the sole modification of the phase shift where the neutrino only interacts with itself to stop free streaming such is the SI-$\nu$ case.
    \item \underline{Additional phase shift from DM-loading}: The phase shift from DM-loading represents an additional contribution where neutrinos (or DR) are scattered by non-relativistic particles like dark matter (or baryons), reducing the sound speed of the DL-$\nu$ fluid below the threshold of a perfect fluid, as illustrated in Fig.~\ref{CsPlot}. The modification of the phase shift due to this effect, as shown in Eq.~\eqref{eq.dkresult}, is independent of the amount of scattering by neutrinos ($f_\nu$) when $f_\chi \lesssim f_\nu$ at matter-radiation equality, and depends linearly on $f_\chi$:
    \begin{equation}
        \Delta\phi_{\rm load} \propto -f_\chi\;.
    \end{equation}
\end{enumerate}

\noindent
 In the subsequent analysis, we probe $\Delta\phi_{\rm load}$ using MCMC analysis with cosmological data. As outlined earlier in this section, $\Delta\phi$ leads to changes in $\theta_s$. Thus, any deviation of $\theta_s$ from the $\Lambda$CDM value will indicate the presence of a phase shift. A positive correlation between $\theta_s$ and $f_\chi$ from the MCMC analysis will suggest the existence of DM-loading effects in our study. Noting that while many of the scenarios described here have been explored in the literature, the physics of DM-loading has been largely overlooked, to our knowledge. Our primary aim with the MCMC analysis is \emph{not} to update bounds with new datasets, but rather to highlight the DM-loading effects that should already exist in these models. For instance, if future data like CMB-S4 shows additional radiation energy, we could compare the fitted values of $\theta_s$ and the suppression of matter power spectrum to see if they align with a non-zero $f_\chi$. A match would strongly support the presence of DR-DM scattering.

 \subsection{Datasets and Methodology}

 We use the combination of the following datasets for our MCMC analysis.
 \begin{itemize}
     \item \underline{Planck}: The Planck 2018 dataset which consists of low-$\ell$ $(\ell < 30)$ TT, EE and high-$\ell$ $(\ell \geq 30)$ TT,TE,EE measurement~\cite{Planck:2019nip}. It also includes the Planck Lensing likelihood~\cite{Planck:2018lbu}.
     \item \underline{BAO}\footnote{ Non-standard phase shift does modify the perturbation template with which BAO peak location is extracted from sky surveys. However, Ref.~\cite{Bernal:2020vbb} showed that these effects do not introduce any significant bias in the conventional BAO analysis and it's safe to use BAO scales derived from those analyses. 
     Note that `DNI' models Ref.~\cite{Bernal:2020vbb} does contain DM-loading effects and particularly `DNI-2' where ~$2\%$ of DM is interacting with neutrinos.
     See also Ref.~\cite{Baumann:2017lmt,Baumann:2017gkg,Baumann:2019keh} for effects of phase shift on the shape and amplitude of the BAO spectrum.}:  The BAO dataset consists of measurements from 6DF
Galaxy survey~\cite{Beutler_2011}, SDSS-DR7 MGS data~\cite{Ross:2014qpa}, and the BOSS measurement of BAO scale and
$f\sigma 8$ from DR12 galaxy sample~\cite{BOSS:2016wmc}.
     \item \underline{Ext (SH0ES + KV450):} The other datasets (EXT) contains the Hubble constant measurement $(H_0 = 73.04 \pm 1.04 {\rm km/s/Mpc})$ from the SH0ES collaboration~\cite{Riess:2021jrx} and matter power spectrum shape measurement from KiDS + Viking 450 (KV450)~\cite{Hildebrandt:2018yau}. For KV-450 we use data up to $k_{\rm max} = 0.3h ~{\rm Mpc}^{-1}$ limiting the analysis from the region where non-linear effects are important.
 \end{itemize}

 We use \texttt{MontePython} to perform the MCMC analysis with the Metropolis–Hastings algorithm~\cite{Audren:2012wb,Brinckmann:2018cvx,Hastings:1970aa}. We use \texttt{GetDist} for analysis and plotting of the MCMC samples~\cite{Lewis:2019xzd}.

 \subsection{Strong $\nu$-DM interaction}

 To demonstrate the $\theta_s$ dependence on DM-loading, we first focus on DL-$\nu$ scenario with a very strong coupling and assume all SM neutrinos to interact with $f_\chi$ fraction of DM\footnote{
 In the ETHOS parameterization, we set \textbf{N\_idr} $= 3.046$, \textbf{N\_ur} $= 0$, \textbf{a\_idm\_dr} = $10^4/f_\chi$ with $T_\nu^2$ dependent interaction and varied $f_\chi$. Note that the exact  $T_\nu$-dependence of the scattering rate $\dot\kappa_{\rm DR-DM}$ does not have any impact as long as $\dot\kappa_{\rm DR-DM}\tau\gg 1$.}. 
 Since neutrinos never free-stream in this model, 
 $(\Delta\phi_{\rm int}-\Delta\phi_{\Lambda{\rm CDM}})$
 is non-zero but fixed. 
   However, the contribution from $\Delta \phi_{\rm load}$ will depend on the corresponding value of $f_\chi$. We performed two types of MCMC runs, one set of runs with $f_\chi$ set to $1.25\%, 2.5\%, 5\%, 10\%$ and another set where we vary $f_\chi$ continuously between $[0,1]$. 
\begin{figure}
    \centering
    \includegraphics[width=0.495\linewidth]{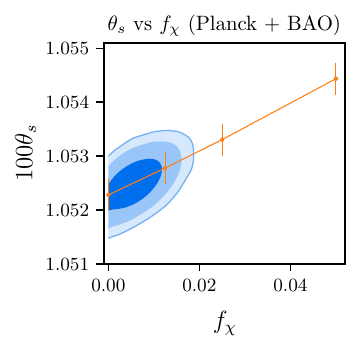} \includegraphics[width=0.495\linewidth]{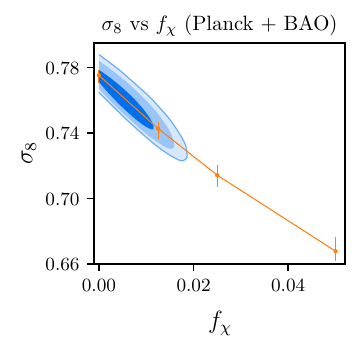}
    \caption{Proof-of-principle demonstration of correlations using strong $\nu$-DM example. Datasets include Planck (with lensing) and BAO. Marginalized contours up to $3\sigma$ are shown in the plots. Mean values for fixed $f_\chi$ runs with $\pm 1\sigma$ error bars are overlaid on top of the contour plots and joined with lines for visual clarity. \textit{Left: }$\theta_s$ vs $f_\chi$ showing positive linear correlation. \textit{Right: } $\sigma_8$ vs $f_\chi$ showing negative linear correlation.}
    \label{fig:fxthetas_fluidnu}
\end{figure}

In Fig.~\ref{fig:fxthetas_fluidnu} we show the marginalized 2D contours on $\theta_s$ and $f_\chi$ for varying $f_\chi$ runs. Overlay-ed on the plot are also the values of $\theta_s$ with $1\sigma$ errorbar for the fixed $f_\chi$ runs. Both these graphs show a linear correlation between $\theta_s$ and $f_\chi$, which is the characteristic signature of DM-loading effects explained in Eq.~\eqref{eq:theta_approx} and~\ref{eq:l_peak} from fitting fixed location of $\ell$-peaks. In the right panel of the plot, we show the decrease in $\sigma_8$ with increasing $f_\chi$, which is another generic signature of interacting dark matter models.

The purpose of this particular analysis is to demonstrate that DM-loading indeed leads to a linear positive (negative) correlation between $\theta_s$ ($\sigma_8$) and $f_\chi$. When assessing the goodness of fit to the Planck+BAO data, models with SI-$\nu$ ($f_\chi=0$) already exhibit a poorer fit compared to the FS-$\nu$ model. The blue contours highlight additional tensions in fitting the data relative to the SI-$\nu$ case, as a large $f_\chi$ is disfavored due to the change of perturbations from dark matter scattering. In the following, we will consider more realistic scenarios where we will try to demonstrate the DM-loading effects.



\subsection{Strong DR-DM interaction}\label{sec:MCMC-DRDM}

We consider a scenario where DR interacts strongly with a fraction of dark matter in the universe. Due to the large interaction strength, the DR propagates as perfect fluid till today unless $f_\chi=0$. This is similar to the DL-$\nu$ case discussed above, but the DR constitutes a small fraction of the total $N_{\rm eff}$. In the MCMC analysis, we fix the free streaming neutrino contribution to $N_\nu = 3.046$ and vary the DR abundance $N_{\rm DR}$\footnote{In terms of ETHOS parameters, we set \textbf{N\_ur} $= 3.046$, \textbf{a\_idm\_dr} = $10^6$ with $T_\nu^2$ dependent interaction and vary \textbf{N\_idr} and $f_\chi$.}. In this scenario, the phase shift relative to $\Lambda$CDM arises from $\phi=\Delta\phi_{\rm int}-\Delta\phi_{\Lambda{\rm CDM}} + \Delta\phi_{\rm load}$, where $\Delta\phi_{\rm int}$ varies with $f_\nu$ and $\Delta\phi_{\rm load}$ varies with $f_\chi$. We provide the mean model parameters obtained from the MCMC study in Appendix~\ref{app.table}, along with the triangular plot and the minimum $\Delta\chi^2$ compared to the $\Lambda$CDM model. In the following discussion, we focus on the parametric dependencies between $\theta_s$, $N_{\rm DR}$, and $f_\chi$.

%
\begin{figure}[t!]
\centering
\begin{subfigure}{.49\textwidth}
\centering
\includegraphics[width=\linewidth]{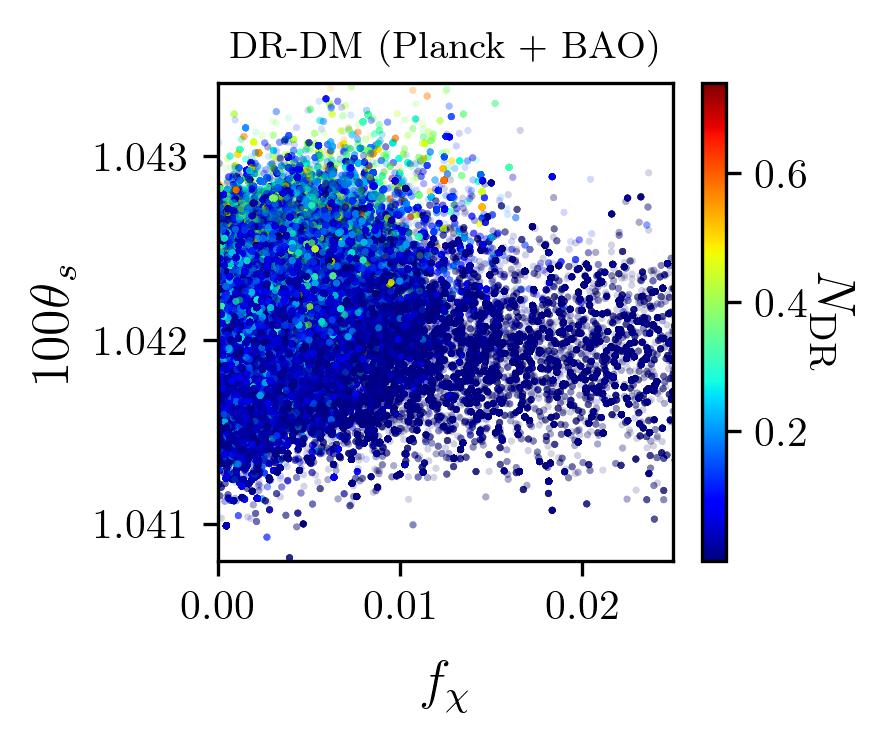}
\end{subfigure}
\begin{subfigure}{.49\textwidth}
\centering
\includegraphics[width=\linewidth]{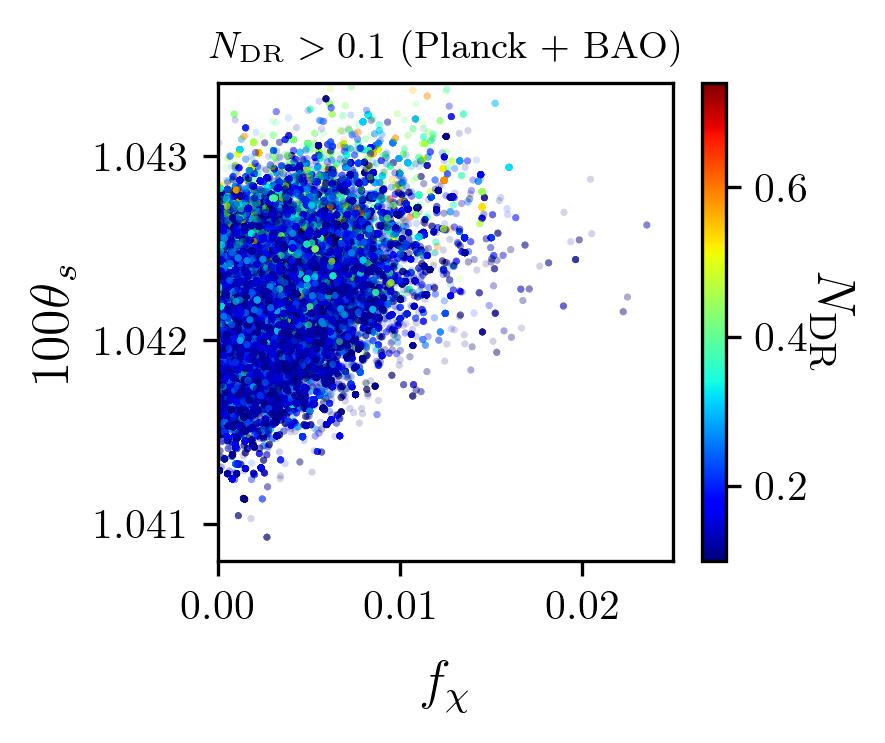}
\end{subfigure}
\begin{subfigure}{.49\textwidth}
\centering
\includegraphics[width=\linewidth]{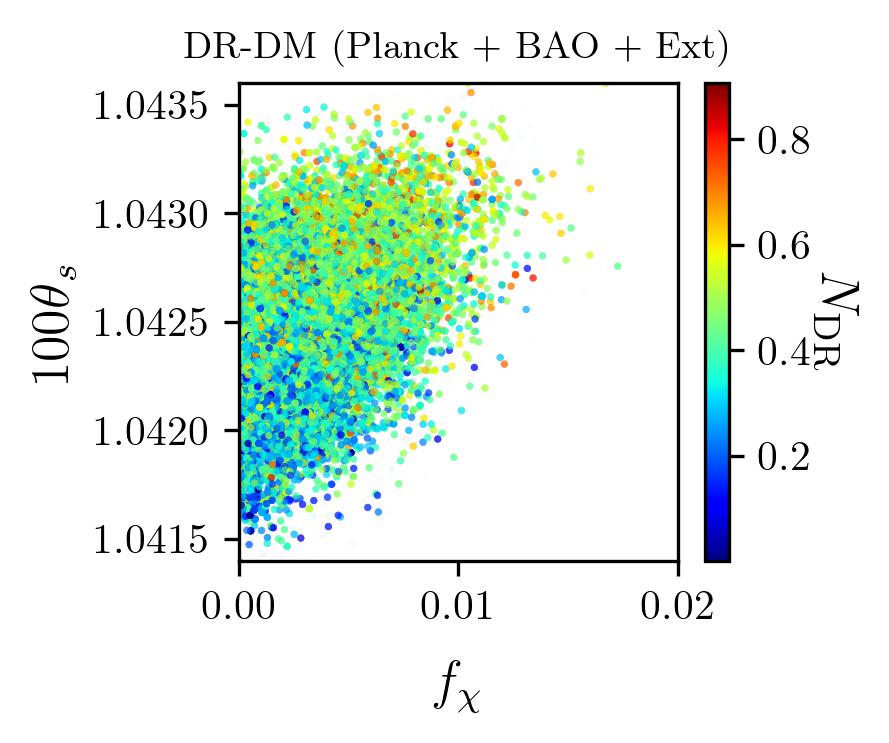}
\end{subfigure}
\begin{subfigure}{.49\textwidth}
\centering
\includegraphics[width=0.825\linewidth]{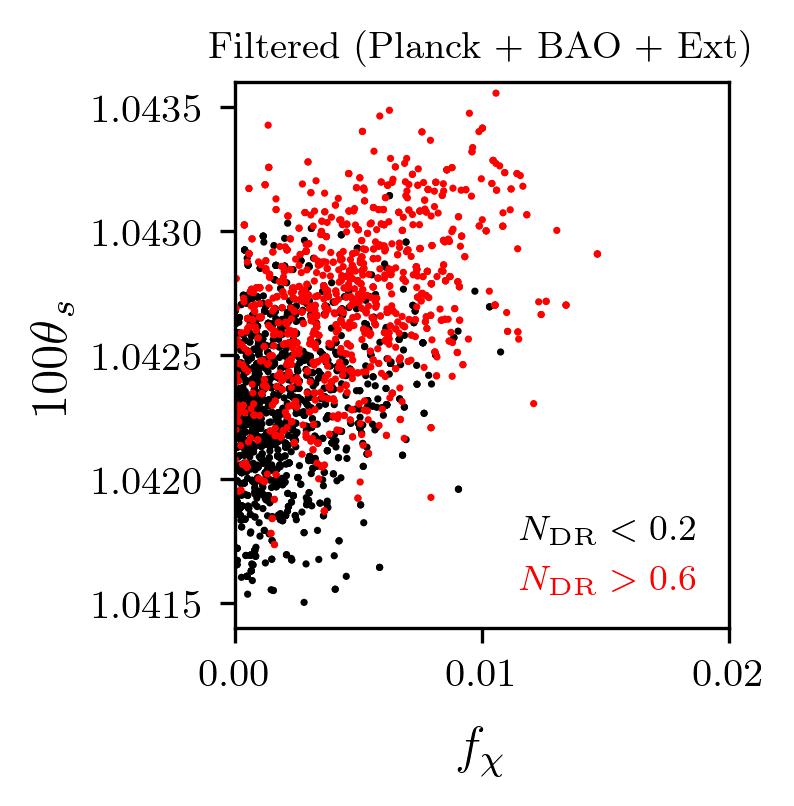}
\end{subfigure}
\caption{
Strong DR-DM scenario: efficient scattering in the dark sector with all neutrinos free-streaming. Plots were obtained by varying both $N_{\rm DR}$ and $f_\chi$. The $f_\chi \rightarrow 0$ limit for $N_{\rm DR} > 0$ corresponds to the SI-DR fluid case. \textit{Top Row:} Planck and BAO datasets. Filter $N_{\rm DR} > 0.1$ applied on the right plot to isolate points corresponding to DL-DR fluid. \textit{Bottom row:} Planck, BAO and Ext~= SH0ES + kv450 datasets. Filter applied on the right plot to isolate two separate bands with high and low $N_{\rm DR}$ respectively. 
}
\label{fig:DMDR3D-thetas}
\end{figure}
\begin{figure}
     \centering
     \includegraphics[width=0.49\linewidth]{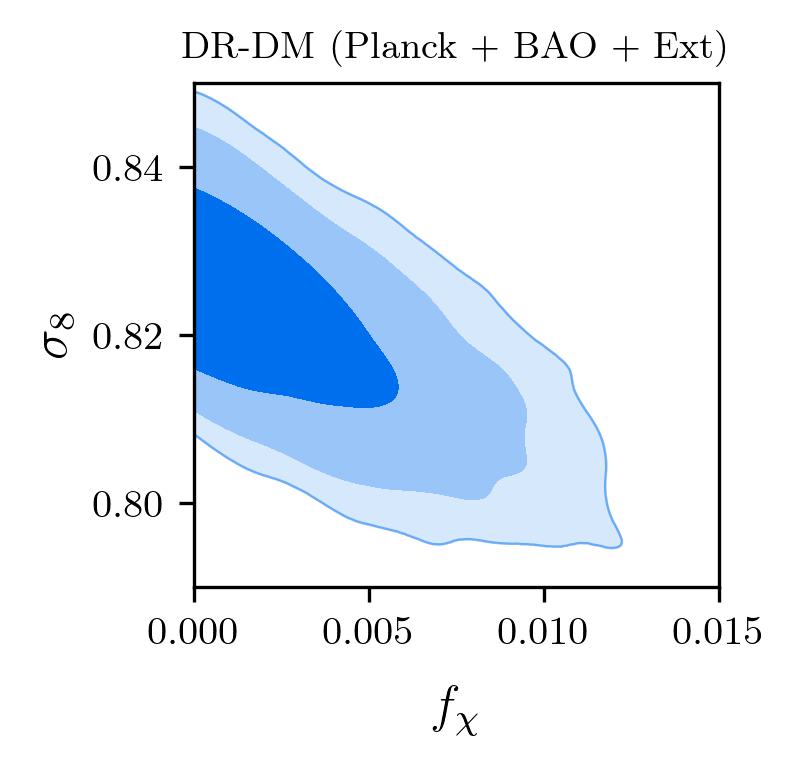}
\caption{
Strong DR-DM scenario:  efficient scattering in the dark sector with all neutrinos free-streaming. Plots were obtained by varying both $N_{\rm DR}$ and $f_\chi$.  Plot shows $\sigma_8$ vs $f_\chi$ contours up to $3\sigma$ for Planck, BAO and Ext~= SH0ES + kv450 datasets. The $f_\chi \rightarrow 0$ limit here corresponds to the SI-DR fluid case, since the bestfit point corresponds to a significantly non-zero value of $N_{\rm DR}$  for this dataset (Appendix.~\ref{app.table})}
     \label{fig:DMDR3D-s8}
 \end{figure}

Fig.~\ref{fig:DMDR3D-thetas} shows the marginalized 2D contours of $\theta_s$ and $f_\chi$ where the points are color-coded with the corresponding $N_{\rm DR}$ values (we will call this kind as 3D plot from now on). By including the SH0ES data that prefers larger $N_{\rm DR}$, as is shown in the \emph{lower-left} plot, a positive correlation between $\theta_s$ and $f_\chi$ shows up for each color of $N_{\rm DR}$ points. The linear relation is even more obvious when we separate the points by different $N_{\rm DR}$ values (\emph{lower-right}), which signifies the presence of DM-loading effects. However, due to the constraint on the $f_\chi$ coming from the physics of dark acoustic oscillations, the increase of $\theta_s$ is limited to $\sim0.1\%$-level correction to $\theta_s$ ($\delta\ell\sim 1$). Additionally, an overall shift in the $\theta_s$ distribution between black and red points at $f_\chi \to 0$ is attributed to the different fractional abundance of free-streaming radiation between these samples. Although we only add fluid-like DR, this raises the total $N_{\rm eff}$ and consequently reduces the energy fraction $f_\nu^{\rm fs}$ in Eq.~(\ref{eq.fsnu}), causing a smaller $\phi$ and larger $\theta_s$ from Eq.~(\ref{eq:l_peak}). The red points therefore distributed with a bit larger $\theta_s$ than the black points.

The upper plots, derived from Planck+BAO data, illustrate a linear correlation between $f_\chi$ and $\theta_s$ for slightly larger values of $N_{\rm DR}$ (ranging from light blue to red). Conversely, smaller values of $N_{\rm DR}$ (darker blue) exhibit no significant correlation between these parameters. This deviation from linear correlation occurs in the small $N_{\rm DR}$ region, where we observe minimal phase shift, allowing for larger values of $f_\chi$ due to reduced effects from dark matter scattering. To highlight the linear correlation between $\theta_s$ and $f_\chi$ for higher $N_{\rm DR}$, we present a 3D plot of the Planck dataset in the upper right panel, focusing solely on MCMC points with $N_{\rm DR} > 0.1$.


In Fig.~\ref{fig:DMDR3D-s8}, we show the marginalized 2D contours for $\sigma_8$ and $f_\chi$ with respect to the self-interacting DR model. As discussed in Ref.~\cite{Chacko:2016kgg}, DM-DR scattering leads to suppression of the matter power spectrum with increasing $f_\chi$, resulting in a reduction of $\sigma_8$, which represents the root mean square of matter fluctuations around the 8$h^{-1}$~Mpc scale. The positive correlation between $\theta_s$ and $f_\chi$, along with the negative correlation between $\sigma_8$ and $f_\chi$, provides a distinct signature of dark matter density scattering with DR or neutrinos. The primary constraint on $f_\chi$ in this analysis comes from the suppression of dark matter perturbations, setting the maximum allowed phase shift $\Delta\phi_{\rm int}$ indicated by the data.




\subsection{Varying the $\nu$-DM interaction}\label{sec:MCMC-NUDM}
In this subsection, we explore scenarios of DL-$\nu$ with varying scattering cross sections to allow neutrinos to decouple at later times. We employ the temperature-independent cross-section detailed in Sec.~\ref{sec.param} and allow for variations in both $f_\chi$ and the interaction strength $y$. 

In the MCMC scan, we fix the effective number of interacting neutrinos at $3.046$\footnote{In terms of ETHOS parameters, N\_idr$=3.046$, N\_ur$=0$, and we vary $f_\chi\times$a\_idm\_dr with $T_\nu$ independent cross-section (nindex\_idm\_dr$=2$, alpha\_idm\_dr$=1$) according to the model outlined in section~\ref{sec.param}.}. This scenario has previously been studied in the context of $H_0$ tension by alleviating the tension with neutrino-induced phase shift~\cite{Ghosh:2019tab}.
The scattering rate between neutrinos and dark matter depends on the product $yf_\chi$. To expedite convergence in our runs, we vary $yf_\chi$ and $f_\chi$ as primary parameters in the MCMC analysis\footnote{Given that ${a_\textbf
{\rm idm\_dr}}$ varies linearly with $y$ according to Eq.~\eqref{eq:ethosmap}, varying $f_\chi a_{\rm idm\_dr}$ is equivalent to varying $yf_\chi$. Thus, we vary $f_\chi a_{\rm idm\_dr}$ in the scan and derive $yf_\chi$ as an output parameter.}.  Additionally, we conduct a series of MCMC analyses with fixed $f_\chi = 10^{-4},\,0.01,\,0.02, \text{and}~0.03$ to highlight the phase shift's dependence on $yf_\chi$. For $f_\chi = 10^{-4}$, the DM-loading effect is negligible and can be considered an approximation to the SI-$\nu$ case. We present more complete results from the MCMC scan in Appendix.~\ref{app.table}. 

\begin{figure}
\centering
\begin{subfigure}{.49\textwidth}
\centering
\includegraphics[width=0.825\linewidth]{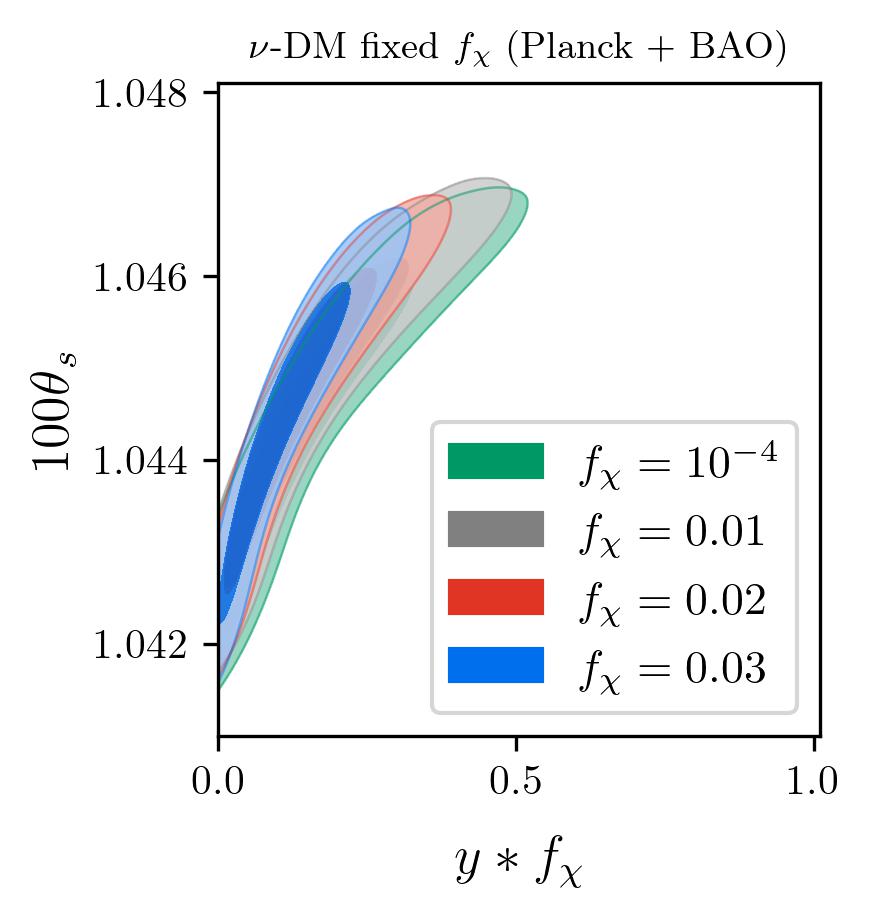}
\end{subfigure}
\begin{subfigure}{.49\textwidth}
\centering
\includegraphics[width=\linewidth]{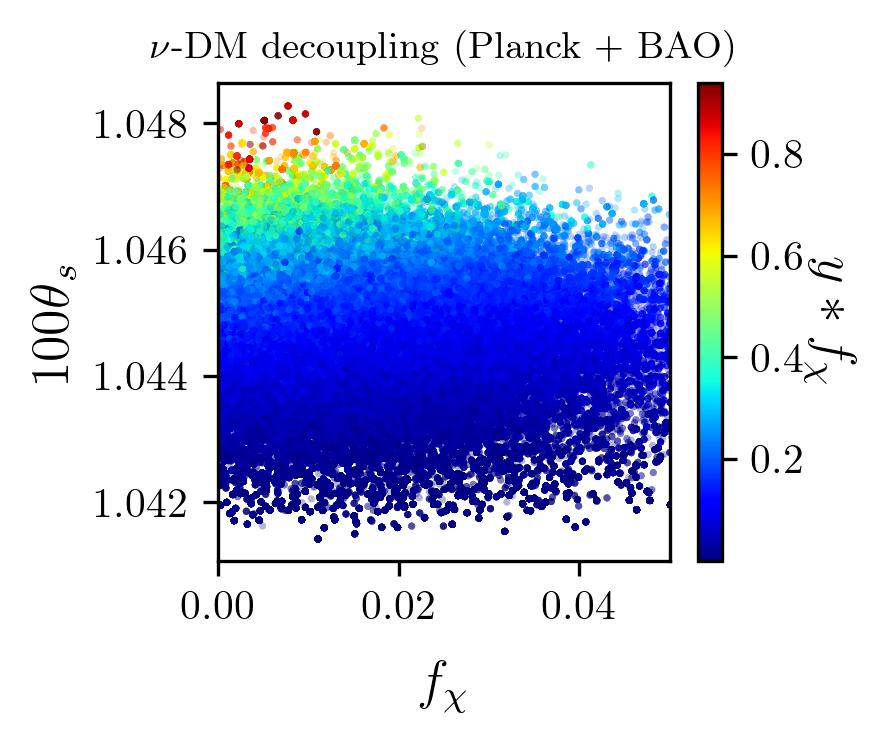}        
\end{subfigure}
\begin{subfigure}{.49\textwidth}
\centering
\includegraphics[width=0.825\linewidth]{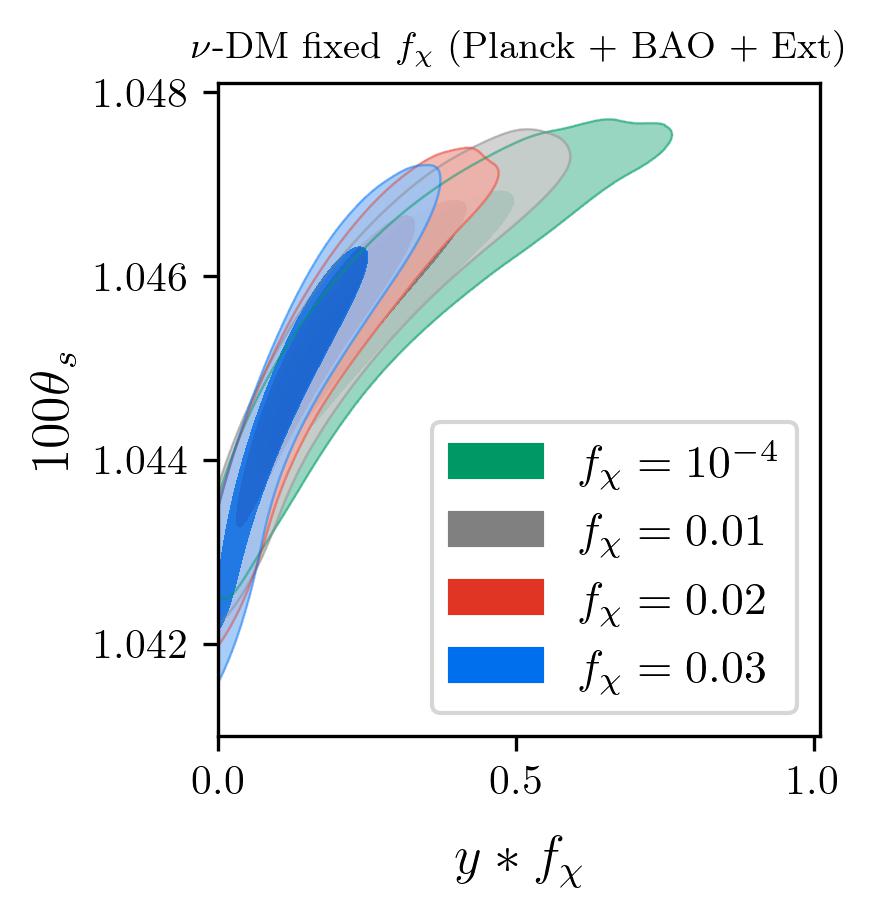}   
\end{subfigure}
\begin{subfigure}{.49\textwidth}
\centering
\includegraphics[width=\linewidth]{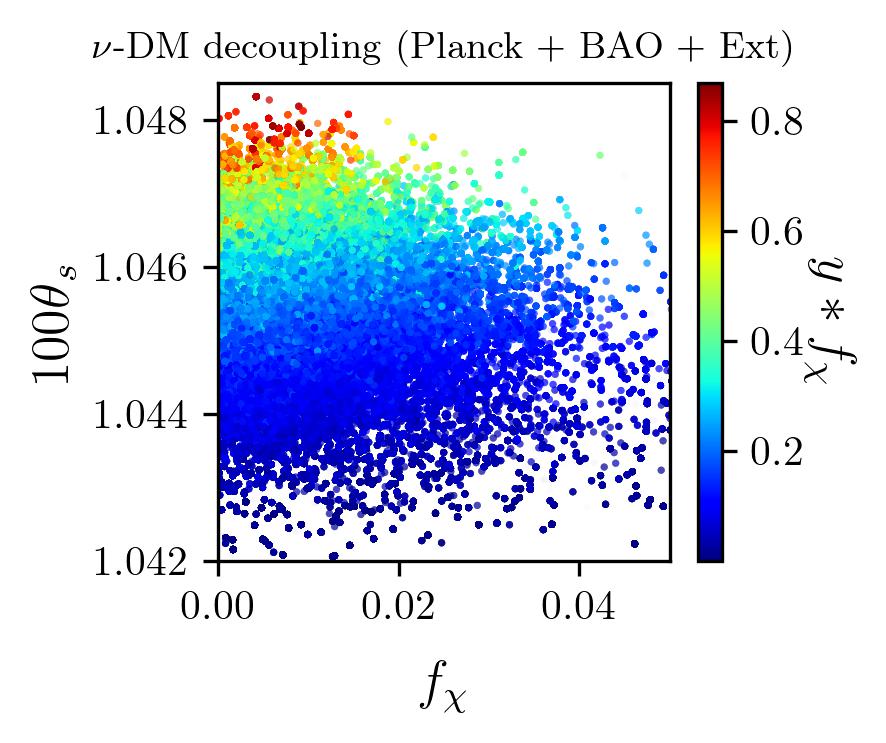}       
\end{subfigure}
\caption{Varying $\nu$-DM interaction: all neutrinos scatter efficiently at early times but free-stream at late times with decoupling time controlled by $y f_\chi$ in the scattering rate Eq.~(\ref{eq:Gamma}). The 2D plots on the left were obtained by varying $y$ for fixed $f_\chi$, showing contours up to $2\sigma$ for a range of $f_\chi$ values. The plots on the right were obtained by varying both $f_\chi$ and $y f_\chi$. \textit{Top row:} Planck and BAO datasets. \textit{Bottom Row:} Planck, BAO and Ext = SH0ES + kv450 datasets.}
\label{fig:fixedfidm1}
\end{figure}
First focusing on the fixed $f_\chi$ runs, the left plots of Fig.~\ref{fig:fixedfidm1} display marginalized 2D contours for $\theta_s$ vs. $yf_\chi$ with Planck+BAO (top) and Planck+BAO+Ext (bottom). The positive correlation between $\theta_s$ and $yf_\chi$ observed with $f_\chi = 10^{-4}$ (green) comes from the phase shift induced by self-interacting neutrinos ($\Delta \phi_{\rm int} - \Delta\phi_{\Lambda{\rm CDM}}$). Larger values of $yf_\chi$ delay the $\nu$-DM decoupling to later redshifts, causing a phase shift that starts at lower $\ell$-modes and results in a more pronounced change in the $\theta_s$ fit. The shift in $\Delta \ell$ due to changes in sound speed is not constant across $\ell$ (Figs.~\ref{largefx} and~\ref{CLTTEE_dNeff}). Due to the domination of matter energy density at later times, this phase shift roughly scales as $\propto \ell$ at lower modes and begins to plateau for $\ell \gtrsim 1000$. This $\ell$-dependence differs from the $\Delta\ell$ change due to $\theta_s$ shift, which behaves as $\Delta\ell \propto \Delta\theta_s \ell$ (Eq.~(\ref{eq:l_peak})). However, partial compensation between these phase shifts occurs because the power spectrum at very high $\ell$-modes is suppressed due to diffusion damping, thus allowing for less constraint on the different phase shifts arising from radiation interaction versus $\theta_s$ shift. These different changes in $\Delta\ell$, nevertheless, still limit how $\Delta \phi_{\rm int}+\Delta\phi_{\rm load}$ from radiation interaction can address the Hubble tension by raising $H_0$~\cite{Ghosh:2019tab}.
The curves for larger $f_\chi$ obtain larger $\theta_s$ from $\Delta\phi_{\rm loading}$ with the same size of coupling.

Large $f_\chi$ additionally brings in more constraints due to the effects of dark matter scattering on the matter power spectrum. For larger $f_\chi$, the $1\sigma$ and $2\sigma$ contours 
 shrink along the $yf_\chi$ direction due to the additional constraints on fractional dark matter scattering that suppresses the matter density perturbations. This also limits the amount of total phase-shift as the extent of the $2\sigma$ contours in the left panels of Fig.~\ref{fig:fixedfidm1} along $\theta_s$ direction is smaller for larger $f_\chi$. This effect is also seen Table~\ref{tab:DMNU-fix-planck} and ~\ref{tab:DMNU-fix-sh0es}, where larger $f_\chi$ results in a smaller $H_0$ due to the lesser extent of the $\theta_s$ shift. Larger $f_\chi$ cases also do not seem to provide a good fit to the data from the $\Delta\chi^2 $ values. The triangle plot for all the parameters for the fixed-$f$ runs are shown in Fig.~\ref{fig:DMNU-fix-planck-triangle} and \ref{fig:DMNU-fix-sh0es-triangle} in Appendix.~\ref{app.table}.

 In the right column of Fig.~\ref{fig:fixedfidm1}, we present 3D plots of $\theta_s$, $f_\chi$, and $yf_\chi$ for the runs with continuously varying $f_\chi$. By focusing on a fixed interaction strength $yf_\chi$ (indicated by fixed color), we observe a positive correlation between $\theta_s$ and $f_\chi$, characteristic of DM-loading. As discussed earlier, $\theta_s$ also increases with $yf_\chi$, which delays the decoupling time. The plot further suggests that $\theta_s$ exhibits a consistent correlation with $f_\chi$ across different values of $yf_\chi$, indicating that the DM-loading phase shift is insensitive to the effective number of scattering neutrinos $f_\nu$. Although we do not directly vary $f_\nu$ in $\nu$-DM scattering, smaller values of $yf_\chi$ effectively correspond to a lower fraction of neutrinos with efficient scattering. Similar to the fixed $f_\chi$ analysis, for larger $yf_\chi$ values the allowed range of $f_\chi$ is smaller due to constraints from the dark matter scattering. Table.~\ref{tab:DMNU-dec} and Fig.~\ref{fig:DMNU-dec-triangle} show the constraints on the parameters for this scenario.

 It is reasonable to ask how our results would change for a temperature-dependent cross section, where the scattering rate scales as $\sigma n_\chi \propto T^n$ for more general $n$ \footnote{The temperature-independent case considered in this section corresponds to $n=3$ since all the scaling comes from the DM number density $n_\chi \propto T^3$.}.  Given that the enhanced phase shift only occurs for perturbation modes that enter the horizon when the $\nu$-DM scattering is efficient, the relevant factor is the decoupling time when the scattering rate falls below the Hubble rate, which scales as $H\propto T^2$. A temperature-dependent cross section can lead to an earlier or later decoupling of the scattering, which changes which $\ell$-peaks experience the enhanced phase shift. The size of this enhanced phase shift due to DM-loading, however, remains unchanged as long as the scattering is efficient. As shown in the right column of Fig.~\ref{fig:fixedfidm1}, when the scattering decouples earlier (smaller $yf_\chi$, indicated by darker blue), the linear dependence of $\theta_s$ on $f_\chi$ disappears. However, for values of $yf_\chi$ that result in efficient scattering, the linear dependence reappears, and $\theta_s$ increases with $f_\chi$ at the same rate for each $yf_\chi$, as seen in the different color bands.

\section{Discussion and conclusion}
In this study, we explore how the phase shift in CMB acoustic oscillations acts as a probe for neutrinos and dark radiation propagation, affecting CMB anisotropy solely through gravitational effects. Specifically, we focus on the enhanced phase shift resulting from radiation-dark matter scattering compared to the self-interacting radiation scenarios.

We first demonstrate the presence of this enhanced phase shift in the TT and EE spectra through the \CLASS~calculations, revealing a linear dependence to the interacting dark matter abundance $\Delta\ell\propto f_\chi$, with $\Delta\ell$ being insensitive to radiation abundance ($f_\nu$ or $N_{\rm DR}$) when radiation's energy density dominates over dark matter around matter-radiation equality. To understand these parametric dependencies, we approximate the evolution of photon and interacting radiation fluctuations using coupled harmonic oscillators. The analysis shows that the delayed radiation sound speed, due to dark matter scattering, enhances the phase shift. Our analytical framework extends to scenarios with multiple dark sectors, each featuring its own interacting DM-dark radiation. In such cases, the total enhanced phase shift is expected to scale proportionally with the sum of interacting dark matter abundance. We also conduct MCMC studies to obtain constraints on the interacting neutrino and dark radiation models with the Planck and BAO data. The effect of the enhanced phase shift from DM-loading shows up as an enhanced $\theta_s$ proportional to $f_\chi$.

The enhanced phase shift amplifies with $f_\chi$ (Fig.~\ref{fig:DMDR3D-thetas}) due to the slowing down of neutrinos and dark radiation by dark matter scattering. From the DM's perspective, interacting radiation delays its structure formation process, resulting in suppression of the matter power spectrum as illustrated in the $\sigma_8$ plot (Fig.~\ref{fig:DMDR3D-s8}). Consequently, cosmological models incorporating dark matter scattering to neutrinos or abundant dark radiation commonly exhibit two characteristics: an enhanced $\theta_s$ and a reduced $\sigma_8$ compared to models lacking efficient radiation-DM scattering. Identifying both of these features in cosmological data could ultimately pave the way for discovering non-minimal dark sectors even without non-gravitational interactions with the Standard Model.



\begin{acknowledgments}
We thank 
Kimberly K. Boddy,
Francis-Yan Cyr-Racine,
Joel Meyers,
Gabriele Montefalcone,
Martin Schmaltz,
Gustavo Marques-Tavares,
Benjamin Wallisch, and Taewook Yon
for the helpful discussions. We especially thank Joel Meyer and Benjamin Wallisch for their valuable feedback on the draft. The research of DH and YT is supported by the National Science Foundation
(NSF) Grant Number PHY-2112540. SG is supported by NSF Grant Number PHY-2112884. YT would like to thank the Aspen Center for Physics (supported by NSF grant PHY-2210452) for hospitality
while this work was in progress.
\end{acknowledgments}
\appendix

\section{An example DL-$\nu$ model and constraints}\label{app.model}
An example of a new physics model featuring DM-$\nu$ coupling is described in~\cite{Ghosh:2017jdy,Ghosh:2019tab}. In this model, the effective DM-$\nu$ coupling with a heavy mediator $\psi$ is generated during electroweak symmetry breaking:
\begin{equation}
    \mathcal{L} \hspace{2.5mm}\supset\hspace{2.5mm} \frac{y_{ij}}{\Lambda}(H^\dagger l_i)(\psi_j\chi)  \hspace{5mm} \Rightarrow\hspace{5mm} \eta_{ij}\nu_i\psi_j\chi \,,\hspace{5mm}\mathrm{where}\hspace{5mm} \eta_{ij} = \frac{y_{ij}v}{\sqrt{2}\Lambda}\,.
\end{equation}
Notice that when expanding the Higgs around its vacuum expectation value and absorbing the phase of Goldstone bosons into left-handed leptons ensures that the operator $H^\dagger l$ only involves neutrino couplings among all SM particles, which is important for circumventing constraints related to charged leptons~\cite{Ghosh:2017jdy}. If the dark matter $\chi$ is a scalar, then the mediator $\psi$ is a fermion, and vice versa. The dimension five coupling can be mediated by massive vector-like fermions $N$:
\begin{equation}
\mathcal{L} \hspace{2.5mm} \supset \hspace{2.5mm} Y_{N,ij} N_i (H^\dagger l_j) + Y_{\bar{N},ij} N^c_i (\psi_j\chi) + M_{N,ij} N_i N^c_j \,,\hspace{5mm}\mathrm{where}\hspace{5mm} \frac{y_{ij}}{\Lambda} \sim 2\frac{Y_{N,ik} Y_{\bar{N},kj}}{M_{N}}\,.
\end{equation}
Further details regarding the charge assignment and flavor symmetry structure of the model can be found in Ref.~\cite{Ghosh:2019tab}. 
The $\nu$-DM scattering remains independent of $T_\nu$ (as illustrated in Sec.~\ref{sec.param} to Sec.~\ref{sec.analytical}) when the dark matter and mediator masses are close $(m_\psi^2-m_\chi^2) \ll 2m_\chi T_\nu$. 
Under these conditions, the leading-order cross-section (zeroth order in $(T_\nu/m_\chi)$) and corresponding interaction parameter $y$ in Eq.~(\ref{eq:u_def}) are:
\begin{equation}
    \sigma = 1.7\times 10^{-6} \left( \frac{\eta}{0.1}\right)^4\left(\frac{\mathrm{GeV}}{m_\chi}\right)^2 \mathrm{GeV}^{-2}\,,\qquad y=1.1\left(\frac{\eta}{0.3}\right)^4\left(\frac{50\,{\rm MeV}}{m_{\chi}}\right)^3\,,
\end{equation}
where benchmark parameters are selected to ensure efficient $\nu$-DM scattering until $z \sim 4 \times 10^4$, impacting modes in the CMB spectra with $\ell \gtrsim 1000$.

Given that the cross-section scales as $m_\chi^{-2}$, efficient $\nu$-DM scattering imposes an upper limit on $m_\chi \lesssim \mathcal{O}(10)$ MeV. Collider constraints to consider for the Higgs and neutrino coupling with light particles $\psi$ and $\chi$ include:
\begin{enumerate}
    \item Invisible Higgs and $Z$ decays: $h(Z) \to \nu \psi \chi$: The branching ratio for the Higgs process is approximately $\mathcal{O}(1\%)$, which is consistent with the current limit of about $18\%$ ($95\%$ C.L.) from the CMS search~\cite{CMS:2022qva}. Invisible $Z$ decay measurements from the LHC put a constraint of $\eta \lesssim 1 (0.2)$ for dark matter and mediator masses of $50(10)~{\rm Mev}$~\cite{Primulando:2017kxf} 
    \item Kaon decay $K \to (\mu,e)\chi\psi$ via virtual $\nu_{\mu,e}$: While there has not been a specific search for this process to the best of our knowledge, some studies have examined Kaon decays into $(\mu,e)\phi\nu$, where $\phi$ is a light mediator with a coupling $g_{ii}\phi\bar\nu_i\nu_i$. For $1 \lesssim m_\phi \lesssim 100$ MeV, the kaon bound requires $g_{\mu\mu,ee} \lesssim (3 \times 10^{-3},10^{-2})$ \cite{Blinov_2019}. A similar bound should also apply to $\eta$, limiting the size below $y \lesssim 0.1$. Alternatively, one can avoid the kaon bound by considering an $\eta$ coupling only to $\nu_\tau$ since bound on $g_{\tau\tau}$ for neutrino self-interaction is rather weak $(g_{\tau\tau} \lesssim 0.3)$~\cite{Blinov_2019}. A more focused study of the kaon bound on the $\eta\nu\psi\chi$ coupling is needed, which we defer to future investigations.
\end{enumerate}

We emphasize that the model involving scattering with all neutrinos was utilized earlier solely to illustrate the DM-loading effect. As discussed in Sec.~\ref{sec.result}, the enhancement of phase shift in the DL-$\nu$ case also occurs with just 1 or 2 neutrinos scattering. Additionally, Appendix~\ref{ap.1nu} demonstrates that the agreement between \CLASS~and the toy model for DL-$\nu$ phase shift enhancement compared to SI-$\nu$ is maintained even with only 1 neutrino scattering. Furthermore, similar phase shift enhancements and $f_\chi$ dependencies are observed in scenarios involving DR-DM interaction, which effectively circumvents experimental constraints relying on non-gravitational SM - dark sector couplings.   

\section{Establishing the toy model}\label{app.toymodel}
The efficient scattering of neutrinos with dark matter is modelled after the Boltzmann system for the photons and baryons under the \textbf{tight coupling approximation}, following the approach used in Ref.~\cite{2020moco.book.....D}. Working in the conformal Newtonian gauge, we follow the convention for the  energy ratio $R_r = 3\rho_m/4\rho_r$, where $(r,m)$ now denotes either $(\gamma,b)$ or $(\nu,\chi)$, to obtain 
\begin{align*}
\ddot{\delta}_r + \mathcal{H}\frac{R_r}{1+R_r}\dot{\delta}_r + \frac{k^2}{3(1+R_r)}\delta_r = 4\ddot{\phi} + 4\mathcal{H}\frac{R_r}{1+R_r}\dot{\phi} - \frac{4k^2}{3}\psi
\end{align*}
Using the definition of the sound speed from Eq.~(\ref{eqn:cs}), it is straightforward to rewrite $R_r$ in terms of $c_r^2$ to obtain the coupled equations
\begin{align}
\ddot{\delta}_\gamma + &\mathcal{H}(1-3c_\gamma^2)\dot{\delta}_\gamma + k^2c_\gamma^2\delta_\gamma = g_{\rm grav}(\tau) \label{TCA1}\\
\ddot{\delta}_\nu + &\mathcal{H}(1-3c_\nu^2)\dot{\delta}_\nu + k^2c_\nu^2\delta_\nu = g_{\rm grav}(\tau) \label{TCA2}\\
 &\textrm{where} \hspace{2mm} g_{\rm grav}(\tau) = 4\ddot{\phi} + 4\mathcal{H}(1-3c_r^2)\dot{\phi} - \frac{4k^2}{3}\psi
\end{align}
where $g_{\rm grav}(\tau)$ denotes the gravitational coupling between the two tightly coupled fluids. This system is still not easy to solve since we would still need to solve the Einstein equations for $\phi$ and $\psi$, which are sourced by the total perturbations. To obtain a closed set of equations in the $\delta_r$, we make four further assumptions.
\newline

\noindent \textit{\textbf{(Assumption 1)} No free-streaming radiation}
\newline
\noindent Assume that all neutrinos are efficiently scattering before recombination. Then all  radiation components are fluid-like and so the total shear vanishes $\sigma = 0$ in the absence of free-streaming radiation components. We can then use the Einstein equation for $\sigma$ to set $\phi = \psi$.  
\newline

\noindent \textit{\textbf{(Assumption 2) } Small matter loading}
\newline
\noindent Suppose that the matter loading effect is small, which occurs deep in the radiation era where $1-3c_r^2 = 1 - \frac{1}{1+R_r} \simeq R_r \ll 1$. Then the Hubble damping term in the Boltzmann equations will be negligible
\begin{align*}
\hspace{1mm}\mathcal{H}(1-3c_r^2)\dot{\delta_r} \hspace{1mm}\ll \hspace{1mm} k^2\delta_r\hspace{1mm}, \ddot{\delta_r}
\end{align*}
Similarly, the Hubble damping can be ignored in the gravitational coupling term 
\begin{align*}
\hspace{1mm}\mathcal{H}(1-3c_r^2)\dot{\phi} \hspace{1mm}\ll \hspace{1mm} k^2\phi\hspace{1mm}, \ddot{\phi}  \hspace{2mm}\Rightarrow \hspace{2mm} g_{\rm grav}(\tau) \simeq 4\ddot{\phi} -\frac{4k^2}{3}\phi
\end{align*}

\noindent \textit{\textbf{(Assumption 3)} Sub-horizon}
\newline
\noindent Consider the Einstein equations for the density and pressure perturbations
\begin{align*}
 -4\pi Ga^2\delta\rho &= k^2\phi + 3\mathcal{H}(\dot{\phi} + \mathcal{H}\phi)\\
4\pi Ga^2\delta P &= \ddot{\phi} + 3\mathcal{H}(\dot{\phi} -\mathcal{H}\omega\phi)
\end{align*}
where we used the second Friedmann equation to write $(2\dot{\mathcal{H}}+\mathcal{H}^2) = -8\pi Ga^2{P} = -3\mathcal{H}^2\omega$. Assume that we only consider $k$-modes that have entered the horizon $k\tau > 1$ during radiation era, where the metric behaves as $\phi \sim 1/\tau^2$. Plugging this ansatz for $\phi$ into the Einstein equations, we find that 
\begin{align}
    4\pi Ga^2(\delta\rho + 3\delta P) = -k^2\phi
\end{align}
The gravitational coupling term also simplifies
\begin{align*}
g_{\rm grav}(\tau) \simeq -\frac{4k^2}{3}\phi = \frac{16\pi Ga^2}{3}(\delta\rho + 3\delta P)
\end{align*}
since the $\ddot{\phi}$ term in $g_{\rm grav}(\tau)$ would be $1/k^2\tau^2$ suppressed compared to $k^2\phi$.
\newline

\noindent \textit{\textbf{(Assumption 4)} Perturbations in radiation only}
\newline
\noindent Staying within the radiation-dominated era, suppose that the metric perturbations are only sourced by perturbations in the radiation component so that we ignore the effect of matter perturbations on the metric. Then
\begin{align*}
g_{\rm grav}(\tau) \simeq& \frac{16\pi Ga^2}{3}\sum_{r=\gamma,\nu}(\delta\rho_r + 3\delta P_r) = \frac{16\pi Ga^2}{3}\sum_{r=\gamma,\nu}(1 + 3\omega_r)\delta\rho_r
\end{align*}
where for adiabatic perturbations, equation (\ref{eqn:ad}) implies that $\delta P_r = \omega_r\delta\rho_r$ in the radiation. The presence of matter in the background serves as a dilution factor to the radiation density as:
\begin{align*}
    \frac{{\rho}_{\mathrm{rad}}(\tau)}{{\rho}(\tau)} = \frac{{\rho}_{\mathrm{rad}}(\tau)}{{\rho}_{\mathrm{rad}}(\tau) + {\rho}_{\mathrm{mat}}(\tau)} = \frac{1}{1 + \frac{a(\tau)}{a_\mathrm{eq}}}
\end{align*}
where ${\rho}_{\mathrm{rad}}, {\rho}_{\mathrm{mat}}$ are the total radiation and matter densities respectively. Using the Friedmann equation 
\begin{align*}
g_{\rm grav}(\tau) \simeq& 2\mathcal{H}^2\frac{1}{1 + \frac{a(\tau)}{a_\mathrm{eq}}}\sum_{r=\gamma,\nu}(1 + 3\omega_r)\frac{{\rho}_{r}}{{\rho}_{\mathrm{rad}}}\frac{\delta\rho_r}{{\rho}_{r}}\\ 
=& 2\mathcal{H}^2\frac{(1 + 3\omega_r)}{1 + \frac{a(\tau)}{a_\mathrm{eq}}}(f_\gamma\delta_\gamma + f_\nu\delta_\nu) = \frac{4\mathcal{H}^2}{1 + \frac{a(\tau)}{a_\mathrm{eq}}}(f_\gamma\delta_\gamma + f_\nu\delta_\nu)
\end{align*}
where $\omega_r =1/3$ and the radiation ratios $f_r = {\rho}_r/{{\rho}_\mathrm{rad}}$ are constant. 
\newline

\noindent Plugging this form of $g_{\rm grav}(\tau)$ back into the tightly coupled equations (\ref{TCA1})--(\ref{TCA2}) and neglecting the Hubble damping terms yields the toy model equations (\ref{toyeqn1})--(\ref{toyeqn2}).
\newline

For the rest of this appendix, we provide more details about the solution in Eq.~(\ref{sec:SInu}) used for the analytical examination of Eqs.~(\ref{toyeqn1}) and (\ref{toyeqn2}). In particular, we provide an analytic expression for $\delta_\gamma(\tau)$ at early times which we can use to approximate the initial phase of the oscillator by matching with the late time solutions. While our primary focus is on the phase shift of the oscillating solution post gravitational driving force decoupling, it is essential to note that initially, the $F_{\rm driv} \sim 4/\tau^2$ term in Eq.~(\ref{eq.Fk}) dominates over the oscillation frequency $k^2c_\gamma^2$. As a result, the early time evolution is governed by
\begin{equation}
\ddot{\delta_\gamma}(\tau)\approx \frac{4}{\tau^2}\delta_\gamma(\tau)\,.
\end{equation}
With initial conditions $\delta_\gamma(\tau_{\rm in})=1$ and $\dot{\delta}_\gamma(\tau_{\rm in})=0$ at $k\tau_{\rm in}=1$, considering the regime $k\tau_{\rm in}\gtrsim 1$ assumed for the toy model, the early-time solution is given by
\begin{equation}
\delta^{\rm early}_{\gamma}\approx \frac{1}{2}(k\tau)^{\frac{1+\sqrt{33}}{2}}\left[(1+\frac{1}{\sqrt{33}})\left(k\tau\right)^{-\sqrt{33}}+(1-\frac{1}{\sqrt{33}})\right]\,.
\end{equation}
The power-law growth $\delta^{\rm early}_\gamma(\tau)\approx\frac{1}{2}(k\tau)^{\frac{1+\sqrt{33}}{2}}$ persists at $k\tau>1$ until
\begin{equation}
\tau_\alpha=\frac{2\alpha}{kc_\gamma}\,.
\end{equation}
After a numerical factor $\alpha$ times the moment when $k^2c_\gamma^2=2F_{\rm driv}(\tau)$, the oscillating solution begins to dominate and can be expressed as:
\begin{equation}\label{sec:SInuAP}
\delta^{\rm late}_{\gamma}(\tau) =  A\cos[kc_\gamma\tau-\phi_{\rm in}]\,.
\end{equation}
Aligning the two approximated solutions $\delta^{\rm early}_\gamma(\tau_\alpha)=\delta^{\rm late}_\gamma(\tau_\alpha)$ and $\dot\delta^{\rm early}_\gamma(\tau_\alpha)=\dot\delta^{\rm late}_\gamma(\tau_\alpha)$ sets the initial phase $\phi_{\rm in}=2\alpha+\tan^{-1}\left(\frac{1+\sqrt{33}}{4\alpha}\right)$. Note that $\phi_{\rm in}$ is a fixed phase introduced to match the later time solution ($\tau>\tau_\alpha$), when the perturbation starts to oscillate as a plane wave. Since $\phi_{\rm in}$ is common for models with and without DM-loading, the specific value of $\phi_{\rm in}$ is unimportant when considering the enhanced phase difference due to DM-loading. We hence drop the phase in Eq.~(\ref{sec:SInu}) in the analytical discussion.

Numerical solutions of Eqs.~(\ref{toyeqn1}) and (\ref{toyeqn2}) for the toy model indicate $\alpha\approx 1.5$ reproduces the oscillation phase in the later time solution of $\delta_\gamma(\tau)$. We therefore use $\tau_\alpha$ with $\alpha\approx 1.5$ to indicate the timescale for the photon perturbation transitioning into a harmonic oscillator and acquiring a phase. Note that $\alpha$ increases with $\Delta\alpha\sim 0.1f_\chi$, which is the enhanced phase shift we discuss in Sec.~\ref{sec.analy} due to the DM-loading effect. In this case, the solution takes the form (with $\alpha\approx1.5$)
\begin{equation}\label{sec:SInuAPDL}
\delta^{\rm late}_{\gamma,{\rm DL-}\nu}(\tau) \approx  A\cos[kc_\gamma\tau-\phi_{\rm in}-2.5\Delta\alpha]\,.
\end{equation}

\section{Additional plots for the phase shift enhancement}\label{ap.1nu}
Here we present more plots comparing the phase shift in the photon transfer function obtained from both \CLASS~and the toy model. 

\begin{figure}[H]
\centering
\begin{subfigure}{\textwidth}
  \centering
  \includegraphics[width=0.495\linewidth]{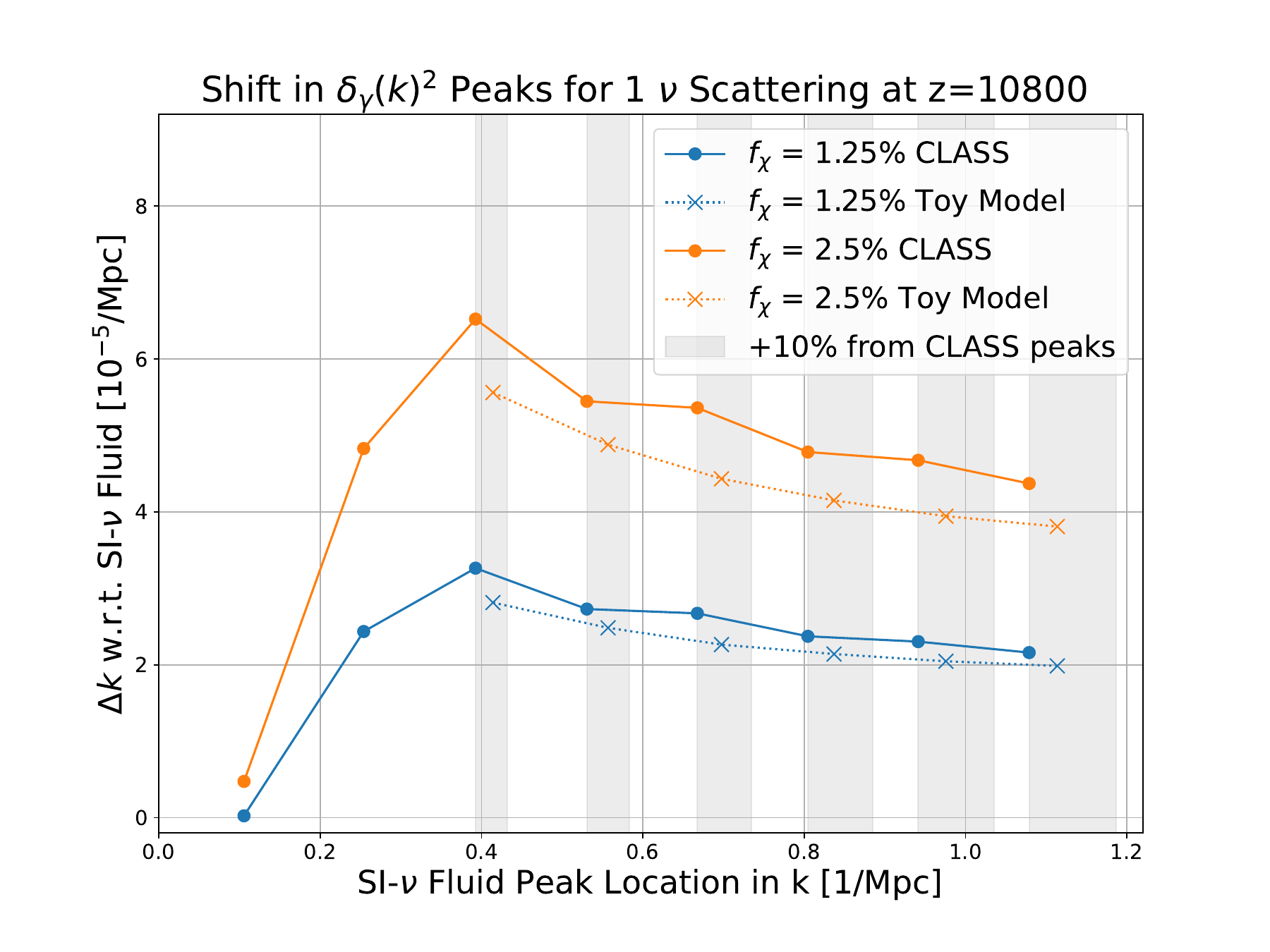}
   \centering
  \includegraphics[width=0.495\linewidth]{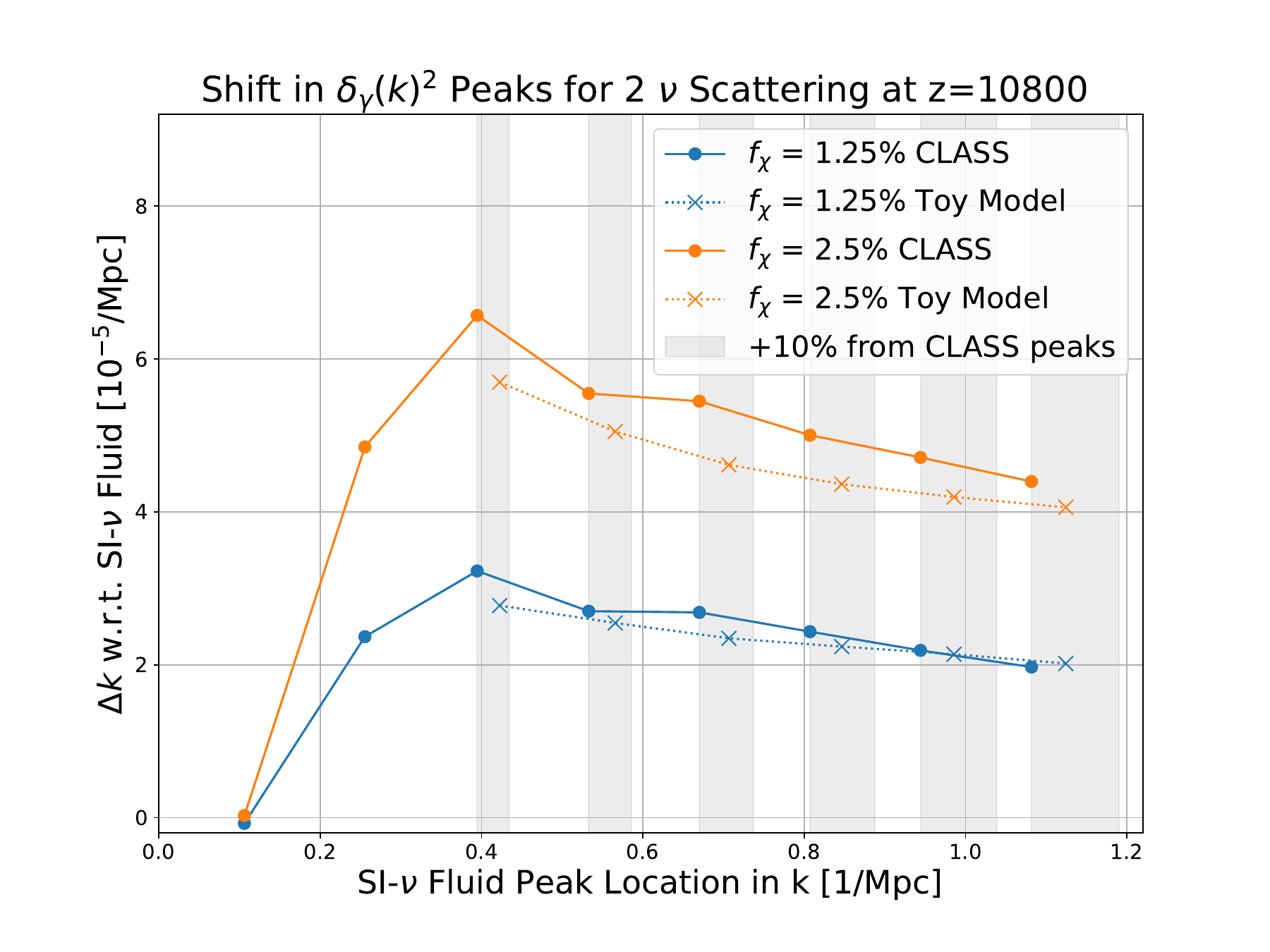}
  \caption{Radiation Era ($z=10800$)}
  \label{RADnu1}
\end{subfigure}
\vfil
\begin{subfigure}{\textwidth}
  \centering
  \includegraphics[width=0.495\linewidth]{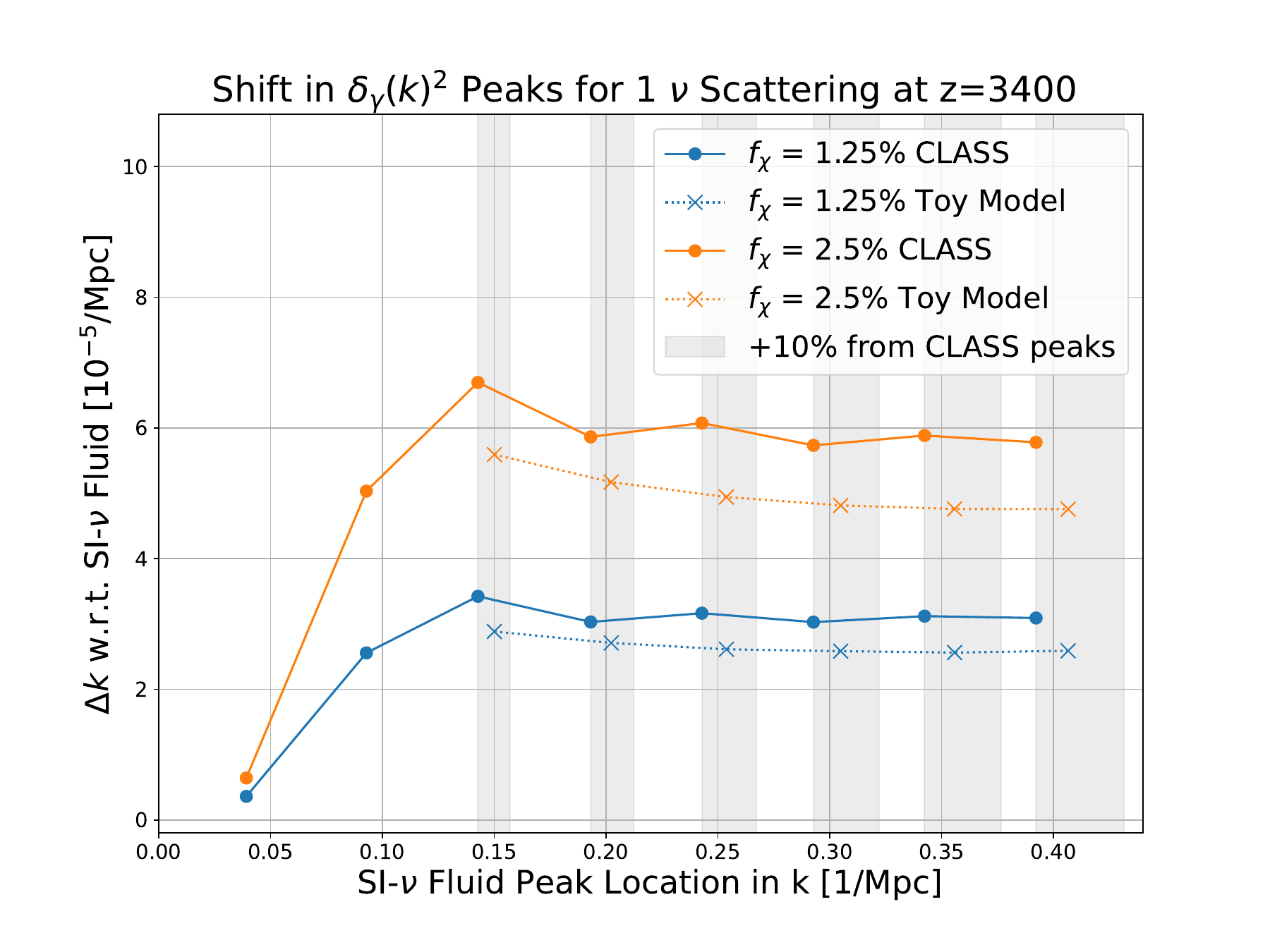}
  \centering
  \includegraphics[width=0.495\linewidth]{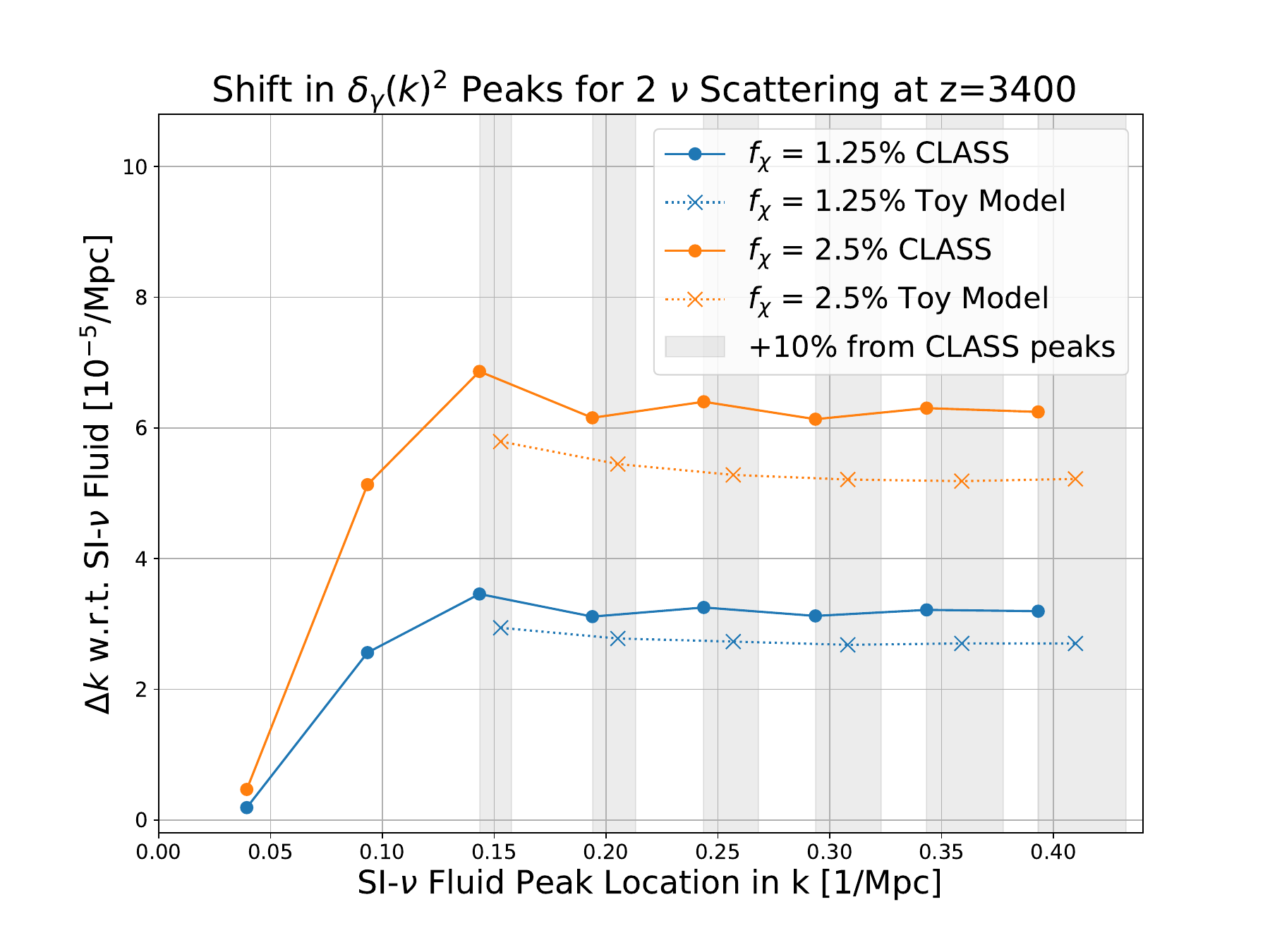}
  \caption{Equilibrium ($z=3400$)}
  \label{EQBnu1}
\end{subfigure}
\vfil
\begin{subfigure}{\textwidth}
  \centering
  \includegraphics[width=0.495\linewidth]{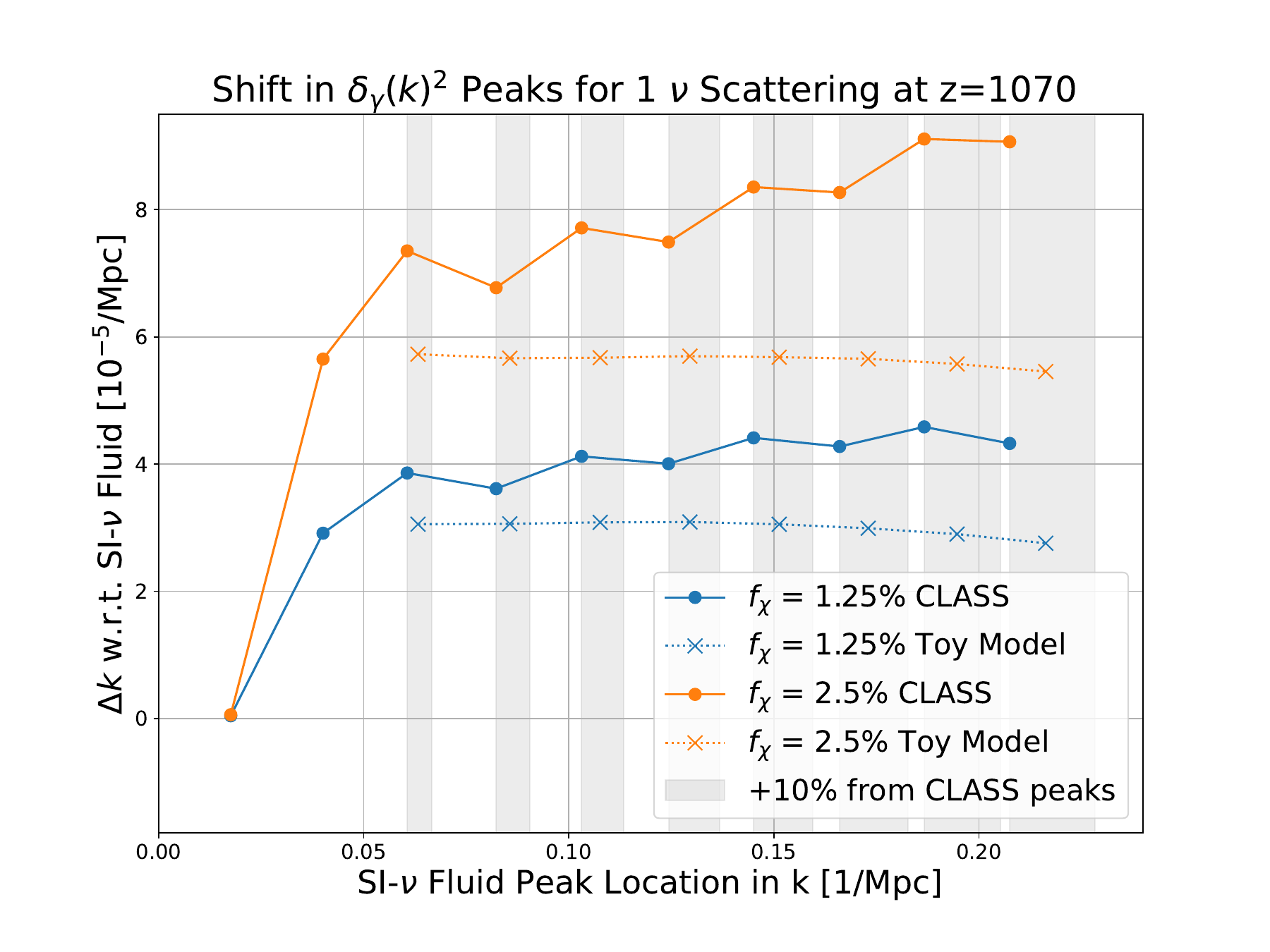}
  \centering
  \includegraphics[width=0.495\linewidth]{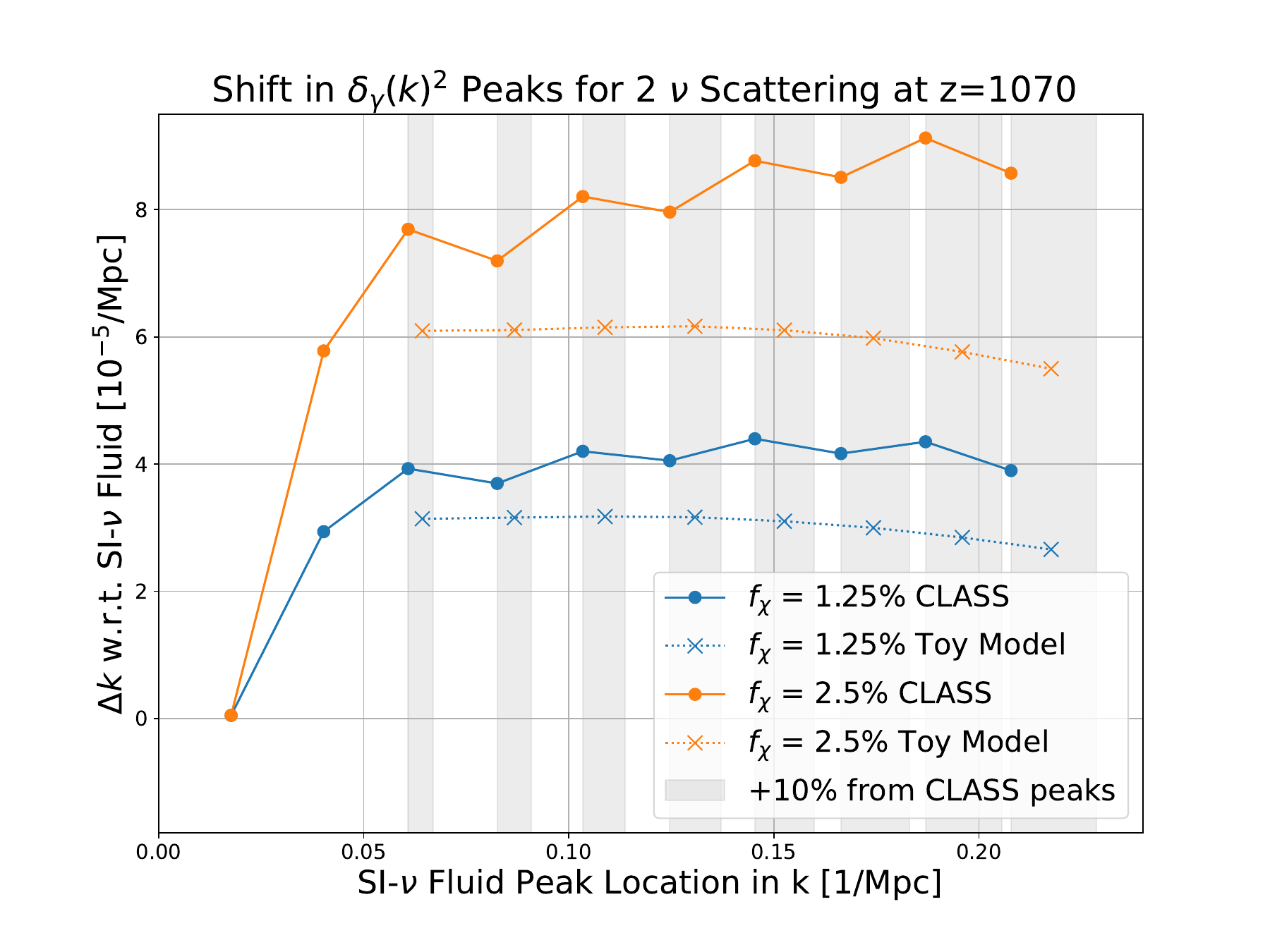}
  \caption{Recombination ($z=1070$)}
  \label{RECnu1}
\end{subfigure}
\caption{\CLASS~(solid) vs toy model (dotted) comparison of $\Delta k$ in photon transfer function for $f_\chi = 1.25\%$ (blue) and $2.5\%$ (orange) DL-$\nu$ with respect to SI-$\nu$. Same as Figs.~\ref{dknu3RAD} and \ref{dknu3EQB}, but for $1$ interacting neutrino (left column) and $2$ interacting neutrinos (right column).}
\label{dknu1}
\end{figure}

Fig.~\ref{dknu1} shows the shift of transfer function peaks for the cases where either one or two neutrinos scatter with DM. For the \CLASS~calculation, the rest of the neutrinos free-stream, while in the toy model only the fraction $f_\nu$ of interacting neutrinos is accounted for. Notably, the toy model peak locations and $\Delta k$ enhancement in DL-$\nu$ with respect to SI-$\nu$ maintains a similar level of agreement with \CLASS~as compared to the three neutrino scattering case presented in the main text (Figs.~\ref{dknu3RAD} and \ref{dknu3EQB}). 

\section{Tables and triangle plots for MCMC analysis}\label{app.table}
In this appendix, we provide the triangle plots and tables of mean values with $1\sigma$ errors for the MCMC analysis done in sections~\ref{sec:MCMC-DRDM} and \ref{sec:MCMC-NUDM}. We also provide best fit values for the scans where we varied $N_{\rm DR}$ and $f_\chi$ (Table~\ref{tab:DMDR}) and $yf_\chi$ and $f_\chi$ (Table~\ref{tab:DMNU-dec}). The $\Delta\chi^2$ with respect to the $\Lambda{\rm CDM}$ for each model fit is also provided in the tables. 

\vspace{5mm}

\begin{table}[H]
    \centering
    \begin{tabular}{|c|>{\centering\arraybackslash}p{20mm}|>{\centering\arraybackslash}p{35mm}|>{\centering\arraybackslash}p{20mm}|>{\centering\arraybackslash}p{35mm}|} \hline  
          &   \multicolumn{2}{|c|}{Planck + BAO}&  \multicolumn{2}{|c|}{Planck + BAO + Ext}\\ \hline 
 Parameter& Best Fit& Mean$\pm\sigma$& Best Fit&Mean$\pm\sigma$\\ \hline  
         $10^{-2}\omega{}_{b }$&   $2.248$&$2.241^{+0.014}_{-0.017}$&   $2.276$&$2.280\pm 0.015$\\ 
         $\omega{}_{cdm }$&   $0.11966$&$0.12115^{+0.00097}_{-0.0024}$&  $0.1263$&$0.1258\pm 0.0027$\\ 
         $n_{s }        $&   $0.9666$&$0.9663\pm 0.0039       $&  $0.9752$&$0.9728\pm 0.0037   $\\  
         $H_0{\rm(km/s/Mpc)}$&   $68.15$&$68.31^{+0.39}_{-0.83}  $&  $71.43$&$71.10\pm 0.82        $\\   
         $10^{-9}A_{s } $&   $2.098$&$2.106^{+0.028}_{-0.033}$&  $2.086$&$2.099^{+0.029}_{-0.034}$\\   
         $\tau{}_{reio }$&   $0.0544$&$0.0557^{+0.0067}_{-0.0078}$&  $0.0573$&$0.0583^{+0.0068}_{-0.0079}$\\    
         $f_\chi       $&    $0.853$&$< 0.248                $&  $0.00162$&$< 0.00370     $\\   
         $N_{\rm DR }$&   $8\times 10^{-5}$&$< 0.0605                 $&   $0.44$&$0.39\pm 0.13  $\\   \hline
         $\sigma_8   $&   $0.817$&$0.802^{+0.027}_{-0.0095} $&  $0.8245$&$0.8208^{+0.0088}_{-0.0077}$\\    
 $100\theta{}_{s }$&  $1.04194$&$1.04202\pm 0.00032  $& $1.04233$&$1.04247\pm 0.00031        $\\\hline
 $\Delta\chi^2 $&   \multicolumn{2}{|c|}{$-1.38      $}& \multicolumn{2}{|c|}{$-11.5  $}\\ \hline
    \end{tabular}
    \caption{Strong DR-DM scenario with varying $N_{\rm DR}$ and $f_\chi$: best fits and mean values to 68\% confidence. The $\Delta \chi^2 = \chi^2 - \chi^2_{\Lambda{\rm CDM}}$ for each fit is also provided. 
    }
    \label{tab:DMDR}
\end{table}

\begin{table}[H]
    \centering
    \begin{tabular}{|c|>{\centering\arraybackslash}p{20mm}|>{\centering\arraybackslash}p{35mm}|>{\centering\arraybackslash}p{20mm}|>{\centering\arraybackslash}p{35mm}|} \hline  
          &   \multicolumn{2}{|c|}{Planck + BAO}&  \multicolumn{2}{|c|}{Planck + BAO + Ext}\\ \hline 
 Parameter& Best Fit& Mean$\pm\sigma$& Best Fit&Mean$\pm\sigma$\\ \hline  
         $10^{-2}\omega{}_{b }$&   $2.245$&$2.244\pm 0.014            $&   $2.248$&$2.260\pm 0.014            $\\    
         $\omega{}_{cdm }$&   $0.1204$&$0.1216\pm 0.0013          $&  $0.1202$&$0.1197\pm 0.0011          $\\  
         $n_{s }        $&   $0.9504$&$0.9488^{+0.0040}_{-0.0073}$&  $0.9466$&$0.9502^{+0.0037}_{-0.0057}$\\  
         $H_0{\rm(km/s/Mpc)}$&   $68.64$&$68.25^{+0.47}_{-0.53}     $&  $69.27$&$69.40\pm 0.47             $\\   
         $10^{-9}A_{s } $&   $2.012$&$2.028^{+0.038}_{-0.047}   $&  $1.990$&$2.009^{+0.034}_{-0.045}   $\\   
         $\tau{}_{reio }$&   $0.0516$&$0.0536\pm 0.0070          $&  $0.0546$&$0.0562\pm 0.0072          $\\   
         $f_\chi       $&    $0.0048$&$< 0.0217                  $&  $0.0037$&$< 0.0111                  $\\    
         $y*f_\chi                  $&   $0.109$&$0.141^{+0.058}_{-0.13}    $&   $0.27$&$0.23^{+0.10}_{-0.16} $\\   \hline
         $\sigma_8   $&   $0.824$&$0.818^{+0.013}_{-0.0084}  $&  $0.826$&$0.819^{+0.010}_{-0.0065}  $\\    
 $100\theta{}_{s }$&  $1.0442$&$1.0444^{+0.0015}_{-0.0011}$& $1.0458$&$1.0454^{+0.0013}_{-0.00082}$\\\hline
 $\Delta\chi^2 $&   \multicolumn{2}{|c|}{$+0.28$}& \multicolumn{2}{|c|}{$-7.4$}\\ \hline
    \end{tabular}
    \caption{Varying $\nu$-DM scenario with varying $yf_\chi$ and $f_\chi$: best fits and mean values to 68\% confidence. The $\Delta \chi^2 = \chi^2 - \chi^2_{\Lambda{\rm CDM}}$ for each fit is also provided. }
    \label{tab:DMNU-dec}
\end{table}

\newpage

\begin{table}[H]
    \centering
    \begin{tabular}{|c|>{\centering\arraybackslash}p{29mm}|>{\centering\arraybackslash}p{29mm}|>{\centering\arraybackslash}p{29mm}|>{\centering\arraybackslash}p{29mm}|} \hline 
 & \multicolumn{4}{|c|}{Mean$\pm\sigma$ for fixed $f_\chi$ (Planck + BAO) }\\ \hline  
          Parameter &  $f_\chi = 10^{-4}$& $f_\chi = 0.01$& $f_\chi = 0.02$&$f_\chi = 0.03$\\ \hline  
         $10^{-2}\omega{}_{b }$&  $2.240\pm 0.014            $&   $2.242\pm 0.013            $& $2.244\pm 0.014            $&$2.247\pm 0.014            $\\ 
         $\omega{}_{cdm }$&  $0.1205\pm 0.0010          $&  $0.1215\pm 0.0011          $& $0.1220^{+0.0013}_{-0.0011}$&$0.1223^{+0.0015}_{-0.0012}$\\    
         $n_{s }        $&  $0.9484^{+0.0044}_{-0.0085}$&  $0.9478^{+0.0039}_{-0.0065}$& $0.9485^{+0.0038}_{-0.0065}$&$0.9495^{+0.0038}_{-0.0071}$\\   
         $H_0{\rm(km/s/Mpc)}$&  $68.55\pm 0.52             $&  $68.36\pm 0.46             $& $68.16\pm 0.44             $&$68.03\pm 0.43             $\\   
         $10^{-9}A_{s } $&  $2.020^{+0.040}_{-0.050}   $&  $2.019^{+0.034}_{-0.045}   $& $2.027^{+0.034}_{-0.041}   $&$2.037^{+0.034}_{-0.043}   $\\    
         $\tau{}_{reio }$&  $0.0524\pm 0.0069          $&  $0.0532\pm 0.0070          $& $0.0536\pm 0.0070          $&$0.0541\pm 0.0070          $\\      
         $y*f_\chi                  $&  $0.152^{+0.060}_{-0.15}    $&   $0.162^{+0.042}_{-0.16}    $& $0.134^{+0.038}_{-0.13}    $&$0.110^{+0.038}_{-0.11}    $\\  \hline
         $\sigma_8   $&  $0.8299\pm 0.0066          $&  $0.8230\pm 0.0062          $& $0.8158\pm 0.0062          $&$0.8089\pm 0.0061          $\\    
 $100\theta{}_{s }$& $1.0443^{+0.0016}_{-0.0012}$& $1.0446^{+0.0014}_{-0.00098}$& $1.0445^{+0.0014}_{-0.00091}$&$1.0443^{+0.0015}_{-0.00094}$\\\hline
 $\Delta\chi^2 $&  $-0.62$& $-0.08$& $+0.32$&$+1.44$\\\hline
    \end{tabular}
    \caption{Varying $\nu$-DM scenario for fixed $f_\chi$ (Planck + BAO): mean values to 68\% confidence. We vary $y$ for each fixed $f_\chi$, but present the $y*f_\chi$ values for meaningful comparison between the cases. The $\Delta \chi^2 = \chi^2 - \chi^2_{\Lambda{\rm CDM}}$ for each fit is also provided.}
    \label{tab:DMNU-fix-planck}
\end{table}
\vspace{10mm}
\begin{table}[H]
    \centering
    \begin{tabular}{|c|>{\centering\arraybackslash}p{29mm}|>{\centering\arraybackslash}p{29mm}|>{\centering\arraybackslash}p{29mm}|>{\centering\arraybackslash}p{29mm}|} \hline 
 & \multicolumn{4}{|c|}{Mean$\pm\sigma$ for fixed $f_\chi$ (Planck + BAO + Ext)}\\ \hline  
          Parameter &  $f_\chi = 10^{-4}$& $f_\chi = 0.01$& $f_\chi = 0.02$&$f_\chi = 0.03$\\ \hline  
         $10^{-2}\omega{}_{b }$&  $2.258\pm 0.014            $&   $2.259\pm 0.013            $& $2.263\pm 0.013            $&$2.265\pm 0.014            $\\ 
         $\omega{}_{cdm }$&  $0.11919\pm 0.00097        $&  $0.11988\pm 0.00099        $& $0.1203^{+0.0012}_{-0.0010}$&$0.1203^{+0.0016}_{-0.0012}$\\  
         $n_{s }        $&  $0.9482^{+0.0039}_{-0.0049}$&  $0.9504^{+0.0037}_{-0.0051}$& $0.9520^{+0.0035}_{-0.0059}$&$0.9542^{+0.0037}_{-0.0077}$\\    
         $H_0{\rm(km/s/Mpc)}$&  $69.64\pm 0.45             $&  $69.34\pm 0.43             $& $69.11\pm 0.41             $&$68.96\pm 0.41             $\\  
         $10^{-9}A_{s } $&  $1.992^{+0.033}_{-0.041}   $&  $2.010^{+0.032}_{-0.042}   $& $2.024^{+0.031}_{-0.043}   $&$2.042^{+0.037}_{-0.045}   $\\   
         $\tau{}_{reio }$&  $0.0541\pm 0.0071          $&  $0.0566\pm 0.0072          $& $0.0572\pm 0.0070          $&$0.0576^{+0.0068}_{-0.0077}$\\  
         $y*f_\chi                  $&  $0.29^{+0.11}_{-0.17}      $&   $0.231^{+0.090}_{-0.15}    $& $0.174^{+0.076}_{-0.15}    $&$0.121^{+0.048}_{-0.12}    $\\ \hline
         $\sigma_8   $&  $0.8266\pm 0.0063          $&  $0.8194\pm 0.0061          $& $0.8121\pm 0.0059          $&$0.8048\pm 0.0063          $\\  
 $100\theta{}_{s }$& $1.0456^{+0.0011}_{-0.00082}$& $1.0454^{+0.0012}_{-0.00080}$& $1.0451^{+0.0014}_{-0.00081}$&$1.0446^{+0.0017}_{-0.00098}$\\\hline
 $\Delta\chi^2 $&  $-6.86$& $-5.74$& $-5.50$&$-3.86$\\\hline
    \end{tabular}
    \caption{Varying $\nu$-DM scenario for fixed $f_\chi$ (Planck + BAO + Ext): mean values to 68\% confidence. We vary $y$ for each fixed $f_\chi$, but present the $y*f_\chi$ values for meaningful comparison between the cases. The $\Delta \chi^2 = \chi^2 - \chi^2_{\Lambda{\rm CDM}}$ for each fit is also provided.}
    \label{tab:DMNU-fix-sh0es}
\end{table}

\newpage

\begin{figure}[H]
    \centering
    \includegraphics[width=\linewidth]{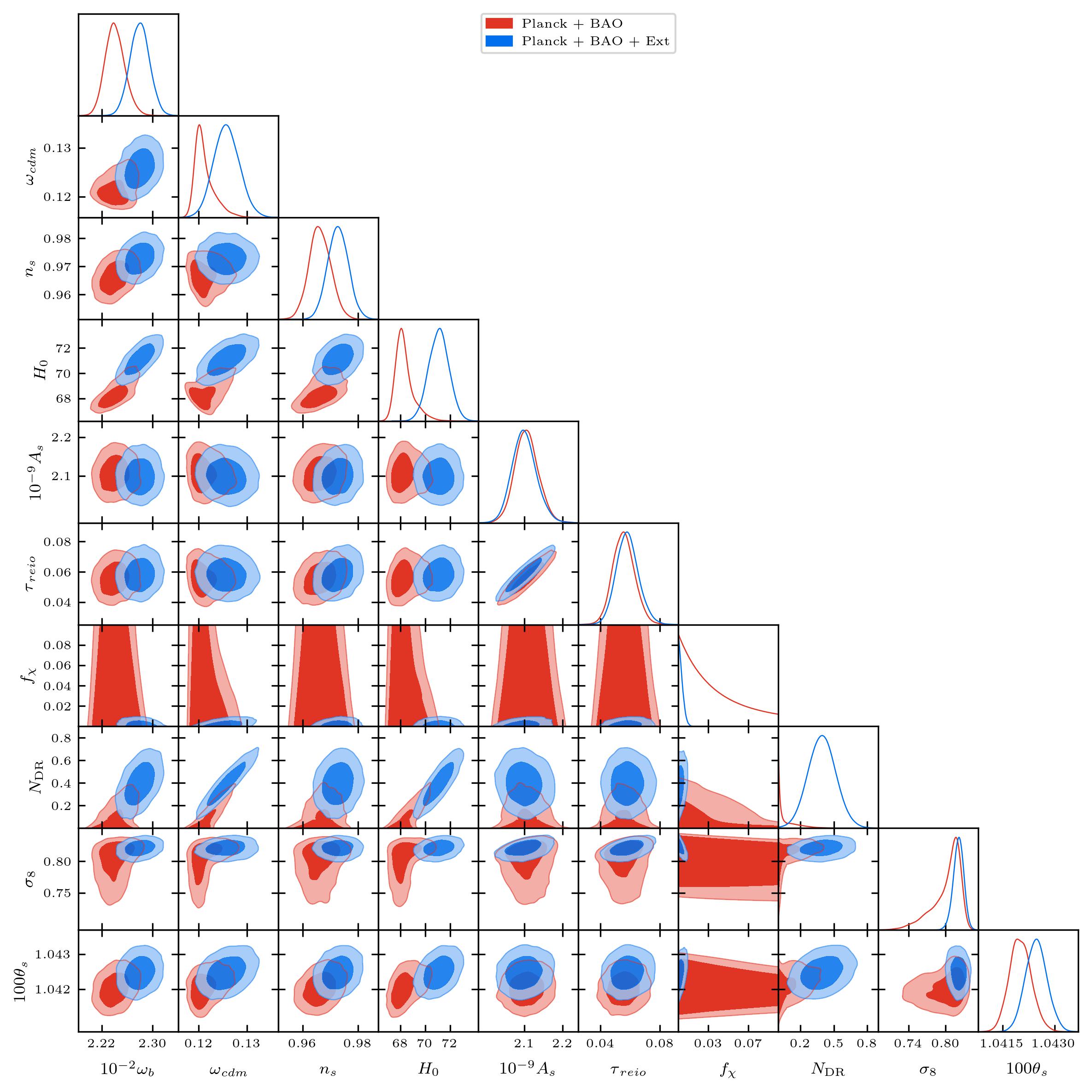}
    \caption{Strong DR-DM scenario with varying $N_{\rm DR}$ and $f_\chi$}
    \label{fig:DMDR-triangle}
\end{figure}

\newpage

\begin{figure}[H]
    \centering
    \includegraphics[width=\linewidth]{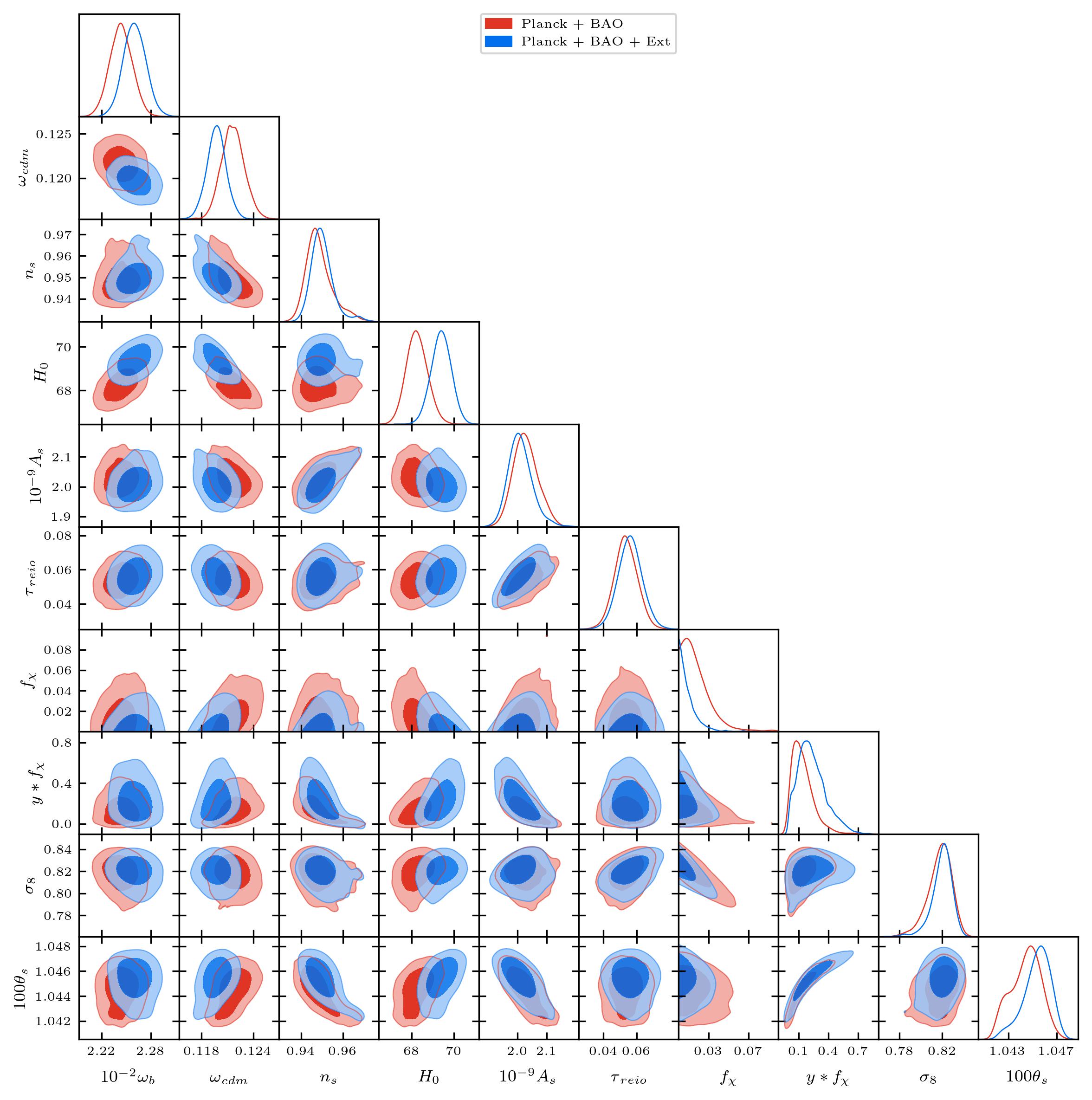}
    \caption{Varying $\nu$-DM scenario with varying $yf_\chi$ and $f_\chi$.}
    \label{fig:DMNU-dec-triangle}
\end{figure}

\newpage

\begin{figure}[H]
    \centering
    \includegraphics[width=\linewidth]{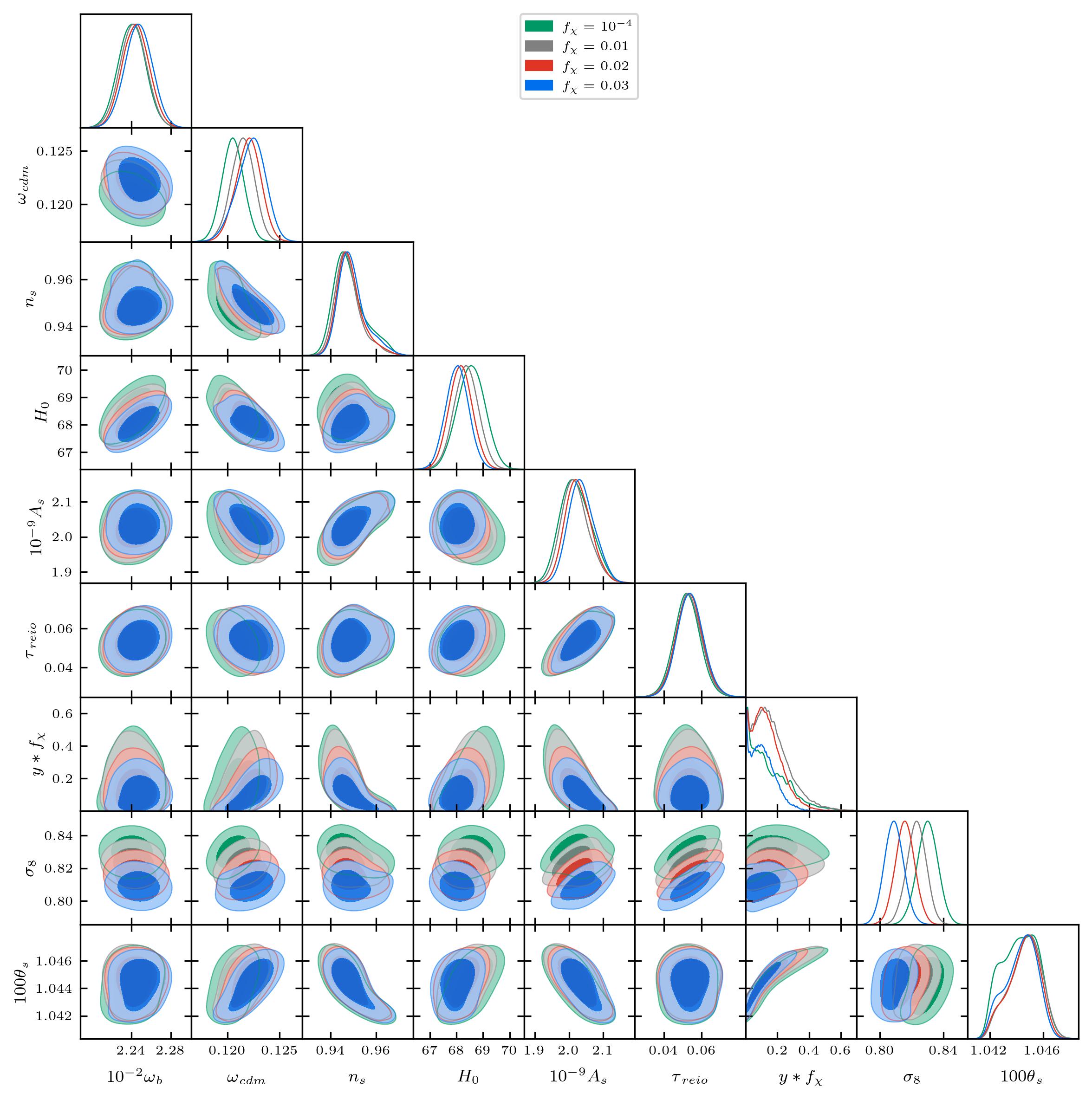}
    \caption{Varying $\nu$-DM scenario for fixed $f_\chi$ (Planck + BAO). Here $y$ is varied for each fixed $f_\chi$, but we present the $y*f_\chi$ values for meaningful comparison between the cases. }
    \label{fig:DMNU-fix-planck-triangle}
\end{figure}

\newpage

\begin{figure}[H]
    \centering
    \includegraphics[width=\linewidth]{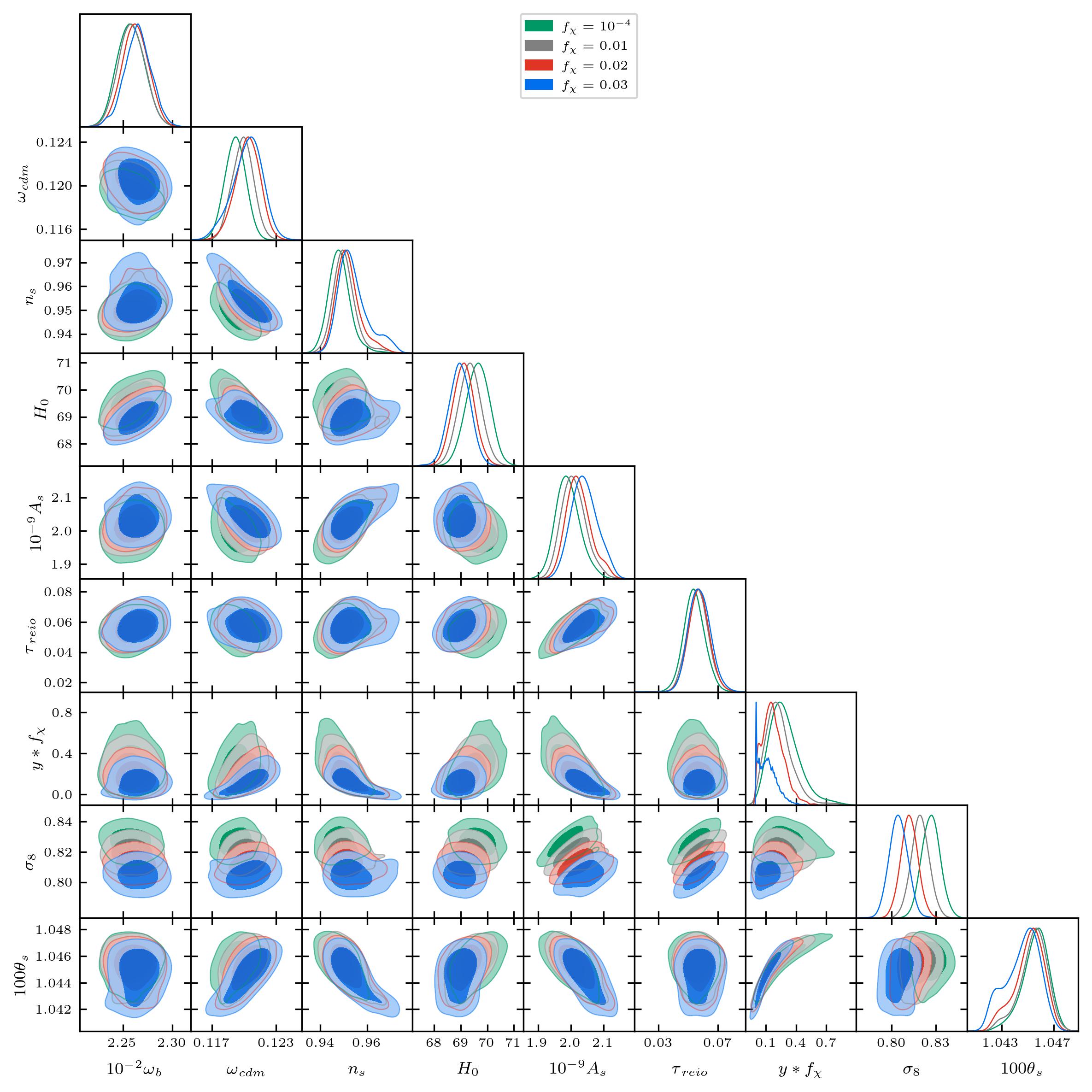}
    \caption{Varying $\nu$-DM scenario for fixed $f_\chi$ (Planck + BAO + Ext).  Here $y$ is varied for each fixed $f_\chi$, but we present the $y*f_\chi$ values for meaningful comparison between the cases. }
    \label{fig:DMNU-fix-sh0es-triangle}
\end{figure}

\newpage

\bibliography{PS-ref}
\bibliographystyle{JHEP}

\end{document}